\documentclass[fleqn,usenatbib]{mnras}

\usepackage{newtxtext,newtxmath,xcolor}

\usepackage[T1]{fontenc}

\DeclareRobustCommand{\VAN}[3]{#2}
\let\VANthebibliography\thebibliography
\def\thebibliography{\DeclareRobustCommand{\VAN}[3]{##3}\VANthebibliography}

\usepackage{graphicx}	% Including figure files
\usepackage{amsmath}	% Advanced maths commands
\usepackage{lineno}
\usepackage{physics, amssymb}	% Extra maths symbols
\usepackage{lipsum}
\usepackage{mathtools}
\usepackage{cuted}
\usepackage{fancyvrb}
\usepackage[normalem]{ulem}

\usepackage{eso-pic}% http://ctan.org/pkg/eso-pic
\AddToShipoutPictureBG*{%
  \AtPageUpperLeft{%
    \hspace{0.92\paperwidth}%
    \raisebox{-4.5\baselineskip}{%
      \makebox[0pt][r]{\textnormal{DES-2023-0809}}
}}}%

\AddToShipoutPictureBG*{%
  \AtPageUpperLeft{%
    \hspace{0.92\paperwidth}%
    \raisebox{-5.5\baselineskip}{%
      \makebox[0pt][r]{\textnormal{FERMILAB-PUB-25-0011-PPD}}
}}}%

\let\mdet\relax
\newcommand{\mfrac}{\texttt{mfrac}}
\newcommand{\mdet}{\textsc{Metadetection}}
\newcommand{\mcal}{\textsc{Metacalibration}}
\newcommand{\bfd}{\texttt{BFD}}
\newcommand{\anacal}{\texttt{AnaCal}}
\newcommand{\gold}{\texttt{Gold}}
\newcommand{\gaia}{\textit{Gaia}}
\newcommand{\nsidehi}{131072}
\title[DES Y6 \mdet\ Shape Catalogue]{Dark Energy Survey Year 6 Results: Cell-based Coadds and \mdet\ Weak Lensing Shape Catalogue}
\author[Yamamoto, Becker, Sheldon, Jarvis et al.]{
\parbox{\textwidth}{
\Large
M.~Yamamoto,$^{1,2}$\thanks{E-mail: masaya.yamamoto@princeton.edu}
M.~R.~Becker,$^{3}$
E.~Sheldon,$^{4}$
M.~Jarvis,$^{5}$
R.~A.~Gruendl,$^{6,7}$
F.~Menanteau,$^{6,7}$
E.~S.~Rykoff,$^{8,9}$
S.~Mau,$^{10,8}$
T.~Schutt,$^{10,8,9}$
M.~Gatti,$^{5,11}$
M.~A.~Troxel,$^{2}$
A.~Amon,$^{1}$
D.~Anbajagane,$^{11}$
G.~M.~Bernstein,$^{5}$
D.~Gruen,$^{12}$
E.~M.~Huff,$^{13}$
M.~Tabbutt,$^{14}$
A.~Tong,$^{5,2}$
B.~Yanny,$^{15}$
T.~M.~C.~Abbott,$^{16}$
M.~Aguena,$^{17}$
A.~Alarcon,$^{18}$
F.~Andrade-Oliveira,$^{19}$
K.~Bechtol,$^{14}$
J.~Blazek,$^{20}$
D.~Brooks,$^{21}$
A.~Carnero~Rosell,$^{22,17,23}$
J.~Carretero,$^{24}$
C.~Chang,$^{25,11}$
A.~Choi,$^{26}$
M.~Costanzi,$^{27,28,29}$
M.~Crocce,$^{30,18}$
L.~N.~da Costa,$^{17}$
T.~M.~Davis,$^{31}$
J.~De~Vicente,$^{32}$
S.~Desai,$^{33}$
H.~T.~Diehl,$^{15}$
S.~Dodelson,$^{25,15,11}$
P.~Doel,$^{21}$
C.~Doux,$^{5,34}$
A.~Drlica-Wagner,$^{25,15,11}$
A.~Fert\'e,$^{9}$
B.~Flaugher,$^{15}$
P.~Fosalba,$^{30,18}$
J.~Frieman,$^{25,15,11}$
J.~Garc\'ia-Bellido,$^{35}$
E.~Gaztanaga,$^{30,36,18}$
G.~Giannini,$^{24,11}$
G.~Gutierrez,$^{15}$
W.~G.~Hartley,$^{37}$
K.~Herner,$^{15}$
S.~R.~Hinton,$^{31}$
D.~L.~Hollowood,$^{38}$
K.~Honscheid,$^{39,40}$
D.~Huterer,$^{41}$
E.~Krause,$^{42}$
K.~Kuehn,$^{43,44}$
O.~Lahav,$^{21}$
M.~Lima,$^{45,17}$
J.~L.~Marshall,$^{46}$
J. Mena-Fern{\'a}ndez,$^{47}$
R.~Miquel,$^{48,24}$
J.~J.~Mohr,$^{49,12}$
J.~Muir,$^{50,51}$
J.~Myles,$^{1}$
R.~L.~C.~Ogando,$^{52}$
A.~Pieres,$^{17,52}$
A.~A.~Plazas~Malag\'on,$^{8,9}$
A.~Porredon,$^{32,53}$
J.~Prat,$^{25,54}$
M.~Raveri,$^{55}$
M.~Rodriguez-Monroy,$^{35}$
A.~Roodman,$^{8,9}$
S.~Samuroff,$^{20,24}$
E.~Sanchez,$^{32}$
D.~Sanchez Cid,$^{32,19}$
V.~Scarpine,$^{15}$
I.~Sevilla-Noarbe,$^{32}$
M.~Smith,$^{56}$
M.~Soares-Santos,$^{19}$
E.~Suchyta,$^{57}$
G.~Tarle,$^{41}$
V.~Vikram,$^{58}$
N.~Weaverdyck,$^{59,60}$
P.~Wiseman,$^{61}$
and Y.~Zhang$^{16}$
\begin{center} (DES Collaboration) \end{center}
}
\vspace{0.4cm}
\\
\parbox{\textwidth}{
The authors' affiliations are shown in Appendix \ref{sec:affiliations}.
}
}

\date{Accepted XXX. Received YYY; in original form ZZZ}

\pubyear{2025}

\begin{document}

\label{firstpage}
\pagerange{\pageref{firstpage}--\pageref{lastpage}}
\maketitle

\begin{abstract}
We present the \mdet\ weak lensing galaxy shape catalogue from the six-year Dark Energy Survey (DES Y6) imaging data. This dataset is the final release from DES, spanning 4422 deg$^2$ of the southern sky. We describe how the catalogue was constructed, including the two new major processing steps, cell-based image coaddition and shear measurements with \mdet. The DES Y6 \mdet\ weak lensing shape catalogue consists of 151,922,791 galaxies detected over $riz$ bands, with an effective number density of $n_{\rm eff}$=8.22 galaxies per arcmin$^2$ and shape noise of $\sigma_{\rm e} = 0.29$. We carry out a suite of validation tests on the catalogue, including testing for PSF leakage, testing for the impact of PSF modeling errors, and testing the correlation of the shear measurements with galaxy, PSF, and survey properties. In addition to demonstrating that our catalogue is robust for weak lensing science, we use the DES Y6 image simulation suite to estimate the overall multiplicative shear bias of our shear measurement pipeline. We find no detectable multiplicative bias at the roughly half-percent level, with $m = (3.4 \pm 6.1) \times 10^{-3}$, at $3\sigma$ uncertainty. This is the first time both cell-based coaddition and \mdet\ algorithms are applied to observational data, paving the way to the Stage-IV weak lensing surveys.

\end{abstract}


%%%%%%%%%%%%%%%%%%%%%%%%%%%%%%%%%%%%%%%%%%%%%%%%%%

%%%%%%%%%%%%%%%%% BODY OF PAPER %%%%%%%%%%%%%%%%%%

\section{Introduction} \label{sec:intro}
% citation: \citep{lamport94}
% \matt{needs to be rephrased to be more inclusive of BFD}

Weak gravitational lensing -- a small deflection of light due to an intervening gravitational potential between its source and an observer --  is a well-established and competitive observational probe to measure how the structure in the Universe has evolved \citep*{2022PhRvD.105b3514A, y3_1x2pt_2, hscy3_1x2pt, hscy3_dalal, kids1000_1x2pt}.  Such analyses measure the tiny statistical patterns in aligned or correlated shapes of a vast number of galaxies.  To do so requires high-accuracy measurements of their shapes that have been extensively validated (for a review, see \citealt{2018ARA&A..56..393M}). As galaxy samples become larger with surveys such as Euclid\footnote{\url{ http://sci.esa.int/euclid}} (\citealt{2011arXiv1110.3193L}), the Vera C. Rubin Observatory Legacy Survey of Space and Time\footnote{\url{ http://www.lsst.org}} (\emph{LSST}: \citealt{2009arXiv0912.0201L, 2019ApJ...873..111I}), and the \emph{Nancy Grace Roman} Space Telescope\footnote{\url{https://roman.gsfc.nasa.gov}} (\emph{Roman}: \citealt{2015arXiv150303757S}), our systematic uncertainty requirements are more stringent, demanding improved shape measurement techniques.

The measured galaxy shapes can contain physical variance (intrinsic scatter) and non-cosmic distortions due to the atmosphere and instrument, thereby complicating the measurement of the true shear signal. To accurately measure a large number of galaxy shapes, one must accurately model the point spread function (PSF) in each image, otherwise the inferred shear field can be significantly skewed (e.g., \citealt{2010MNRAS.404..350R}). We assess the shear systematics by defining multiplicative, $m$, and additive bias, $c$, which modulate the relationship between observed and true shear (\citealt{2006MNRAS.368.1323H}):
\begin{equation}
    \gamma^{\rm obs}_{i} = (1+m_i)\gamma^{\rm true}_{i} + c_i.
\end{equation}
Non-zero $m$ and $c$ terms can be introduced due to imperfect (i) PSF modeling and (ii) shear estimation. During PSF modeling, sources such as inaccurate characterization of detector effects, impurity in stellar samples, and mismodeling of atmospheric effects can cause the misestimation of PSF models, and the PSF size and ellipticity errors result in $m$ and $c$ respectively. Even if PSF models are accurate, misevaluation of the PSF for individual galaxies during shear estimation can lead to significant multiplicative bias. We will explore this further in this paper. Furthermore, in the low signal-to-noise (S/N) regime, the pixel noise causes prominent ``noise bias'' (e.g., \citealt{2003MNRAS.343..459H, 2012MNRAS.427.2711K, 2012MNRAS.424.2757M, 2012MNRAS.425.1951R}), because shape estimation is a non-linear function of pixel intensities. Other well-known significant shear biases include ``model bias'' due to the mismatch of assumed and the actual galaxy light profile during shape estimation (e.g., \citealt{2014MNRAS.441.2528K, 2010A&A...510A..75M, 2010MNRAS.404..458V}) and ``selection bias'' that happens when the galaxy selections depend on shear (e.g., \citealt{2003MNRAS.343..459H, bernstein_jarvis_2002}).

In order to account for these biases, modern lensing analyses calibrate the shears that have been estimated. In this step, one measures the shear response $\partial e_i/\partial \gamma_j$, where the measured galaxy ellipticity ($e_i$) responds to an applied shear ($\gamma_j$). By determining the shear response for an ensemble of galaxies, one can calibrate the aforementioned biases over a wide range of galaxy populations. In the Shear TEsting Programme (STEP; \citealt{2006MNRAS.368.1323H, 2007MNRAS.376...13M}) and the GRavitational lEnsing Accuracy Testing challenges (GREAT; \citealt{2010MNRAS.405.2044B, 2013ApJS..205...12K, 2015MNRAS.450.2963M}), the lensing community has dedicated efforts to develop new unbiased shear estimation and calibration algorithms and mitigate/self-calibrate these biases for the Stage-III surveys to satisfy their systematics budgets such as the Hyper-Suprime Cam Subaru Strategic Program\footnote{\url{http://hsc.mtk.nao.ac.jp/ssp/}} (HSC: \citealt{2018PASJ...70S...4A}), the Kilo-Degree Survey\footnote{\url{http://kids.strw.leidenuniv.nl/}} (KiDS: \citealt{2013ExA....35...25D}) and the Dark Energy Survey\footnote{\url{http://www.darkenergysurvey.org/}} (DES: \citealt{2005astro.ph.10346T}).
We briefly introduce a few more recent algorithms that are relevant to this paper. The first widely utilized technique is \mcal ~\citep{2017ApJ...841...24S, 2017arXiv170202600H}. This method artificially applies a shear to each object and numerically measures the response. It can empirically correct for ``noise'', ``model'', and ``selection'' biases to a few percent (\citealt*{2017ApJ...841...24S, y3-shapecatalog}). Additionally, Bayesian Fourier Domain (\bfd; ~\citealt{bernstein_bfd1, bernstein_bfd2}) can independently self-calibrate the biases by constructing templates from deep-field galaxies and building the probability distribution of Fourier-space moments for wide-field galaxies. We also point out that \anacal, which computes the shear response of basis functions and propagates the shear response of basic modes to derive the shear response of the final observable, has shown the potential to calibrate shear calibration biases to a sub-percent level \citep{anacal_1, anacal_2}.

With this progress, a number of challenges remain, which motivate the advancements in this work. In PSF modeling, \cite{2015ApJ...807..182M} identified the differential chromatic refraction (DCR) and the wavelength dependence of seeing as primary drivers of PSF misestimation. These chromatic effects will be especially essential to be accounted for in Stage-IV surveys. Further, they have already been detected in previous DES analyses (\cite{2021MNRAS.501.1282J}, \cite*{y3-shapecatalog}; hereafter GS20). In shear estimation, as the number of single-epoch (SE) images increases for a given region on the sky, naive image coaddition introduces SE image edges into the final coadd image, which produce discontinuities in the coadded PSF. These PSF discontinuities can bias subsequent shape measurements performed on the coadd image \citep{2023OJAp....6E...5M}. This effect motivated the techniques that use galaxies detected in coadd images but with measurements taken from the SE images. Unfortunately, this approach has two main limitations. First, the computational cost of the shear measurement increases linearly with the number of SE images. While the DES has only $\sim$10 SE images overlapping a given point on the sky in a given band, Stage-IV surveys will, in some cases, have $\cal{O}$(100) SE images \citep{2009arXiv0912.0201L}. Second, detecting objects is a weak lensing shear-dependent operation, and ignoring this dependence causes significant biases in shear measurements (e.g., \citealt{2003MNRAS.343..459H, 2019A&A...624A..92K, 2020ApJ...902..138S}). Finally, as the depths of surveys increase, the fraction of objects that are blended increases as well. The increased blend fraction coupled with detection biases can cause catastrophic biases in shear measurements for future surveys (\citealt{2020ApJ...902..138S}; hereafter S20).

To address these issues, we have pursued a new approach to shear measurement which proceeds in two stages. First, we employ a \textit{cell-based coadding} algorithm to reprocess the entire DES survey. A \textit{cell-based coadd} in this work is defined as a small coadd image, approximately one arcminute on a side, within which none of the input SE images have edges. This kind of coadd has a well-defined PSF model \citep{2023OJAp....6E...5M, shearoncoadds} and so is suitable for weak lensing shape measurements. We lose some survey depth, unfortunately, in the process of constructing cell-based coadds since not all input SE images fully overlap a given region on the sky \citep{shearoncoadds}. This trade-off is discussed in more detail below. Second, we run the \mdet\ weak lensing shear estimation algorithm over the cell-based coadd images. The application of \mdet\ on simulated Rubin data by \cite{sheldon_mdet_rubin} has demonstrated that \mdet\ can produce unbiased shear measurements in the presence of detection and blending at Rubin LSST ten-year depths \citep{2009arXiv0912.0201L} to a fractional accuracy better than $\sim0.1\%$. While this level of accuracy is not required for DES Y6 cosmological analyses, we aim for $m$ around 1.3\% or smaller. This level, according to the LSST DESC Science Requirements Document \citep{lsst_srd}, would be sufficient for LSST Year 1 analysis and certainly for DES Year 6 analysis. The previous DES Y3 \mcal\ pipeline \citep*{y3-shapecatalog} produced significant enough biases due to blending to shift the Y3 cosmological contours by $\approx1\sigma$ \citep{2022MNRAS.509.3371M, 2022PhRvD.105b3514A} if not corrected.

As for additive bias, previous DES analyses have found non-negligible additive bias in DES data \citep*{y3-shapecatalog} that is beyond cosmic variance and above what was measured in image simulations \citep*{2022MNRAS.509.3371M}. The origin of this bias is unknown, and \cite{2022PhRvD.105b3514A} chose to subtract it from the data before measuring the shear two-point correlation functions. This procedure is valid under the assumption that the bias is actually constant and when the two-point function measurements are restricted to angular scales where the two-point function amplitudes are significantly larger than the additive signal squared. For non-constant additive signals that are driven by PSF modeling and deconvolution, one can use the data themselves to measure the additive bias signals. For the DES Y3 shear catalog, these biases were shown to be small and could thus be safely ignored. Alternatively, one can use the additive bias measurements to directly marginalize over them during the cosmological analysis \citep{hscy3_1x2pt}. For DES Y6, we will employ similar procedures to account for any residual additive biases, and will also attempt to reduce them using new PSF models with chromatic dependence \citep{y6psf}.

In this paper, we present the DES Y6 shape \mdet\ catalogue and our implementation of the two-stage pipeline described above. We document extensive validation tests based both on DES data and on image simulations (where we can test absolute calibration). The tests with DES data are phrased as empirical `null tests,' and are used to search for unexpected systematics.

For the image simulations, we use an updated version of the DES Y3 image simulation code \citep{2022MNRAS.509.3371M}, to be presented by \citealt*{y6imagesims}. Finally, this paper is further accompanied by \citealt{y6psf}, which describes the upgraded PSF modeling with {\sc Piff} (PSFs in the Full FOV; \citealt{2021MNRAS.501.1282J}) that includes color-dependent modeling. This work reports several diagnostic tests on the models created for Y6 data, showing better PSF models in all $griz$ bands compared to Y3 \citep{2021MNRAS.501.1282J}. Finally, the DES collaboration has a second shear catalogue in preparation using the \bfd\ technique \citep*{y6bfd}, and this catalogue will eventually act as a strong cross-check on this work.

This paper is organized as follows. We describe the DES data in Sec.~\ref{sec:data}. Section~\ref{sec:coadd} describes the cell-based coadd process, while  Sec.~\ref{sec:mdet} describes the \mdet\ process. Sec~\ref{sec:psfsyst} \& \ref{sec:shearsyst} describe our data-based validation tests on the catalogue. Sec.~\ref{sec:imagesims} describes tests of our methods in simulations. We conclude in Sec.~\ref{sec:conclusion}.

\section{Data} \label{sec:data}

\subsection{DES Data} \label{subsec:des}

The Dark Energy Survey images were taken with the Dark Energy Camera (\citealt{2015AJ....150..150F}) mounted on the Blanco telescope at the Cerro Tololo Inter-American Observatory \citep{diehl2012dark}. The survey was operated over six seasons from August 15, 2013 to January 9, 2019 and imaged an area of approximately 5000 deg$^2$ in the southern sky. The Y6 footprint is approximately the same as the Y3 footprint, but deeper up to $i_{\rm AB} \sim 23.4$, which is the median depth for extended objects at S/N $\sim$ 10 \citep{y6gold}, instead of $i_{\rm AB} \sim 23.0$ for Y3 \citep{y3gold}. The total DES survey resulted in 76,217 pointings spread across the $grizY$ photometric bands. This data was released as Data Release 2 (DR2: \citealt{2021ApJS..255...20A}). From DR2, the survey detected about 691 million astronomical objects. The DES collaboration curated the highest-quality objects into the DES Y6 Gold catalogue (hereafter \gold; \citealt{y6gold}).

We will use the following data products to build our shear catalogue. These data were generated during the standard DES processing described in \citet{desdm}.
\begin{itemize}
    \item The SE images: The $griz$-band calibrated images from the DES processing pipeline, created by DES Data Management (DESDM) \citep{desdm}.

    \item The {\sc Pixmappy}\footnote{\url{https://github.com/gbernstein/pixmappy}} astrometric solutions: These are important because they incorporate multiple astrometric effects, including tree rings \citep{treerings} and chromatic distortions \citep{des_astrometry}.

    \item The {\sc Piff} PSF models: These are updated to account for chromatic effects for the DES Y6 analysis \citep{y6psf}. We will describe how we incorporate the color dependence of the PSF model into \mdet\ in Sec.~\ref{sec:coadd}.

    \item The Forward Global Calibration Method (FGCM) \citep{fgcm} photometric zero points from the DES \gold\ catalogue.

    \item The background images, bad pixel masks, and weight maps: These are produced by DESDM for each single-epoch image.
\end{itemize}

\subsection{Mock Catalogues} \label{subsec:mocks}
In some validation tests of the \mdet\ shear catalogue, we require covariance matrices for statistics extracted from complex operations on the data. These covariance matrices need to include both galaxy shape noise and the underlying sample variance contributions from large-scale structure. For these kinds of tests, mock shape catalogues were generated from the shear maps in gravity-only N-body simulations. We make use of {\sc CosmogridV1} \citep{2022PhRvD.105h3518F}. The dataset is described in \cite{2023JCAP...02..050K}. We use 200 independent realisations at the fixed cosmology ($\Omega_{\rm m} = 0.26$, $\sigma_8 = 0.84$, $\Omega_{\rm b} = 0.0493$, $n_{\rm s} = 0.9649 $, $h = 0.673$). The simulations provide 69 full-sky lens planes between redshift 0 and redshift 3.5 with HEALPix\footnote{http://healpix.sourceforge.net} \citep{2005ApJ...622..759G} {\tt nside}=2048; the lens planes are converted into convergence planes under the Born approximation (e.g., Eq. 2 from \citealt{Fosalba2015}), and shear planes are obtained from the convergence maps using a full-sky generalisation of the \cite{KaiserSquires} algorithm \citep*{y1massmaps, y3-massmapping}. Mock DES Y6 shear fields are obtained by integrating the shear planes assuming preliminary DES Y6 redshift distributions; last, we create mock shear catalogues by sampling the simulated shear fields at the DES Y6 \mdet\ galaxy positions and adding DES Y6 shape noise. This last step is performed by randomly rotating the DES Y6 galaxy ellipticities and adding them to the simulated shear field sampled at each galaxy position. Since we can cut four DES Y6 footprints out of each simulation, this procedure provides 800 independent shear catalogues that can be used to infer covariance matrices and perform statistical tests.

\subsection{Catalogue Blinding} \label{subsec:blind}
In order to prevent accidental unblinding of cosmological information during testing, we applied a random scaling factor to the catalogue. To do this, we followed \citealt*{2018MNRAS.481.1149Z, y3-shapecatalog} to first transform ellipticity $\vb{e}$ into |$\vb{\eta}$| $\equiv$ 2arctanh(|$\vb{e}$|) and multiplying |$\vb{\eta}$| by a blinding factor $f$ (0.9<$f$<1.1). This random factor was removed once we completed the initial validation of our catalogue using the null tests detailed below.

\section{Cell-based Coadding Algorithms} \label{sec:coadd}

The DES Y6 shape catalogue is generated in two steps: first, we build cell-based coadds, as described in this section, and second we measure shapes with \mdet\ (Section~\ref{sec:mdet}). The cell-based coadding method is the precursor of that developed for Rubin LSST, presented in \citet{sheldon_mdet_rubin}. The versions of key software packages used for each step of the pipeline are listed in Appendix~\ref{app:software}.

\subsection{Coadd Image Geometry}
To produce cell-based coadds for DES, we take advantage of the definition of a DES coadd tile as a 10,000 $\times$ 10,000 pixel image with an associated coordinate transformation from pixel locations to sky locations (i.e., its WCS transformation). Approximately 10,000 of these coadd tiles cover the entire DES survey footprint. Each coadd tile slightly overlaps adjacent tiles and comes with a specified range in right ascension and declination that it uniquely covers on the sky. Within each coadd tile, we define many cell-based coadds of size 200 $\times$ 200 pixels. These regions are large enough to limit the loss of depth due to excluding partial overlaps of the SE images, but small enough that we can approximate the PSF and WCS as constant over the cell \citep{shearoncoadds,sheldon_mdet_rubin}. We lay out the cell-based coadds so that they start 50 pixels within the edge of a coadd tile and have unique regions that are 100 pixels on each side. Each cell-based coadd has a border of 50 pixels that overlaps with adjacent cells. Cell-based coadds near the border of the coadd tile have their unique regions extended to the edge of the tile.

Detection and measurement are run on the 200 $\times$ 200 pixel cell-based coadds, including their outer boundaries/buffers. Duplicates are rejected from adjacent cells using the unique cell-based coadd regions and then across coadd tiles using the unique coadd tile region. Objects detected in a cell-based coadd with their centres within the cell-based coadds unique region are kept and the rest are removed as duplicates. A similar procedure is then repeated across coadd tiles.

\begin{figure}
	\includegraphics[width=\columnwidth]{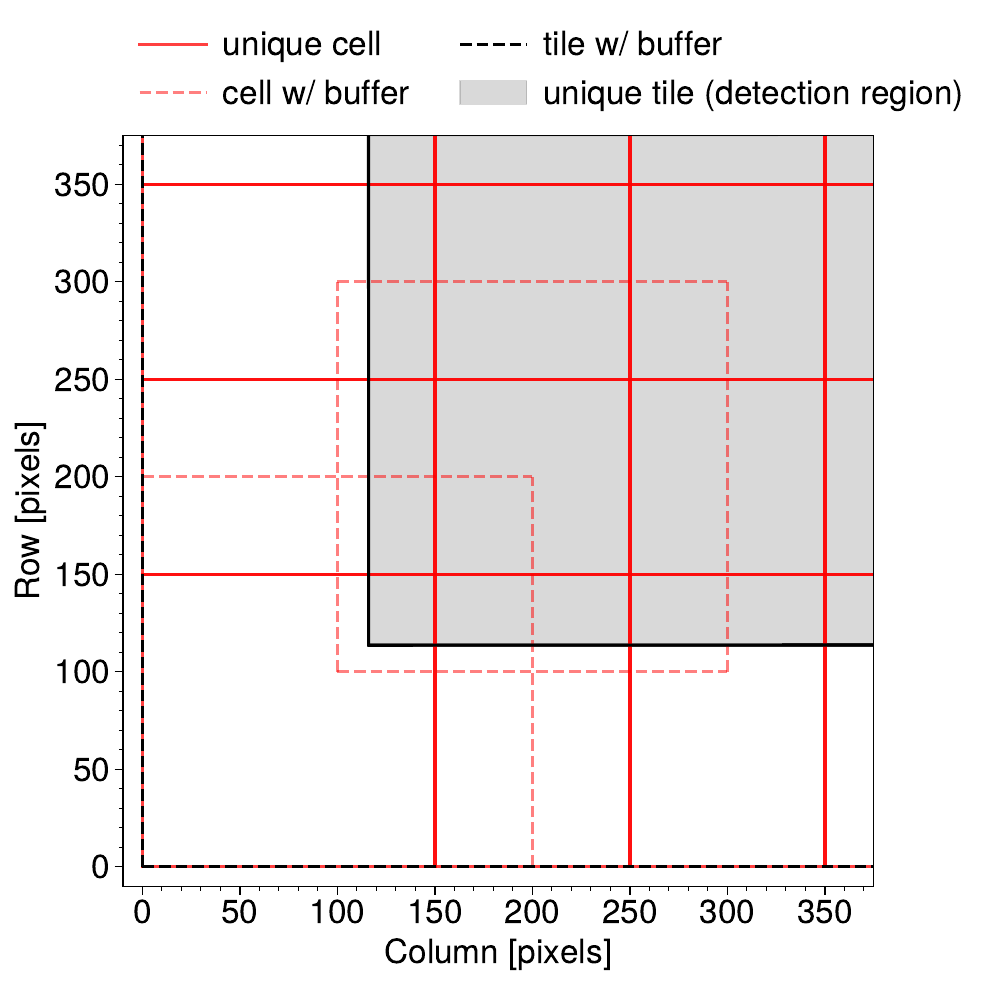}
    \caption{The cell-based coadd and DES coadd tile geometry for a corner of the DES coadd tile. The black dashed line shows the outer boundary of the 10k$\times$10k DES coadd tile. The inner black solid line shows the \textit{unique region} of the coadd tile, which is defined on the sphere in right ascension and declination. Adjacent coadd tiles overlap the same area, but share adjacent unique region boundaries. The red solid lines show the unique regions for the cell-based coadds. In the interior of the coadd tile, these regions are 100$\times$100 pixels. The red dashed line (shown only for two of the cell-based coadds along the diagonal) shows the outer boundary of each cell-based coadd. Adjacent cell-based coadds with their outer boundaries overlap by 50 pixels. At the edge of the coadd tile, the unique region for the cell-based coadd is extended all the way to the outer boundary of the coadd tile.}
    \label{fig:tile_geom}
\end{figure}

The net result of these definitions is that each cell-based coadd in the survey covers a unique footprint on the sky and has a predefined WCS solution derived from its location on a coadd tile and the coadd tile's WCS transformation. Figure~\ref{fig:tile_geom} shows the image geometry.

\subsection{Cell-based Coadd Construction and Data Products}
The individual cell-based coadds are constructed as follows. We first find all of the SE images that do not have an edge in the cell. We cut a border of 48 pixels from the edges of the CCDs due to larger astrometric residuals \citep{treerings} and tape bumps \citep{2006SPIE.6276E..08D}. For these intersection tests, we allow for SE images to have edges within one pixel of the edge of the 200 $\times$ 200 cell-based coadd with its buffer. This slightly looser criterion allows for more SE images to be used while avoiding biases near the centers of any object we keep (which is at least $\sim$49 pixels away). On average, about a quarter of the images that overlap any given point on the sky are rejected in this step. This loss of images results in a loss of depth of a few tenths of a magnitude, consistent with estimates in \cite{shearoncoadds}. Second, we use the bad pixel masks\footnote{These masks are in the form of a set of flags for each pixel marking the locations of pixels that should be ignored due to artifacts in the data.} from DESDM to find all of the pixels that we could not observe due to effects like bleeds\footnote{Bleeds are the overflow of charges in the detector readout direction.}, cosmic rays, bad columns, etc. Following previous DES shear measurement pipelines, we rotate the bad pixel mask by 90 degrees and apply it via a logical \texttt{OR} to the original bad pixel mask. This process helps to cancel additive biases in the final shape measurement. Then, we apply a two-dimensional Clough-Tocher interpolation \citep{ct1,ct2} from scipy \citep{scipy} to these pixels using their surrounding data. Images with a missing pixel fraction higher than 10\% were removed from the coadd. Third, we use the {\sc pixmappy} WCS solutions for each SE image to map the SE pixels to the cell-based coadd pixels using a Lanczos-3 interpolant. Finally, all of the SE image contributions to the cell-based coadd are averaged together with inverse-variance weighting determined by the maximum of the weight map over the region that intersects the cell-based coadd. We process all of the cell-based coadds in a given coadd tile in a single job for I/O efficiency. We further build a Clough-Tocher interpolant of the relevant WCS transformations in order to increase computational efficiency.

In the coadd process, we generate three other specialized data products.
\begin{enumerate}
    \item Noise image: We use the empirically measured sky noise level to generate noise images for each of the input SE images. We then coadd these noise images in the same way as the data, producing a Monte Carlo realization of the background pixel noise in the cell-based coadd image. The coadding process introduces subtle correlations in the background pixel noise, but we account for this by using the Monte Carlo noise images with \mdet.

    \item Masked fraction image: We generate a masked fraction image (\mfrac\ image), which measures the fraction of the input images in each pixel which were masked and then interpolated. To make this image, we construct an image of zeros (no interpolation) and ones (interpolated) for each input SE image according to which pixels were interpolated. We then coadd these images using the same Lanczos-3 interpolation with the same weights as the image and noise coadds above. Negative values in the warped masked fraction images are clipped to zero before coadding.

    \item PSF image: We generate a PSF image for each cell-based coadd. The algorithms for handling the PSF coadding are critical to not generating multiplicative shear biases \citep{2023OJAp....6E...5M}. In particular, the procedure adopted in this work is to pick a single pixel center in the cell-based coadd as the location about which we will compute the PSF \citep{shearoncoadds}. This pixel location is mapped back to the SE image location for each input SE image. We then draw the SE image PSF models at the SE image location and coadd them with the same weights and interpolation as the image and noise coadds above. This coadding process slightly broadens the PSFs due to the smearing of the interpolation \citep[see][for more details]{shearoncoadds}. Within each cell, we treat the PSF as constant spatially and chromatically. The DES Y6 PSF and astrometric models have chromatic dependence. Namely, the PSF and astrometric solution vary as a function of the color of the input object. In this work, we take a simplified approach of using the median galaxy color for the PSF models. We used a color more appropriate for stars for the {\sc pixmappy} astrometric solutions. However, this difference is small relative to the PSF models and does not affect our final results. We estimate the impact of these approximations using the image simulations as described in Sec.~\ref{sec:imagesims}, where we show them to be negligible at the precision of the DES dataset.
\end{enumerate}

\begin{figure*}
	\centering
    \includegraphics[width=0.65\columnwidth]{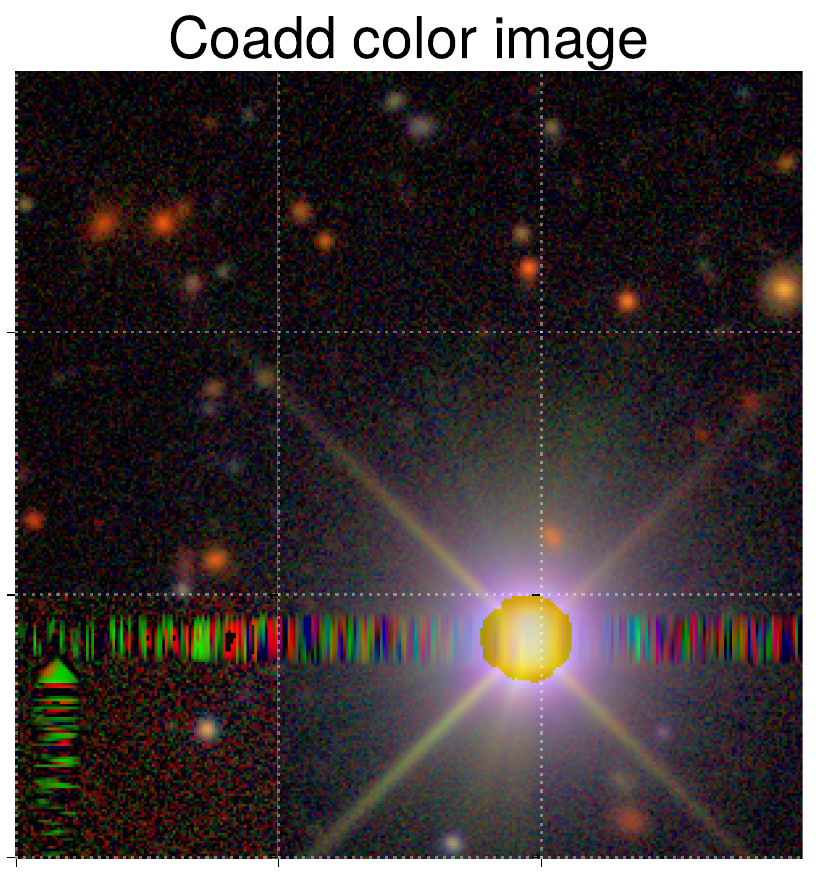}~
    \includegraphics[width=0.65\columnwidth]{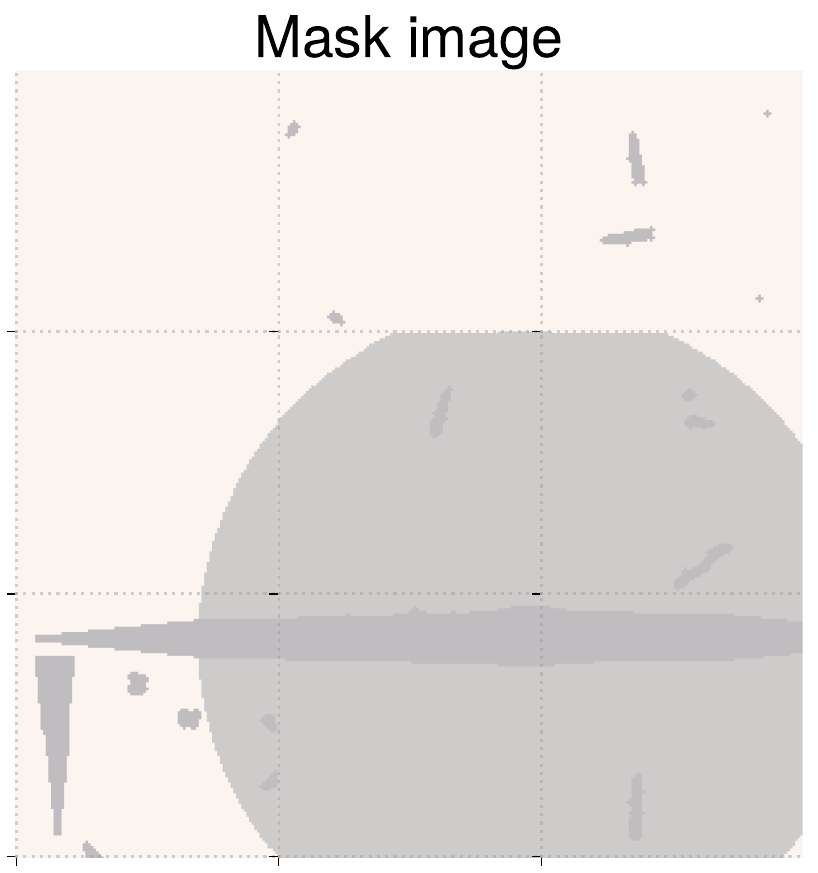}~
    \includegraphics[width=0.65\columnwidth]{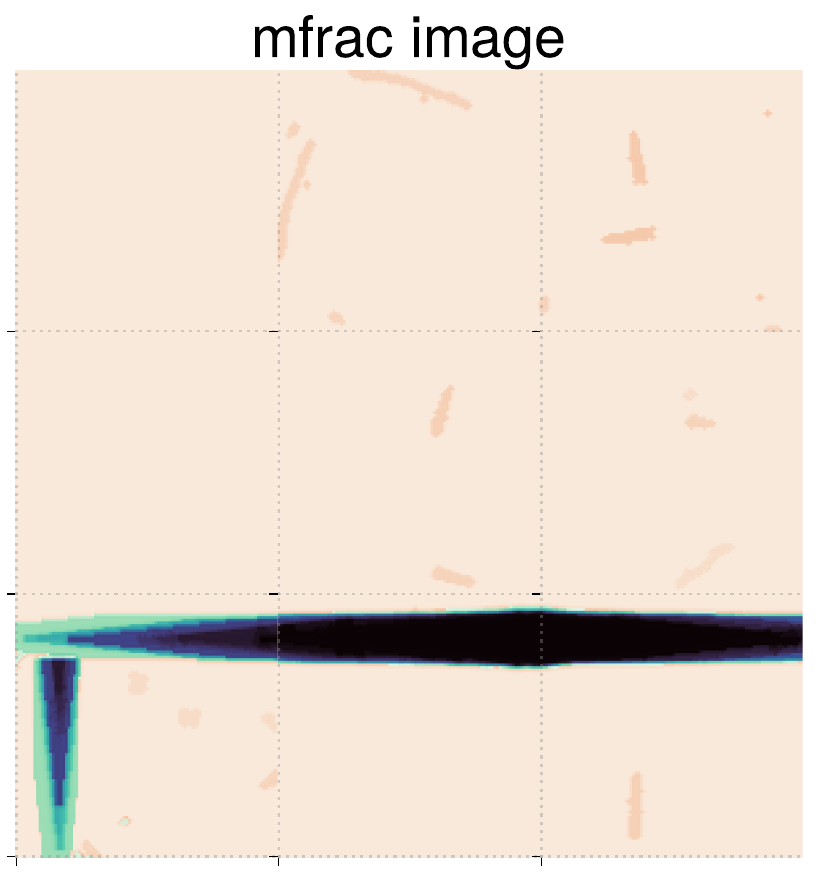}~
    \includegraphics[width=0.11\columnwidth]{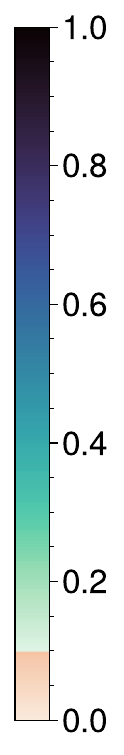}
    \caption{\textbf{\textit{Left}}: Example false-color image of nine coadd cells (with borders indicated as white dotted lines) made from $gri$ bands. The inner 100 $\times $ 100 pixels are cut out from the whole coadd cell of 200 $\times $ 200 pixels by removing the 50 pixels buffer region around the cell. \textbf{\textit{Middle}}: The mask image over the same region of the sky, where the grey region is excluded from the detection and further analysis. \textbf{\textit{Right}}: The \mfrac\ image over the same region of the sky, averaged over $riz$-bands. For each detection object, the average masked fraction is computed from the \mfrac\ coadd image, and any objects with \mfrac\ > 0.1 are excluded from our analysis. The interpolation and masked fraction differs for the bottom left cell relative to the two other cells on the bottom row due to the fact that 90-degree rotations are not applied to the input pixel masks near bright stars. See Sec.~\ref{sec:coadd} for more details. }
    \label{fig:cells}
\end{figure*}

Figure~\ref{fig:cells} shows example color images of nine adjacent cells from the DES Y6 analysis on the left, the mask in the center, and the masked fraction images on the right. We applied the cell-based coadding algorithm to the $griz$ bands and have produced a $gri$ false-color image for this figure. In this figure, we have cut the 50-pixel edge buffers and stitched the inner cell images together. We can see that there are subtle shifts in the background noise as the input SE images change from cell to cell. We also see some residual striping from the interpolation applied to the input SE images. This effect is accounted for using a cut in the masked fraction around each detection when running \mdet, as described below. The masked fraction image shows that in regions where the interpolant is applied to nearly every input SE image, and thus the output cell-based coadd image is not usable, the masked fraction approaches unity. Any source detected in or near these highly masked regions will have a large masked fraction and is excluded from use in the catalogue. The bottom left cell in Figure~\ref{fig:cells} shows the effects of the 90-degree rotation applied to the pixel masks. This rotation has not been applied to all cells in the bottom row of the images because the cells on the right are marked as being near a bright star, and we do not rotate pixel masks near bright stars since that data will be excluded anyway in our final survey mask as discussed below.

\section{\textsc{Metadetection} Formalism \& Shape Catalogue} \label{sec:mdet}

In this section, we overview the basic concepts of the \mdet\ algorithm and describe our object measurement method. \mdet\ is an updated algorithm similar to \mcal, which was used in our previous DES Y3 measurements. We discuss the masking of objects and object selections performed on the shape catalogue to enable the unbiased measurement of cosmic shear and present summary statistics relevant to the weak lensing analysis. The versions of key software packages used for each step of the pipeline are listed in Appendix~\ref{app:software}.

\subsection{Shear Calibration Formalism} \label{subsec:mdet}

Here we summarise the basic concept of shear calibration and how the ``raw'' ellipticity measurement of galaxy shapes is self-calibrated by the \mdet\ algorithm.

Consider a measurement of galaxy ellipticity ($e$). In the limit of small gravitational shear and convergence ($\gamma<<1$, $\kappa<<1$), one can Taylor-expand $e$ around zero shear $\gamma=0$. One then obtains an expression of the ellipticity in terms of the intrinsic galaxy shape and shear,
\begin{equation}\label{eqn:shear_cal}
    \vb{e} = \vb{e}|_{\vb{\gamma}=0} + \frac{\partial \vb{e}}{\partial\vb{\gamma}}\bigg|_{\vb{\gamma}=0}\vb{\gamma} + \ldots,
\end{equation}
where $\partial \vb{e}/\partial\vb{\gamma}|_{\vb{\gamma}=0}$ is what we call shear response $R$. Due to shape noise, we must average over many galaxy shapes to obtain a sufficiently unbiased estimate of the inverse of the shear response. The average shear estimate is expressed as
\begin{equation}\label{eqn:ensemble}
    \langle\vb{\gamma} \rangle \approx \langle\vb{e}\rangle / \langle \vb{R}\rangle,
\end{equation}
assuming that galaxies are intrinsically randomly oriented such that $\langle\vb{e}\rangle|_{\vb{\gamma}=0}=0$. For higher-order statistics, like two-point functions, previous work has shown that it is sufficient at the precision of DES data to compute a mean scalar response over the catalogue (or tomographic bin) and correct each individual object shape by this mean response.

Of course, we have not specified how to compute the response to shear $R$. The shear response can quantitatively be considered as how the observed shapes of galaxies respond to an applied gravitational shear. It is important to note that the shear response $R$ depends on the intrinsic properties of the object we are measuring (in addition to detection itself). From Eqn~\ref{eqn:shear_cal}, we can define the full shear response matrix (a derivative of spin-2 objects) as
\begin{equation}\label{eqn:responsematrix}
    \vb{R} \equiv \frac{\partial \vb{e}}{\partial \vb{\gamma}}\bigg|_{\vb{\gamma}=0} =
    \begin{pmatrix}
        \partial e_{1}/\partial \gamma_{1} & \partial e_{2}/\partial \gamma_{1} \\
        \partial e_{1}/\partial \gamma_{2} & \partial e_{2}/\partial \gamma_{2}
    \end{pmatrix}.
\end{equation}

\mcal\ (\citealt{2017arXiv170202600H, 2017ApJ...841...24S}) was the first algorithm to implement this concept into practice to self-calibrate ``raw'' galaxy shapes. \mcal\ operates on existing object detections and includes corrections for non-detection dependent selection effects implicitly in the ensemble average in Eqn.~\ref{eqn:ensemble}. This technique deconvolves the PSF from the image, applies a small artificial shear, and then reconvolves the image with a slightly larger PSF. To this artificially sheared image, we add a noise image put through the same process but with an opposite shear applied.\footnote{In detail, we rotate the noise image by 90 degrees, run an identical shearing process as the original image, and then rotate the noise image back by 90 degrees.}. This correction accounts for the effect of sheared and correlated background noise on the shape measurement \citep{2017ApJ...841...24S}. Without it, the technique will produce catastrophic biases at the order of $\sim10\%$. Finally, from the measured shapes on artificially sheared images, we can use a two-sided finite difference formula to compute the shear response
\begin{equation}
    R_{ij} = \frac{e^{+}_i - e^{-}_i}{\Delta \gamma_j},
\end{equation}
where the subscript $i$ denotes one of the two shear components and the superscript +/- denotes the sign of the applied shear. In total, we create five images (unsheared, sheared in $+\gamma_1, -\gamma_1$, $+\gamma_2$ and $-\gamma_2$ directions with the amplitude of 0.01) to compute the signal and response. We use the {\sc GalSim} package\footnote{\url{https://github.com/GalSim-developers/GalSim}} \citep{2015A&C....10..121R} and the high-level routines from the \textsc{ngmix} package\footnote{\url{https://github.com/esheldon/ngmix}} for these image operation steps.

In previous DES shear analyses, the technique above was applied at the locations of objects detected in the main survey \gold\ catalogue. However, since object detection is shear-dependent especially in the presence of blending, \mcal\ causes a selection/detection bias which amounts to percent-level multiplicative bias in the shape measurement (see S20 for more detail). Instead, for \mdet, we apply the \mcal\ artificial image shearing procedure to an entire cell-based coadd. This step generates five 200 $\times$ 200 cell coadds, one for each artificial shear, as described above. We then apply detection using the \textsc{sep} package \citep{sep} to each cell-based coadd. We use the same settings as the main survey\footnote{\url{https://github.com/esheldon/sxdes}}. Finally, we compute the shear response by averaging over the catalogues of detections, after the same selections have been applied to all the catalogues, like so
\begin{equation}\label{eqn:response}
    \langle R_{ij} \rangle = \frac{\langle e^{+}_i \rangle - \langle e^{-}_i \rangle}{\Delta\ \gamma_j}.
\end{equation}
The difference between this procedure and \mcal\ is important. \mdet\ cannot measure an individual object's response. Further, because object detection depends on shear, one cannot match the five different sets of detections against one another\footnote{The number of objects in each catalogue may differ.}. The response above includes both the ``shear'' and ``selection'' responses used with the DES Y3 shear catalogues \citep*{y3-shapecatalog}. Finally, we note that separating sources into tomographic bins by photometric redshift is a selection cut that is done on the source magnitudes. Thus tomographic selection must be included in the selection cuts used to compute the response in eqn.~\ref{eqn:response}, producing a different response estimate for each tomographic bin.

\subsection{Masking} \label{sec:masking}

There are two levels at which we want to remove data from our processing: problem areas in individual images, which we interpolate before image combination, and problem areas on the sky, where we remove objects from the catalogue.  In \S \ref{subsec:pixmasks} we describe the pixel masks (i.e., image level), and in \S \ref{subsec:objmask} we describe the object masks (i.e., catalog-level).

\subsubsection{Pixel Masks} \label{subsec:pixmasks}

Before image combination, we mask and interpolate artifacts in single epoch images, such as bad columns, cosmic rays, and saturated regions. During the \mdet\ processing, we further mask bright stars with apodization to avoid FFT artifacts during the deconvolution, shearing, and reconvolution processes following \cite{sheldon_mdet_rubin}. We also apply an apodized mask around the edge of each cell coadd to prevent artifacts from bright sources landing on the edge of a cell, as described in detail below.

We assemble an initial bright star catalogue using \gaia\ data release 2 \citep{gaia2}. Each \gaia\ star was masked using a circle with a magnitude-dependent radius in arcseconds given by
\begin{equation}
\mathrm{log}_{10}(R_{\gaia}) =  0.004432 \mathrm{G}^2 - 0.2257 \mathrm{G} + 2.996,
\end{equation}
where $\mathrm{G}$ is the magnitude in the \gaia\ band. We further place a lower bound on the radius of 5 arcseconds.
The image was set to zero inside the circular masked area, with the edges ``apodized'' to transition smoothly between unity on the outside to zero on the inside of the mask. This smooth transition was parameterized using the cumulative integral of a triweight kernel, which is a function of two parameters, $m$ and $h$,
and is defined for a point $x$ with quantity $y = (x-m)/h$ as
\begin{equation}
K(x, m, h) = \begin{cases}
0 & y < -3 \\
(-5y^7 / 69984 \\
+ 7y^5 / 2592 \\
- 35y^3 / 864 & -3 \le y \le 3 \\
+ 35y / 96 \\
+ 1 / 2) \\
1 & y > 3
\end{cases}
\end{equation}
This kernel smoothly transitions from zero to unity over a range of $6h$, centered on the location $m$.  We chose $h$ to be 1 pixel, which should vary more slowly than the PSF profile without significant additional expansion of the mask.
We also set $m+3h$ to the radius of the star mask hole so that the mask reaches unity at its nominal size. 

This same apodization kernel was used around the edge of each cell coadd. For this step, we applied the kernel along the four edges of the cell-based coadd so that it reached zero at the edge and unity at $6$ pixels inside the edge (again using $h=1$). 

After object detection and measurement, the \gaia\ star hole masks were expanded by 16 pixels, and any object with its center inside the \gaia\ mask hole was flagged to be cut. This step is needed to avoid excessive apodization effects on the object properties. Finally, we also applied the same masking and apodization to the \mfrac\ and noise images as appropriate throughout the \mdet\ measurement process.

\subsubsection{Object Level Masks} \label{subsec:objmask}

After the production of the full catalogue of measurements, we additionally cut objects based on measurement flags and \mfrac\ (see \S \ref{subsec:measurement}). We also apply a spatial mask in order to remove problematic regions of the sky. This mask, which is created in the {\sc HealSparse} format \footnote{\url{https://github.com/LSSTDESC/healsparse}}, incorporates all the \gaia\ pixel level masks described above plus the following additional sources. All masking is done at a HEALPix resolution of \texttt{nside=\nsidehi} except where noted otherwise.

\begin{enumerate}
    \item \mdet\ footprint: An overall footprint for the initial \mdet\ catalogue was produced by keeping areas corresponding to all cell-based coadds that had data in all four $griz$ bands and had no failures in running \mdet. In this step, we retain the union of only the unique cell-based coadd and unique coadd tile regions as described in Sec.~\ref{sec:coadd}.

    \item Foreground mask: The foreground mask contains regions and objects that are close to bright and extended sources. The foreground object mask generated includes very bright \gaia\ stars (G < 11.5), with a different radius relation from that used for our pixel level \gaia\ masking, Yale bright stars, 2MASS stars (5<J<12), Globular clusters, and a region near the LMC (see Sec~5.2 of \citealt{y6gold}).

    \item \gaia\ mask: We masked all stars in the \gaia\ catalogue using the parametric radius-magnitude relationship described above with the additional mask radius expansion of 16 pixels. This mask matches what was done in the pixel-level masking but at a higher spatial resolution since we use a HEALPix \texttt{nside=\nsidehi} realization of each masked star hole.

    \item DES stars: We augmented the \gaia\ star catalogue with high-confidence DES stars, defined as \texttt{MASH}$ == 0$ and $r < 21$ \citep{y6gold}.  These objects were masked with a fixed circular radius of 5 arcseconds.  Many of these stars are also in \gaia\, and thus may already have a larger mask applied.

    \item HyperLEDA mask: After application of the foreground mask, we visually found additional unmasked regions near bright and extended objects at low redshift that caused spurious detections and contaminated the light profile of neighboring objects. We found that masking objects from the {\sc HyperLEDA} (\citealt{2003A&A...412...45P}) significantly reduced spurious detections and poor measurements. The {\sc HyperLEDA} catalogue contains galaxies brighter than $B$-mag $\sim 18$ with coordinates, diameter, and axis ratio. Each object was masked with a circular, $B$-magnitude dependent circular mask
    with a radius in degrees $0.147 - 0.00824 * B$. Finally, for a handful of extremely large, low redshift objects, we identified spurious detections due to resolved structure within the object. Thus, we produced by hand special elliptical masks around those objects (NGC0055, NGC0253, NGC0300, IC1613) utilizing the diameter and axis ratio information of those objects in the {\sc HyperLEDA} catalogue with some manual tuning.
\end{enumerate}

\begin{figure*}
	\includegraphics[width=\textwidth]{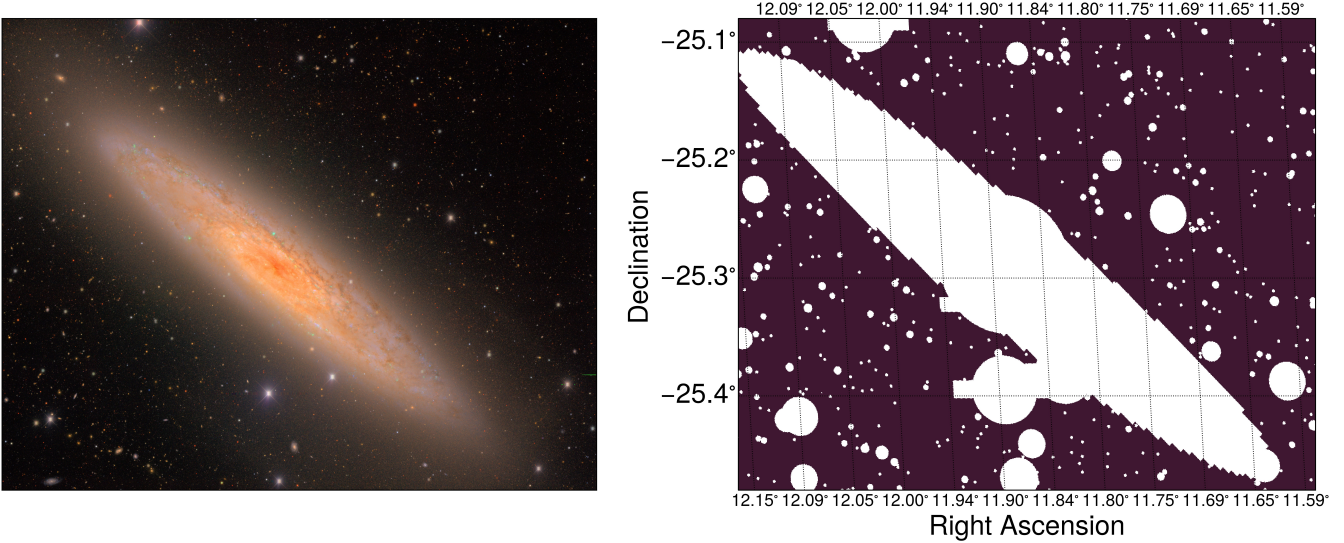}
    \caption{Example image of our combined pixel-level and object-level mask and its associated coadd image near the local galaxy NGC0253. Several features are clearly apparent, including bright star masks (e.g., the mask hole right below the center of the large galaxy) and missing cell-based coadds due to bleeds from bright stars (e.g., rectangular missing regions sticking out of the bright star mask hole).}
    \label{fig:masks}
\end{figure*}

Figure \ref{fig:masks} shows an example of the cumulative mask near a local, extended bright galaxy. For the pixel-level mask, we removed a total area of 13.88 deg$^2$. For all object-level masks, we remove a total area of 489.5 deg$^2$ with contributions from 4.85 deg$^2$ for \gold\ footprint mask, 427.27 deg$^2$ for foreground mask, 137.84 deg$^2$ for the \gaia\ mask, 73.69 deg$^2$ for DES stars mask, 30.18 deg$^2$ for {\sc HyperLEDA} mask. These numbers do not account for the overlap region each mask contains, which is significant for some regions of the sky. In the end, we are left with 4421.52 deg$^2$ footprint for DES Y6 \mdet\ shear catalogue with both pixel-level and object-level masks applied.

\subsection{Object Identification \& Measurement} \label{subsec:measurement}

We detect and measure objects in the five different cell-based coadds \mdet\ produces as follows. For detection, we use the \textsc{sep} package \citep{sep}, a Python-based fork of {\sc SExtractor} \citep{SourceExtractor1996, sextractor}. We use the same configuration setting as in \citealt{2021ApJS..255...20A} as contained in the \textsc{sxdes}\footnote{\url{https://github.com/esheldon/sxdes}} package. Given the coadd images for each band, we form the multi-band $riz$ coadd image through a straight addition of the coadds without any PSF matching, and run the detection on an inverse-variance weighted coadd image\footnote{Early tests found that including the $g$-band in \mdet\ reduced the precision of the shear estimates.}.

For each detection in each of the five images, we make three main measurements. All measurements were done with the \textsc{ngmix} software package.\footnote{\url{https://github.com/esheldon/ngmix}} First, we measure the average masked fraction per object. To make this measurement, we compute a single masked fraction image for the cell-based coadd across all of the bands used for lensing. In this work, we use an inverse-variance weighted average image of the cell-based coadd masked fraction images from $riz$ bands. Then for each detection, we compute the average value of this single masked fraction image with a two-arcsecond FWHM Gaussian weight function about the detection's center. As for each object property measurement, we performed a Gaussian forward model fit across the $riz$ band images individually. Table~\ref{tab:priors} shows the list of parameters to fit each object and their priors.
\begin{table}
    \centering
    \begin{tabular}[width=\columnwidth]{l l l}
    \hline\hline
    Parameter & &  Prior \\
    \hline
    Centroid offset & Gaussian & $\sigma_{x,y}$ = 0.263\\
    Shape & Gaussian &$\langle g\rangle = 0.0$, $\sigma_{|g|} = 0.3$\\
    Galaxy size & Two-side Erf & min = -10.0, $\sigma_{\rm min}$ = 0.03, \\
     & & max = $1.0\times10^{6}$, $\sigma_{\rm max}$ = $1.0\times10^{5}$\\
    Total flux & Two-side Erf & min = $-1.0\times10^{4}$, $\sigma_{\rm min}$ = 1.0, \\
     & & max = $1.0\times10^{9}$, $\sigma_{\rm max}$ = $0.25\times10^{8}$\\
    \hline\hline
    \end{tabular}
    \caption{List of prior values and distributions used for the Gaussian forward model fit.}
    \label{tab:priors}
\end{table}
This fit follows a similar procedure as the DES Y3 \mcal\ shear catalogue \citep*{y3-shapecatalog}, including the use of \texttt{uberseg} \citep{2016MNRAS.460.2245J} to remove some of the effects of light from neighboring objects. We refer to this measurement as the \texttt{gauss} quantities, and the shape from this measurement is used as the catalogue shape measurement with \mdet. We additionally use a specialized Fourier-space method, detailed in Appendix~\ref{app:pgauss}, to measure a two-arcsecond FWHM Gaussian weighted flux for each detection on a PSF-deconvolved image. This second set of measurements does not use \texttt{uberseg}. This second set of measurements is referred to as the \texttt{pgauss} quantities. These quantities include fluxes in each of the $griz$ bands along with an inverse variance-weighted size measure. Both \texttt{gauss} and \texttt{pgauss} measure pre-PSF properties of the object so that, e.g., point-like objects would have zero size in the absence of other effects.

While we do not detect with the $g$-band cell-based coadd, we use the detected object position to measure a $g$-band flux in addition to $riz$ fluxes. Measurements of the reconvolved PSF model of each band's cell-based coadd are made with the \texttt{pgauss} method as well, but without PSF deconvolution. The default parameters and priors for all measurements are detailed in the \texttt{metadetect}\footnote{\url{https://github.com/esheldon/metadetect}} and \texttt{pizza-cutter-metadetect}\footnote{\url{https://github.com/beckermr/pizza-cutter-metadetect}} packages. The versions of key software packages used for each step of the pipeline are listed in Appendix~\ref{app:software}.

\subsection{Object Selection Criteria} \label{subsec:mdetcuts} %Cuts
We consider several criteria to select the \mdet\ shape catalog, after applying the masks described in Sec.~\ref{sec:masking}. All selections are applied in each of the five \mdet\ catalogues, corresponding to the five \mdet\ shears.

%\begin{enumerate}[I]
    \subsubsection{Star/galaxy separation}
\begin{figure}
	\includegraphics[width=\columnwidth]{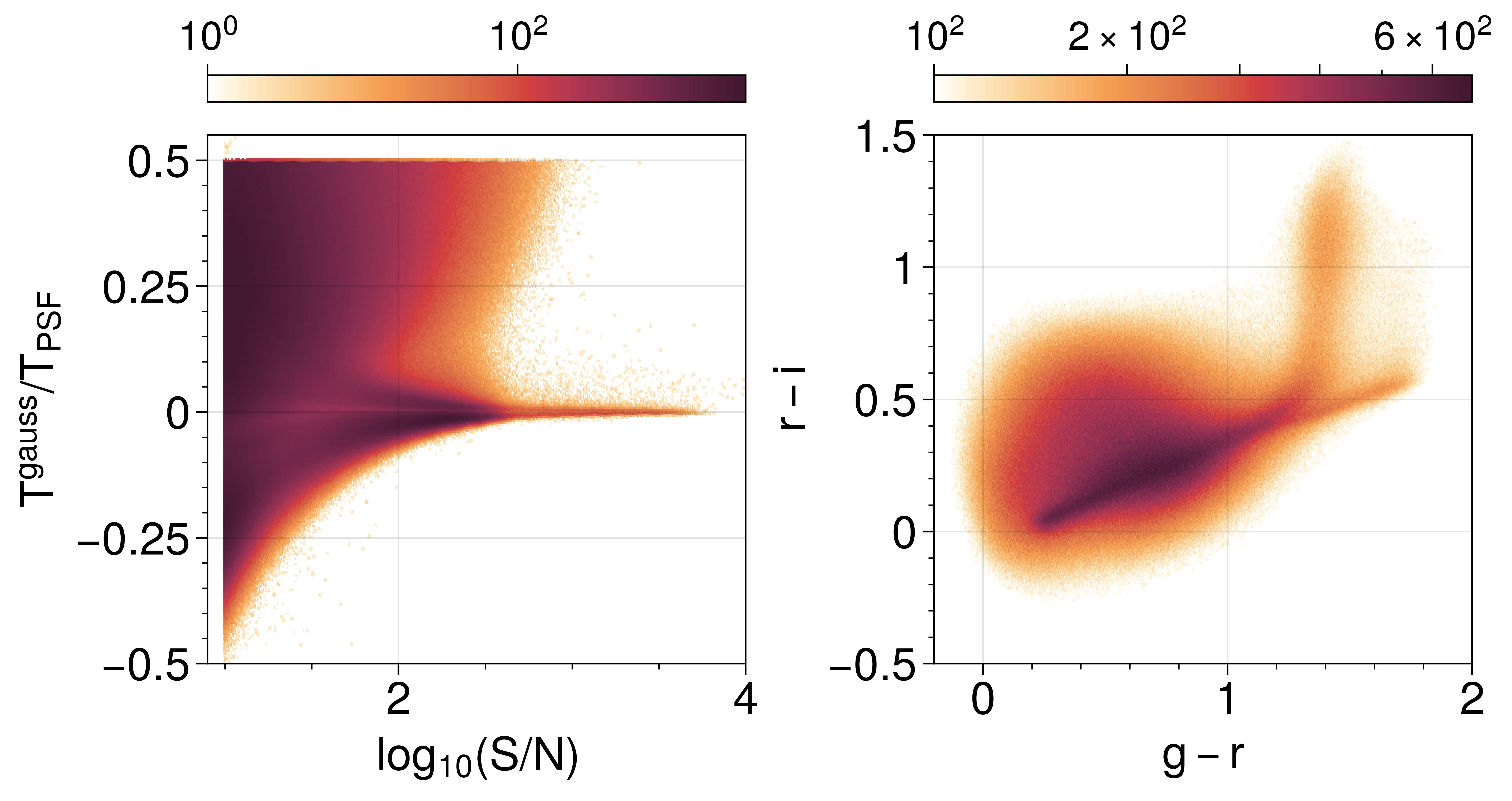}
    \caption{Log density of objects excluded by the star-galaxy separation criteria, showing object S/N vs size ratio ($T/T_{\rm PSF}$; \textbf{\textit{left}}) space and color space ($g-r$ vs $r-i$; \textbf{\textit{right}}). All other selection criteria have been applied. Objects are chosen from 100 randomly selected patches out of 200 over the footprint. Well-measured stars populate the plot near $T/T_{\rm PSF}=0$, but not exactly on $T/T_{\rm PSF}=0$, due to how the shape and size priors are set for the Gaussian fit.}
    \label{fig:sg_sep}
\end{figure}

    \begin{itemize}
        \item $T^{\rm gauss}/T^{\rm gauss}_{\rm PSF} > 0.5$ -- We employ a star-galaxy separation using object size ratio ($T^{\rm gauss}/T^{\rm gauss}_{\rm PSF}$), which is a measure of how well an object is resolved. While high S/N stars populate at $T^{\rm gauss}/T^{\rm gauss}_{\rm PSF}$=0.0, low S/N stars have larger error bars and hence their measurement deviates from $T^{\rm gauss}/T^{\rm gauss}_{\rm PSF}$=0.0. Thus, we select objects whose size ratio is larger than $0.5$. Figure~\ref{fig:sg_sep} shows the impact of this star-galaxy separation cut on the population. Table~\ref{tab:cuts_fraction} shows the fraction of our estimated rate of stellar contamination in our catalogue, and Appendix~\ref{app:stellar} describes how we estimate the stellar contamination in our catalogue in more detail. The fractions in this table are underestimated by a factor of several due to the fact that the star-galaxy separation in the \gold\ catalogue is poor at faint magnitudes. See Appendix~\ref{app:stellar} for further discussion. 

        \item $S/N^{\rm gauss} > 10$ -- A standard \mdet\ cut to avoid noisy shape measurements for low S/N detections and to reject low S/N stars that otherwise pass the $T^{\rm gauss}/T^{\rm gauss}_{\rm PSF} > 0.5$ cut.
        
    \end{itemize}

    \subsubsection{Size selection}
    \begin{itemize}
        
        \item $T^{\rm gauss}$ < 20 -- We visually inspected the large-size objects and found a large fraction of them to be image artifacts or small galaxies whose measurement is impacted by the fluxes from nearby objects. While some objects were real, the fraction at $T^{\rm gauss}$ > 20 was large enough that we decided to remove all of them.
    \end{itemize}

    \subsubsection{Flux/color \& heavily interpolated objects }
        \begin{itemize}
            \item In order to obtain reliable photometric redshifts of the sources, we make magnitude\footnote{Magnitudes were estimated from the \texttt{pgauss} fluxes with an inverse hyperbolic sine function \citep{asinh_mag}. The magnitudes were further corrected for the extinction using $E(B-V)$ values from the reddening map of \citet{sfd98}, which is at the HEALPix resolution of {\tt nside}=4096. } cuts on each bandpass to exclude excessively faint sources following \cite*{y3_source_redshift}. These cuts are $g < 26.5$, $r < 26.5$, $i < 24.7$\footnote{This selection is to reduce COSMOS20 photometric redshift outliers \citep{cosmos20} with which our redshift calibration is anchored \citep{y6_source_redshift}.}
            , and $z < 25.6$.
            \item In order to reject objects that have odd colors that appear to be the result of flux measurement failures, we make cuts on measured galaxy color.  $|g-r| < 5$, $|r-i| < 5$, $|i-z| < 5$.
             \item $\mfrac < 0.1$ -- The weighted masked fraction computed as described above. Tests in our image simulations suggested that utilizing $\mfrac<0.1$ is sufficient not to introduce shear calibration biases.
        \end{itemize}

    \subsubsection{Junk detections}
    \begin{itemize}
        \item $T^{\rm pgauss} < 1.6 - 3.1 \times T^{\rm pgauss}_{\rm err}$ -- We have identified a population of false detections that fall in a specific region of ($T^{\rm pgauss}$, $T^{\rm pgauss}_{\rm err}$) space. We verified visually that this cut removes those objects. These detections typically occur around bright stars where the background subtraction leaves halos of stellar light in the image. This population does not appear in $T^{\rm gauss}$ space.
        \item $T^{\rm gauss}\times T^{\rm gauss}_{\rm err} < 1$ or $T^{\rm gauss}/T^{\rm gauss}_{\rm err} > 10$ -- This cut was based on the similar cut in \gold\ \citep{y6gold} to avoid junk objects in cluster fields (i.e., super-spreader objects). We visually confirmed that most of these objects appear to be junk as well.
    \end{itemize}

% \end{enumerate}

\begin{table}
    \centering
    \caption{\label{tab:cuts_fraction}The fraction of objects removed with each criterion, computed as those in \textit{noshear} \mdet\ catalogue. We also show the fraction of stars within matched samples that are identified in \gold\ sample (\texttt{MASH}$ == 0$ OR $1$) within 0.263 arcseconds. Before ``mask'' selection, ``basic'' selections (\texttt{flags}$ == 0$ and $S/N^{\rm gauss} > 5$) have been applied in order to save disk space.}
    \begin{tabular}[width=\columnwidth]{ccccc}
        \hline\hline
             Selection & Fraction of removed objects & Fraction of stars \\ \hline
             Mask & 14.3\% & 7.06\% \\
             Star-galaxy separation & 73.6\% & 0.0369\%\\
             Size & 0.128\% & 0.0369\%\\
             Flux/color, mfrac & 3.81\% & 0.0319\%\\
             Junk & 0.295\% & 0.0318\%\\ \hline
             All & 78.3\% & 0.0318\% \\
        \hline\hline
    \end{tabular}
\end{table}

Readers can refer to Appendix~\ref{app:junks} for a more detailed discussion of the junk detection cuts and postage stamp images of randomly selected objects that were rejected. Table.~\ref{tab:cuts_fraction} presents the fraction of objects removed from the \textit{noshear} catalogue at each selection step. After all of the selection cuts and masks are applied on our \mdet\ catalogue, the total number of objects is 151,922,791.

\begin{figure*}
	\includegraphics[width=0.8\textwidth]{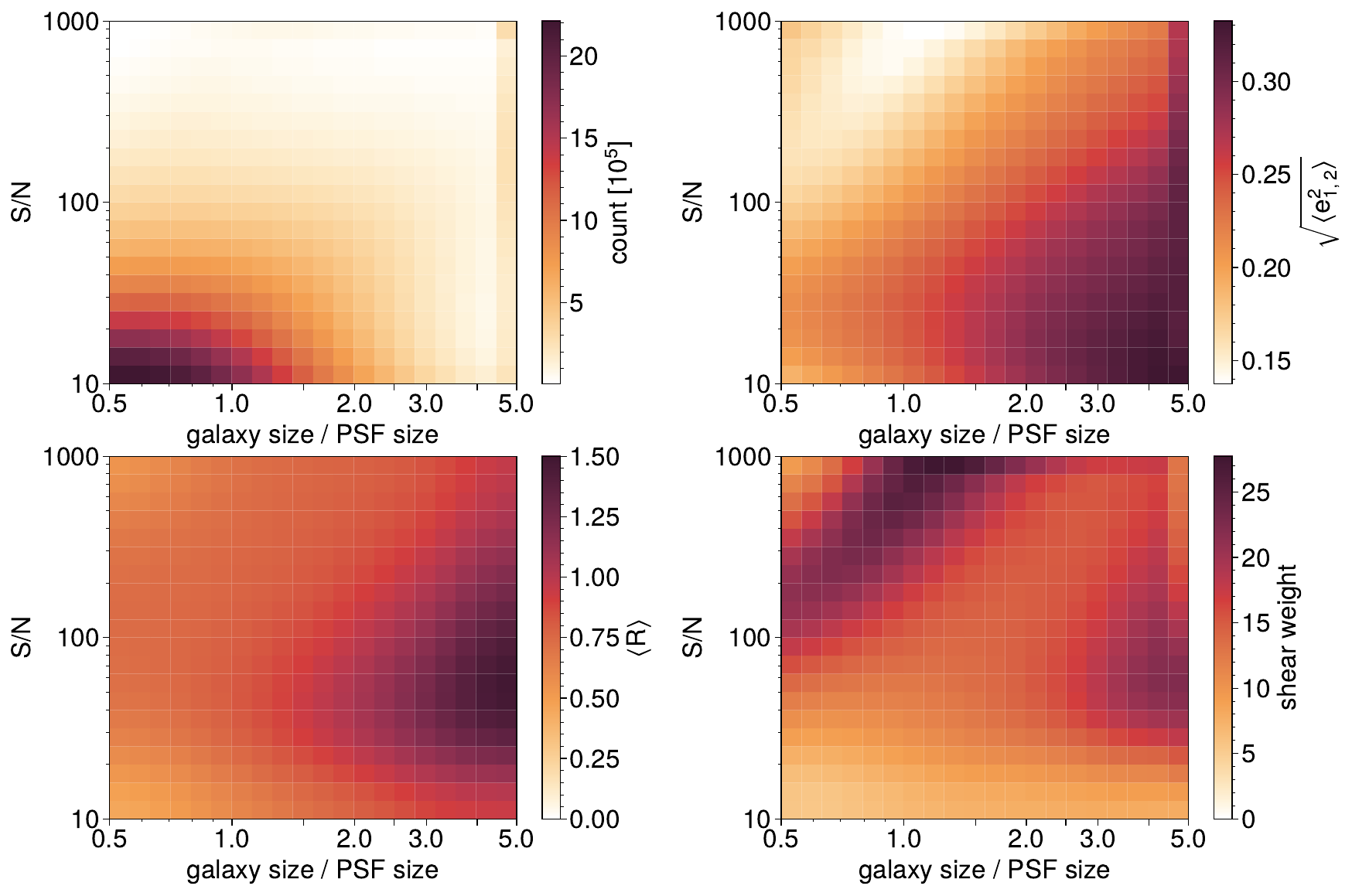}
    \caption{Various statistics (number count, a root-mean-square of measured shape, shear response, shear weight) as a function of object S/N and size ratio (galaxy size/PSF size). Objects are binned into a grid of signal-to-noise and size ratio using 20 logarithmic bins with a limit of 10<S/N<1000 and 0.5<galaxy size/PSF size<5.0. The objects whose S/N is larger than 1000 are allocated in the last bin, and the same goes for the objects whose size ratio is larger than 5.0. The measured shapes for each sheared image are averaged to compute the shear response in each bin. This grid of shear response is smoothed by a Gaussian kernel of $\sigma$=2.0 to lower the noise in each bin. The shear weight is then computed from this smoothed response grid using Eqn.~\ref{eqn:shearweight}.
    }
    \label{fig:shearweight}
\end{figure*}

\subsection{Shear Catalogue Statistical Weights} \label{subsec:weight}

We define the weights as the inverse variance of the measured ellipticities in each grid bin corrected by the shear response,
\begin{equation}\label{eqn:shearweight}
    w_{i}(T/T_{\rm PSF}, S/N) = \sigma^{-2}_{i,\gamma}(T/T_{\rm PSF}, S/N) = \sigma^{-2}_{i,e} \langle \boldsymbol{R_{i,\gamma}} \rangle^{2},
\end{equation}
where the subscript $i$ denotes the index of each bin in S/N and size ratio grid, $\sigma^{2}_{i,e}$ is the shear variance in the grid cell, and $R_{i,\gamma}$ is the shear response in the grid cell. The actual response correction for shear calibration is done via Eqn.~\ref{eqn:response} for the given sample under consideration, using the statistical weights in the catalogue averages to compute a weighted average.

The variance of measured ellipticities (intrinsic and measurement-related shape noise) is computed as
\begin{equation}
    \sigma^2_{i,e}(T/T_{\rm PSF}, S/N) = \frac{1}{2} \left[\frac{\Sigma(e_{i,1})^2}{n_{i,\rm gal}} + \frac{\Sigma(e_{i,2})^2}{n_{i,\rm gal}}\right],
\end{equation}
where $n_{i,\rm gal}$ is the number of galaxies in each grid cell.

The distributions of number counts, shape variance averaged in two components, shear response, and shear weight are shown in Fig.~\ref{fig:shearweight}. We note here that the grid of shear response is smoothed with a Gaussian kernel of $\sigma=2.0$ to reduce the impact of shot noise of the response itself. We noticed objects with moderate S/N and large size having larger shape noise compared to Y3. We viewed those objects and found they are largely in a blended system. Nonetheless, the distribution of the shear weight is similar to that of the \mcal\ shape catalogue in Y3 \citep*{y3-shapecatalog}.

\begin{figure*}
    \centering
    \includegraphics[width=\columnwidth]{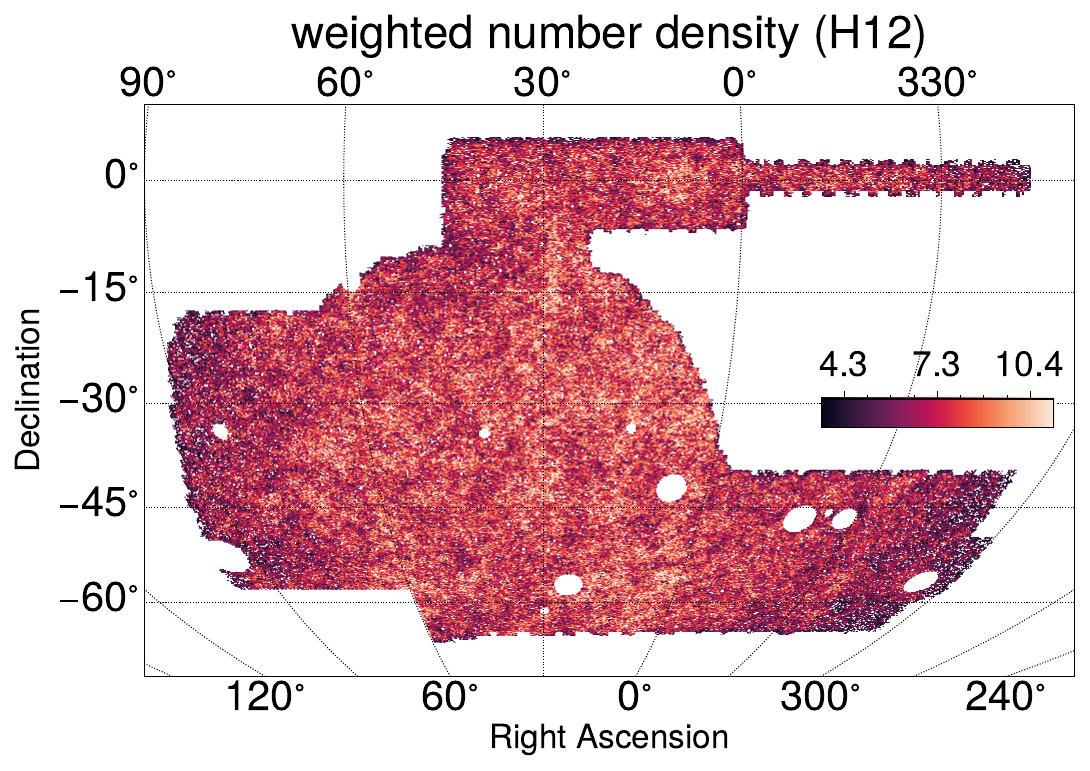}~
    \includegraphics[width=\columnwidth]{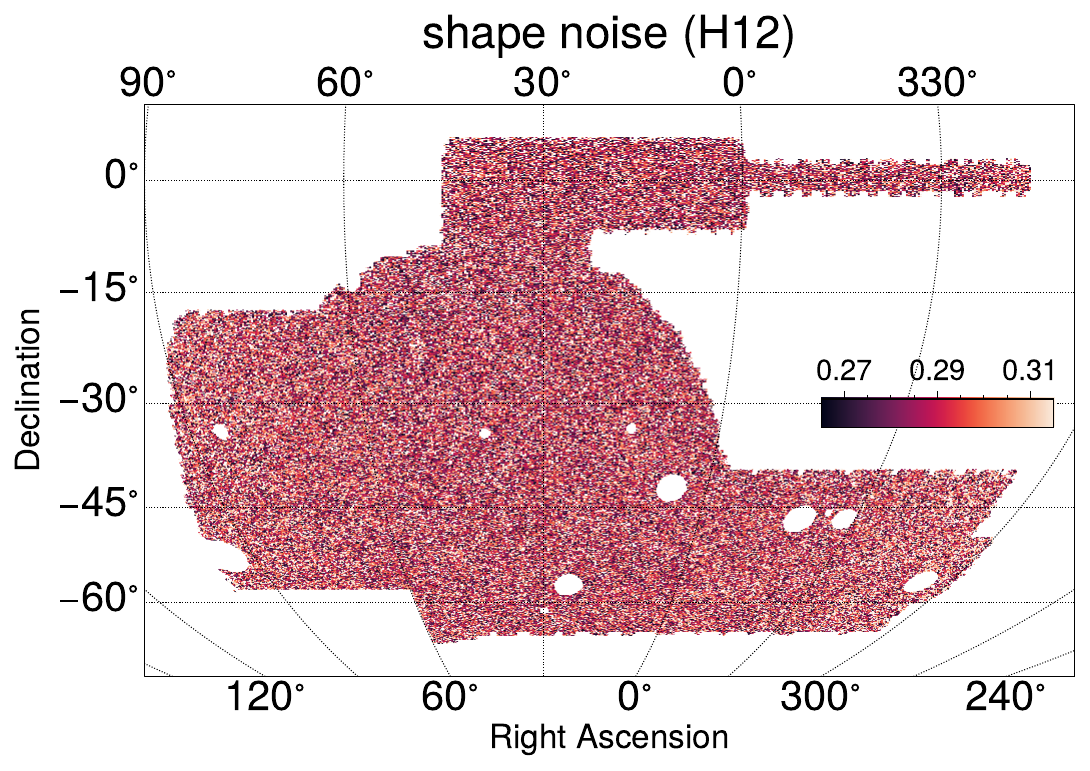}
    \caption{
        \textbf{\emph{Left}}: weighted galaxy number density in units of gal/arcmin$^2$ using Eq.~\ref{eqn:neff_h12}. The number density
        is computed in a HEALPix grid \citep{2005ApJ...622..759G, 2019JOSS....4.1298Z} of
        resolution 10 ({\tt nside}$\,=2^{10}=1024$). \textbf{\emph{Right}}: object shape noise computed
        with Eq.~\ref{eqn:sige_h12} in a HEALPix grid of resolution 10 ({\tt nside}$\,=2^{7}=1024$).
    }
    \label{fig:neff_shape_noise}
\end{figure*}

\subsection{Summary Statistics of the Catalogue} \label{subsec:stat_power}
Here we present summary statistics for the catalogue after the selections described in Sec.~\ref{subsec:mdetcuts}. As we have described earlier, for \mdet\ we compute the shear response averaged over the catalogue rather than each object using Eqn.~\ref{eqn:response}. The shear response over the catalogue is $\langle R \rangle = 0.817$. 
% The magnitude limit of our catalogue is $\sim$ 0.5 magnitude shallower than that of \gold\ catalogue \citep{y6gold} in $i$-band, resulting in $m_i \sim  22.8$. This number very approximately represents the median depth of the catalogue across the survey at a S/N$\sim10$.
We measure the statistical power of the catalogue by computing the standard error of the mean shear across the catalog, $\sigma_{\rm \gamma} = \sigma_{\rm e}/\sqrt{N_{\rm eff}}$. Following the definitions of \citealt{2012MNRAS.427..146H} (H12) and \citealt{2013MNRAS.434.2121C} (C13), the shape noise ($\sigma_{\rm e}$; standard deviation of intrinsic ellipticities) and the effective number density ($n_{\rm eff}$) are computed as follows. For the C13 definition, we compute
\begin{eqnarray}
    \sigma_{e, \rm C13}^2 &=& \frac{1}{2} \frac{\Sigma w_i^2 (e_{i,1}^2 + e_{i,2}^2 - \sigma_{m,i}^2)}{\Sigma w_i^2 R^2} \label{eqn:sige_c13} \\
    n_{\rm eff, C13} &=& \frac{1}{A} \frac{\sigma_{e, C13}^2 (\Sigma w_i R)^2}{\Sigma w_i^2 (R^2 \sigma_{e, C13}^2 + \sigma_{m,i}^2/2)} \label{eqn:neff_c13},
\end{eqnarray}
where $w_i$ is the shear weight computed in Sec.~\ref{subsec:weight}, $R$ is the global shear response, $e_{i,j}$ is the per-object ellipticity (uncorrected for $R$), and $\sigma_{m,i}^2$ is the variance of the per-object ellipticity. $A$ is the effective area of our footprint. For the H12 definition, we compute
\begin{eqnarray}
    \sigma_{e, \rm H12}^2 &= \frac{1}{2} \left[\frac{\Sigma w_i^2(e_{i,1}-\langle e_1 \rangle)^2}{(\Sigma w_i R)^2} + \frac{\Sigma w_i^2(e_{i,2}-\langle e_2 \rangle)^2}{(\Sigma w_i R)^2}\right] \frac{(\Sigma w_i)^2}{\Sigma w_i^2} \label{eqn:sige_h12} \\
    n_{\rm eff, H12} &= \frac{1}{A} \frac{(\Sigma w_i)^2}{\Sigma w_i^2} \label{eqn:neff_h12}
\end{eqnarray}

We report the effective galaxy number density and the variance of the estimated shear of the objects in our \mdet\ shape catalogue in Table~\ref{tab:neff_sigmae}. One finds that the shape noise in Y6 has increased compared to Y3. We speculate that this increase is due to detecting fainter objects than Y3 and possibly detecting more early-type galaxies that have larger intrinsic scatter. The effective number density increase between Y3 and Y6 is difficult to predict because it depends on the faint-end slope of the luminosity function. However, note that we lose approximately two observations of each object during the cell-coadding process, so that the number density increased less than it might have otherwise. Figure~\ref{fig:neff_shape_noise} shows the weighted number density and shape noise in the H12 definition. The statistical power of the Y6 shape catalogue has increased by 12\% for H12 definition and 10\% for C13 definition, compared to that of Y3 weak lensing shape catalogue \citep*{y3-shapecatalog}.

\begin{table}
    \centering
    \caption{\label{tab:neff_sigmae} Effective number density, shape noise computed using two definitions Eqn.~\ref{eqn:sige_c13} and Eqn.~\ref{eqn:sige_h12}, and weighted residual mean shear in the catalogue. The bottom row presents the same statistics from GS20, which used H12 definition for the final shape catalogue.
    }
    \begin{tabular}[width=\columnwidth]{ccccc}
\hline\hline
      Year & $n^{\rm C13}_{\rm eff}$, $n^{\rm H12}_{\rm eff}$ & $\sigma^{\rm C13}_{\rm e}$, $\sigma^{\rm H12}_{\rm e}$ & $c_1$ & $c_2$ \\ \hline
    Y6 & 7.97, 8.22 & 0.270, 0.289 & $+1.94\times10^{-4}$ & $-0.314\times10^{-4}$ \\
    Y3 & 5.32, 5.59 & 0.255, 0.261 & $+3.52\times10^{-4}$ & $+0.594\times10^{-4}$\\
\hline\hline
    \end{tabular}
\end{table}

\begin{figure*}
	\includegraphics[width=0.8\textwidth]{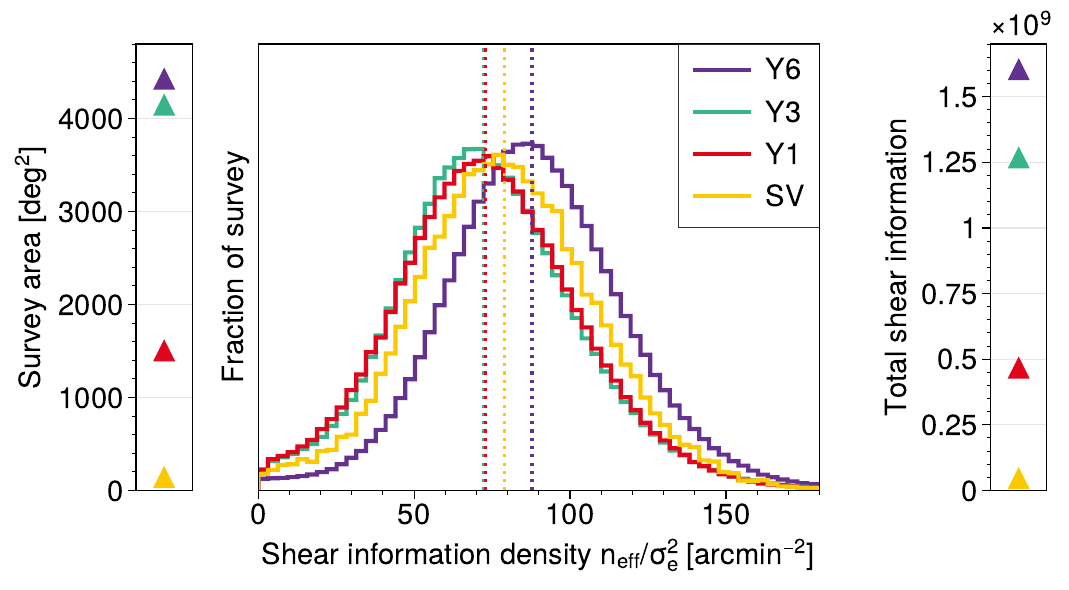}
    \caption{A histogram showing the shear information density ($n_{\rm eff}/\sigma^{2}_e$) in HEALPix pixels (\texttt{nside}=1024) for the DES Y6 \mdet\ catalogue footprint, compared with the same quantity from several catalogue releases from the Dark Energy Survey -- Science Verification (SV, \citealt{2016MNRAS.460.2245J}), Year 1 (Y1, \citealt*{2018MNRAS.481.1149Z}), Year 3 (Y3, \citealt*{y3-shapecatalog}). Each histogram has the same area under the curve, and the median of each histogram is shown as the dotted line for each release. The interquartile range (IQR) for each release is $\{38.22, 38.57, 37.55, 36.81\}$ for SV through Y6. The histogram is accompanied by the panels showing the survey area (left) and the total shear information content of the catalogue (right), which is the sum of pixel areas for each survey times the shear information density. We can see that as the DES survey has progressed, the data set has become more uniform and its total information content has increased substantially.
    }
    \label{fig:neff_sige_sv2y6}
\end{figure*}

Figure~\ref{fig:neff_sige_sv2y6} shows the Y6 catalogue precision compared to other DES weak lensing shape catalogue releases -- Science Verification (SV, \citealt{2016MNRAS.460.2245J}), Year 1 (Y1, \citealt{2018MNRAS.481.1149Z}), Year 3 (Y3, \citealt*{y3-shapecatalog}). The figure presents three key quantities, the survey area, the shear information density, and total shear information content. The histogram shows the number density weighted by the shape noise$^2$ in HEALPix pixels with \texttt{nside=1024}, which we call shear information density per unit area. The total shear information is approximately the survey area $\times$ shear information density. The inverse of the square root of this number is the error on the mean shear across the survey. The figure shows several major achievements by the DES collaboration. The survey has increased in area substantially, increased in depth (increasing the shear information density), and become more homogeneous. These changes are a testament to the hard work of dozens of scientists working on weak lensing shape measurement in the DES collaboration over more than a decade.

\subsection{DES Y6 \textsc{Metadetection} Shape Catalogue Public Release} \label{subsec:release}
The full \mdet\ catalogue will be made available for public use with the publication of the key DES Y6 weak lensing cosmological results. This catalogue will include a superset of the catalogue tested in this work. To reproduce the set of objects used in this work, we will provide a flags column for cuts. We caution the reader that manually applying the selection cuts above in Sec.~\ref{subsec:mdetcuts} will result in a slightly different catalogue. These differences come from some cuts being applied at different numerical precision and from some masks being applied at higher spatial resolution than those used to define the final catalogue footprint. Finally, the \mdet\ shape catalogue used for the cosmological analysis will include further selections to account for the quality of the angular clustering sample and photometric redshift distributions used for the 3x2pt analysis.

\section{Correlations with PSF Properties} \label{sec:psfsyst}
In this section, we test for residual PSF contamination in our shear catalogue and for the effects of any residual PSF modeling errors. The PSF modeling errors are estimated from a reserve sample of the stars not used in fitting the PSF models. The reserve sample is 20\% of the total sample of PSF stars. We explore the following items in this section:
\begin{itemize}
    \item dependencies of the mean galaxy shear on PSF model size and shape (Sec.~\ref{subsec:meanshear_psf});
    \item additive biases due to the PSF leakage and modeling errors (Sec.~\ref{subsec:rho});
    \item tangential and cross-component galaxy shear around PSF stars (Sec.~\ref{subsec:tanshear_stars}).
\end{itemize}

\subsection{Mean Shear as a Function of PSF Properties} \label{subsec:meanshear_psf}
\begin{figure*}
    \centering
    \includegraphics[width=\columnwidth]{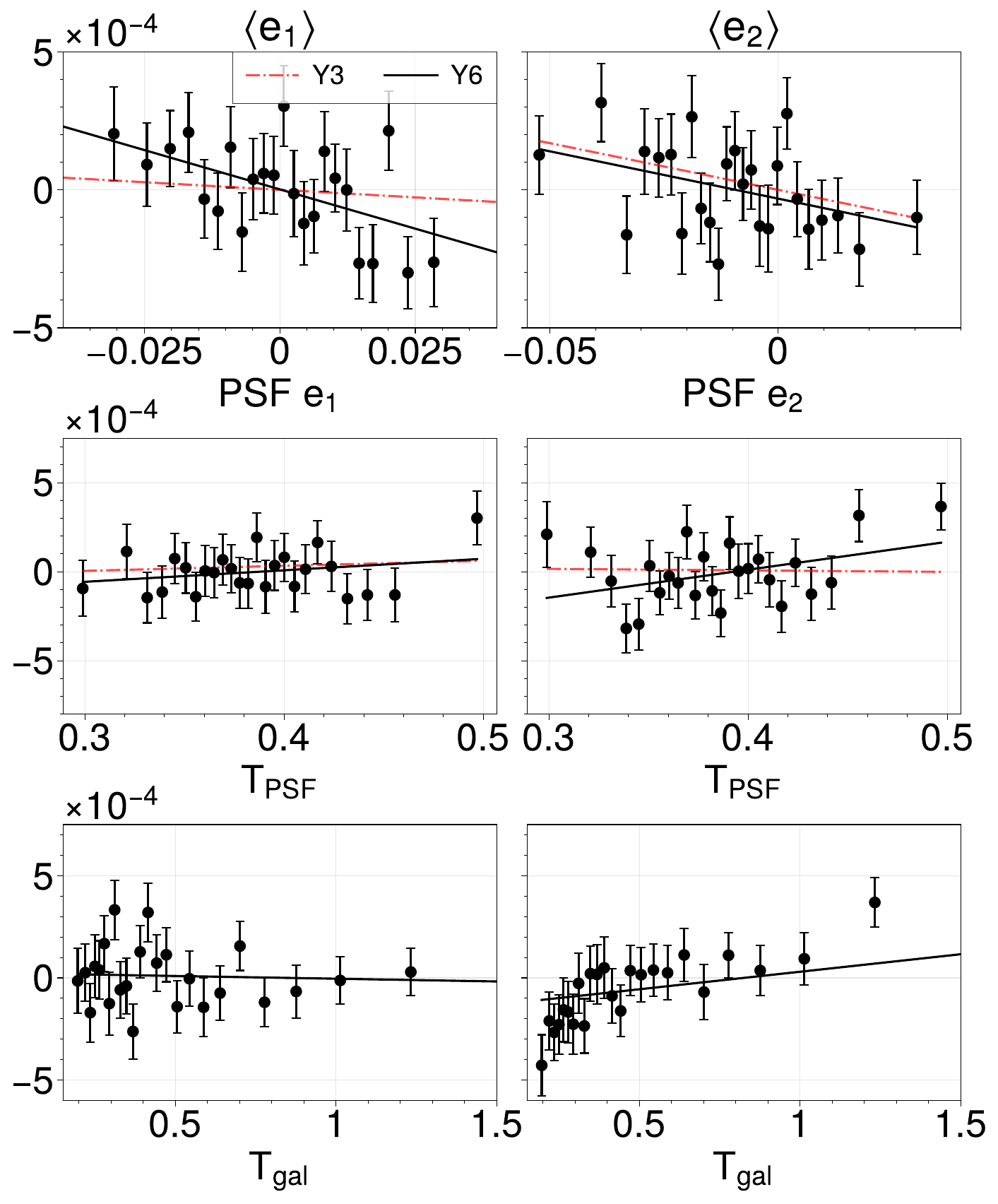}~
    \includegraphics[width=\columnwidth]{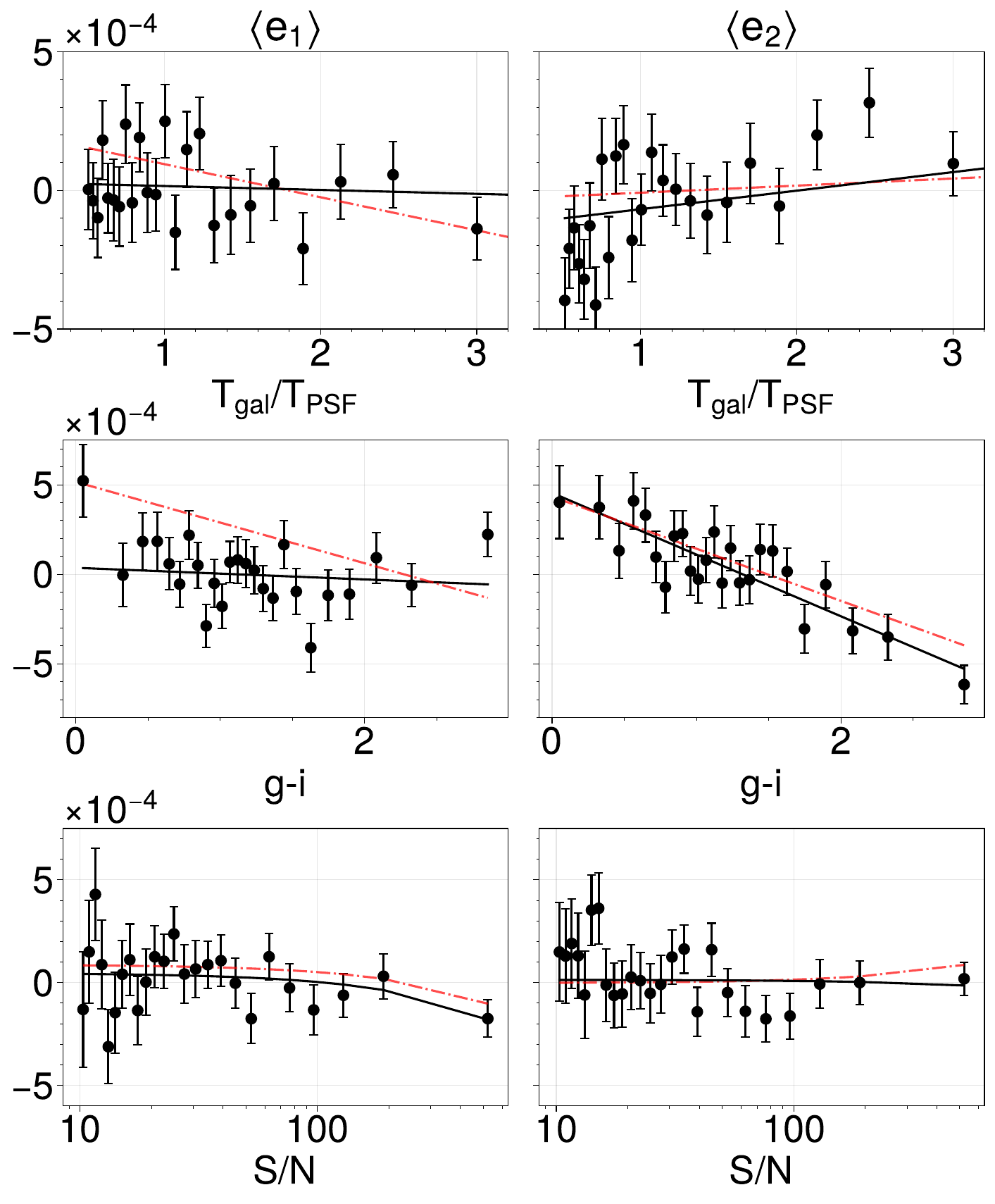}
    \caption{Mean shear ($e_1$, $e_2$) as a function of several PSF and galaxy properties. The mean shear is corrected by the shear response and the global mean shear in the catalogue is subtracted from each data point. The linear fit (slope and offset) to the data points was computed with ordinary least-squares. The error bars were estimated by dividing the sample into 200 patches in the sky and computed via jackknife covariance estimation. \textbf{\emph{Top-Left}}: as a function of PSF shape. \textbf{\emph{Middle-Left}}: as a function of PSF size. \textbf{\emph{Bottom-Left}}: as a function of galaxy size. \textbf{\emph{Top-Right}}: as a function of galaxy and PSF size ratio. \textbf{\emph{Middle-Right}}: as a function of galaxy $g-i$ color. \textbf{\emph{Bottom-Right}}: as a function of galaxy signal-to-noise (S/N).}
    \label{fig:meanshear_psf}
\end{figure*}

\begin{figure*}
	\includegraphics[width=0.88\textwidth]{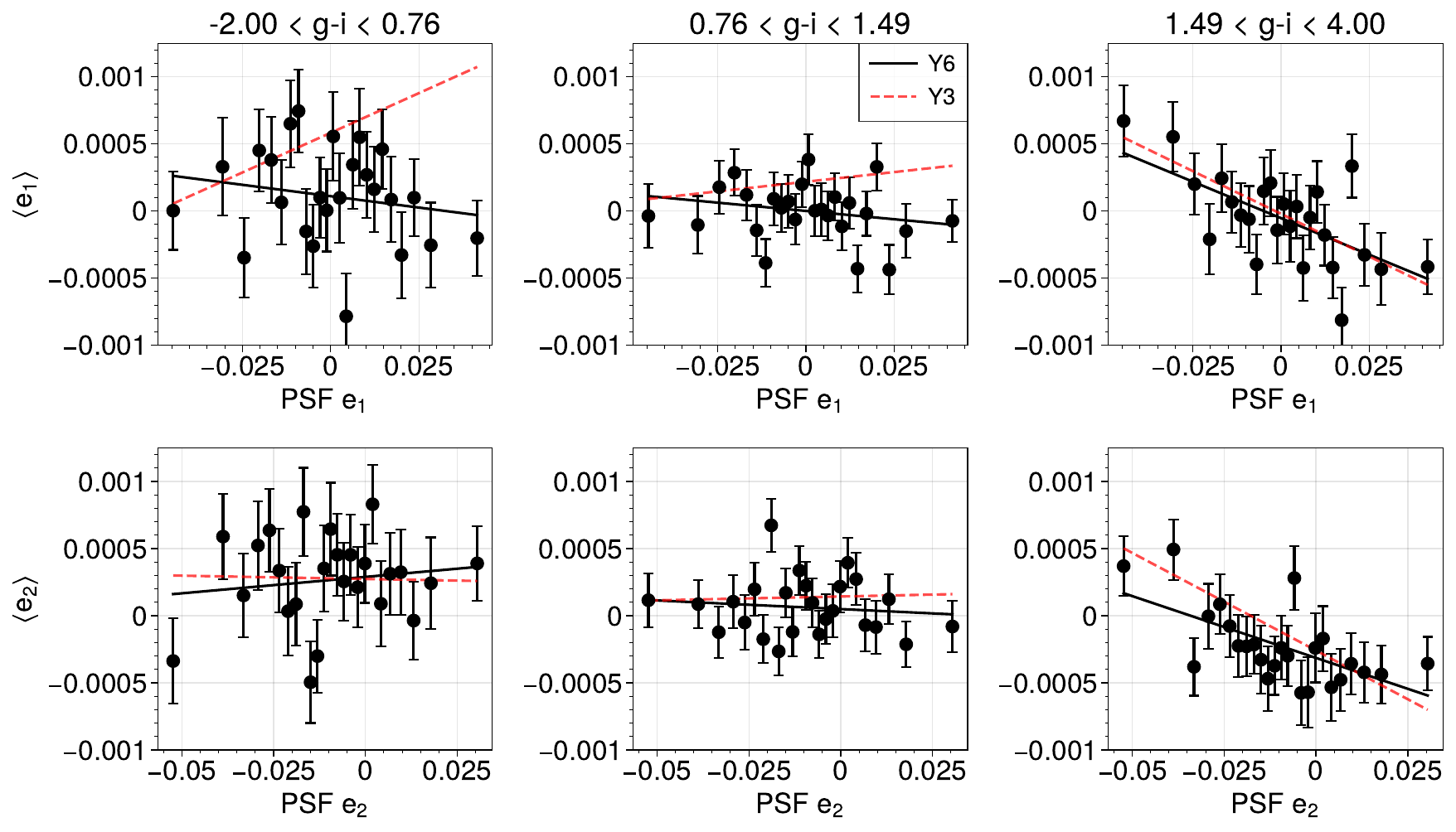}
    \caption{Mean shear (\emph{top} row: $e_1$ and \emph{bottom} row: $e_2$) as a function of PSF shapes where the catalogue is split in 3 ($g-i$) color bins. \textbf{\emph{Left}} column: color range is set between $-2.00 < g-i < 0.76$. \textbf{\emph{Middle}} column: color range is set between $0.76 < g-i < 1.49$. \textbf{\emph{Right}} column: color range is set between $1.49 < g-i < 4.00$. Ordinary least-square fits are performed on the data points to produce the black line, and red dotted line for the linear fit performed on the Y3 shape catalogue.
    }
    \label{fig:mean_shear_color_split}
\end{figure*}

A simple but effective test to ensure the quality of measured shear is to check the dependence of mean shear on the sizes and shapes of the PSF models. The left two columns in Figure~\ref{fig:meanshear_psf} include the PSF model dependency on galaxy shapes. Here, the shear response is re-computed in each bin of the PSF property, and the mean shear is divided by the response. Ideally, the slopes and intercepts of these relations would be zero. In practice, non-zero slopes and offsets can be driven by both mishandling of the PSF by the shear measurement code (PSF leakage) and PSF modeling errors. For our catalog, we find non-zero slopes and offsets as one can see in Fig.~\ref{fig:meanshear_psf} ($\partial e_1/\partial e_{1,\rm PSF}$ = -0.0057$\pm$0.0018 and $\partial e_2/\partial e_{2,\rm PSF}$ = -0.0034$\pm$0.0017). These quantities are about 2$\sigma$ larger for the slope in $e_1$ and statistically consistent in $e_2$ compared to the same test in GS20. This non-zero slope in $e_1$ can be explained by the PSF modeling errors associated with chromatic effects. When galaxy samples are split into three $g-i$ color bins (blue: [-2.00, 0.76], mid: [0.76, 1.49], red: [1.49, 4.00]), the slopes in each color sample are about the same or smaller compared to the samples in Y3 catalogue (Figure~\ref{fig:mean_shear_color_split}). The overall slope is, however, larger for Y6 since the signs of the slopes seen in Fig.~\ref{fig:mean_shear_color_split} for Y3 cancel to produce a smaller overall slope in Fig.~\ref{fig:meanshear_psf} for the full sample. We expect the slope in the middle color bin to be the smallest compared to blue and red sample bins in Y6, because we evaluate the PSF at the center of cell-based coadd at the median galaxy color ($g-i=1.1$) (Sec.~\ref{sec:coadd}) which most closely matches the middle color bin. Moreover, we find that in Y6, the additive bias in each color sample is typically smaller than in Y3, likely due to the chromatic PSF models. See the results in Table~\ref{tab:meanshear_color}.

It is essential to control these chromatic systematics across a wide color range, which we have significantly improved in Y6 vs.~Y3, since color is a close proxy for redshift. In a weak lensing cosmological analysis, even if the effects cancel as they did for the full sample in Y3, residual biases that modulate the cosmological weak lensing signal as a function of redshift would remain.

\begin{table*}
    \centering
    \caption{\label{tab:meanshear_color} Comparison of mean shear ($e_1$ and $e_2$) of the objects in Y3 and Y6 shape catalogue. We split the objects into three color ($g-i$) ranges (blue: [-2.00, 0.76], mid: [0.76, 1.49], red: [1.49, 4.00]). For the Y6 data, we compute the shear response of the objects that are selected to be in each defined color range. For Y3, we used shapes from the Y3 catalogue, and the shear and selection responses were re-computed for each color sample. We find improved additive bias control as a function of color.
    }
    \begin{tabular}[width=\textwidth]{cccccc}
        \hline\hline
             & Color sample & N & $\langle R \rangle$ & $\langle e_1 \rangle$ & $\langle e_2 \rangle$ \\ \hline
             & Blue & 16,043,877 & 0.707 & $(+9.75\pm0.75)\times10^{-4}$ & $(+3.34\pm0.75)\times10^{-4}$ \\
             Y3 & Middle & 48,602,072 & 0.743 & $(+5.79\pm0.39)\times10^{-4}$ & $(+2.02\pm0.39)\times10^{-4}$ \\
             & Red & 34,980,219 & 0.692 & $(+2.84\pm0.48)\times10^{-4}$ & $(-1.80\pm0.48)\times10^{-4}$\\
        \hline
             & Blue & 36,633,208 & 0.728 & $(+2.92\pm0.55)\times10^{-4}$ & $(+2.30\pm0.55)\times10^{-4}$ \\
            Y6 & Middle & 71,910,867 & 0.842 & $(+1.74\pm0.33)\times10^{-4}$ & $(+0.22\pm0.33)\times10^{-4}$ \\
             & Red & 43,326,212 & 0.833 & $(+1.52\pm0.43)\times10^{-4}$ & $(-2.73\pm0.43)\times10^{-4}$ \\
        \hline\hline
    \end{tabular}
\end{table*}

\begin{table}
    \centering
    \renewcommand{\arraystretch}{1.2}
    \caption{\label{tab:abe} Posterior constraints on the coefficients of the PSF contamination model.
    }
    \begin{tabular}[width=\columnwidth]{ccc}
\hline\hline
      parameter & 2.5' < $\theta$ < 250' & 0.25' < $\theta$ < 1000'\\ \hline
      $\alpha^{(2)}$ & $ 0.00980_{-0.00954}^{+0.0108} $ & $0.00682_{-0.00796}^{+0.00801}$ \\
     $\beta^{(2)}$ & $ 0.737_{-0.571}^{+0.576} $ & $1.19_{-0.23}^{+0.23}$ \\
     $\eta^{(2)}$ & $ 2.71_{-7.47}^{+6.90} $ & $-1.46_{-1.37}^{+1.38}$ \\
     $\alpha^{(4)}$ & $ -0.0115_{-0.0307}^{+0.0283} $ & $-0.00747_{-0.02491}^{+0.02488}$ \\
     $\beta^{(4)}$ & $ 1.94_{-0.268}^{+0.285} $ & $2.14_{-0.21}^{+0.21}$ \\
     $\eta^{(4)}$ & $ 3.10_{-2.60}^{+2.67} $ & $0.461_{-0.703}^{+0.702}$ \\
     $\eta^{(24)}$ & $ -1.27_{-0.677}^{+0.675} $ & $-0.103_{-0.100}^{+0.100}$ \\
     $\eta^{(42)}$ & $ -11.8_{-18.3}^{+20.2} $ & $-0.646_{-1.728}^{+1.754}$ \\
     $\chi^{2}$ & $218.62$ & $176.17$ \\
     $\chi^{2}_{\rm reduced}$ & $0.88$ & $0.71$ \\
     % $\alpha^{(2)}$ & $0.0104_{-0.0100}^{+0.0100}$ & $0.00682_{-0.00796}^{+0.00801}$ \\
     % $\beta^{(2)}$ & $0.737_{-0.570}^{+0.566}$ & $1.19_{-0.23}^{+0.23}$ \\
     % $\eta^{(2)}$ & $2.44_{-7.11}^{+7.04}$ & $-1.46_{-1.37}^{+1.38}$ \\
     % $\alpha^{(4)}$ & $-0.0128_{-0.0292}^{+0.0290}$ & $-0.00747_{-0.02491}^{+0.02488}$ \\
     % $\beta^{(4)}$ & $1.95_{-0.272}^{+0.274}$ & $2.14_{-0.21}^{+0.21}$ \\
     % $\eta^{(4)}$ & $3.15_{-2.59}^{+2.60}$ & $0.461_{-0.703}^{+0.702}$ \\
     % $\eta^{(24)}$ & $-1.28_{-0.664}^{+0.663}$ & $-0.103_{-0.100}^{+0.100}$ \\
     % $\eta^{(42)}$ & $-10.8_{-18.9}^{+18.9}$ & $-0.646_{-1.728}^{+1.754}$ \\
     % $\chi^{2}$ & $218.62$ & $176.17$ \\
     % $\chi^{2}_{\rm reduced}$ & $0.88$ & $0.71$ \\
\hline\hline
    \end{tabular}
\end{table}

\subsection{Additive Biases Due to PSF Leakage and Modeling errors}
\label{subsec:rho}

In this section, we constrain any additive biases incurred from how the PSFs are used by \mdet\ (PSF leakage errors) and any PSF misestimation (PSF modeling errors). We decompose the observed shape of a galaxy into its intrinsic shape, the real shear signal, the systematic contribution from the PSF, and noise.
\begin{equation}\label{eqn:psfcont_xi}
    g^{\rm obs} = g^{\rm int} + g + \delta e^{\rm sys}_{\rm PSF} + \delta e^{\rm noise}
\end{equation}

Following \citealt{2008A&A...484...67P, 2010MNRAS.404..350R, 2016MNRAS.460.2245J}, we can decompose $\delta e^{\rm sys}_{\rm PSF}$ into terms that scale like the PSF model shape, the PSF star image shape, and the product of the PSF size residual times the PSF star shape. Recently, \cite{2022arXiv221203257Z, 2023MNRAS.520.2328Z} pointed out that higher-order radial moments that scale as the radius to the fourth power, as opposed to the radius squared, are also important in this expansion and such terms were used the HSC Y3 analysis \citep{2022PASJ...74..421L}. We follow \cite{2023MNRAS.520.2328Z} and include spin-0 and spin-2 fourth-order moments as well (see Sec. 4 of \cite{y6psf} for the definitions of the PSF moments).
% \begin{equation}
%     e^{\rm (4)}_{\rm PSF} = \frac{M_{31} + iM_{13}}{T^2} - 3e^{\rm (2)}_{\rm PSF}, T^{\rm (4)}_{\rm PSF} = \frac{M_{22}}{M_{11}}.
% \end{equation}

With these contributions, the total PSF contamination term is
\begin{eqnarray}\label{eqn:model}
    \delta e^{\rm sys}_{\rm PSF} = \alpha^{(2)} e_{\rm PSF} + \beta^{(2)} \Delta e_{\rm PSF} + \eta^{(2)}(e_{*}\Delta T_{\rm PSF}/T_{\rm PSF}) \nonumber \\
    + \alpha^{(4)} e^{(4)}_{\rm PSF} + \beta^{(4)}\Delta e^{(4)}_{\rm PSF} + \eta^{(4)}(e^{(4)}_{*}\Delta T^{(4)}_{\rm PSF}/T^{(4)}_{\rm PSF}) \nonumber \\
    + \eta^{(24)}(e_{*}\Delta T^{(4)}_{\rm PSF}/T^{(4)}_{\rm PSF}) + \eta^{(42)}(e^{(4)}_{*}\Delta T_{\rm PSF}/T_{\rm PSF})
\end{eqnarray}
where $\Delta e_{\rm PSF} = e_{*} - e_{\rm PSF}$, $e_{\rm PSF}$ is the ellipticity of the PSF model, $e_{*}$ is the PSF ellipticity measured directly measured from stars, $\Delta T_{\rm PSF}=T_{*} - T_{\rm PSF}$, $T_{\rm PSF}$ is the PSF model size and $T_{*}$ is the PSF size measured from the star. The superscript $(4)$ on the ellipticity and size indicates a moment with a fourth-order radial term. If the galaxy shear estimator correctly uses the PSF model (i.e., no PSF leakage), we expect $\alpha^{(2)}$ and $\alpha^{(4)}$ to be zero. The impact of PSF modeling errors corresponds to non-zero detections $\beta$- and $\eta$-like terms, assuming the PSF modeling residuals themselves are non-zero.

\begin{figure}
    \includegraphics[width=\columnwidth]{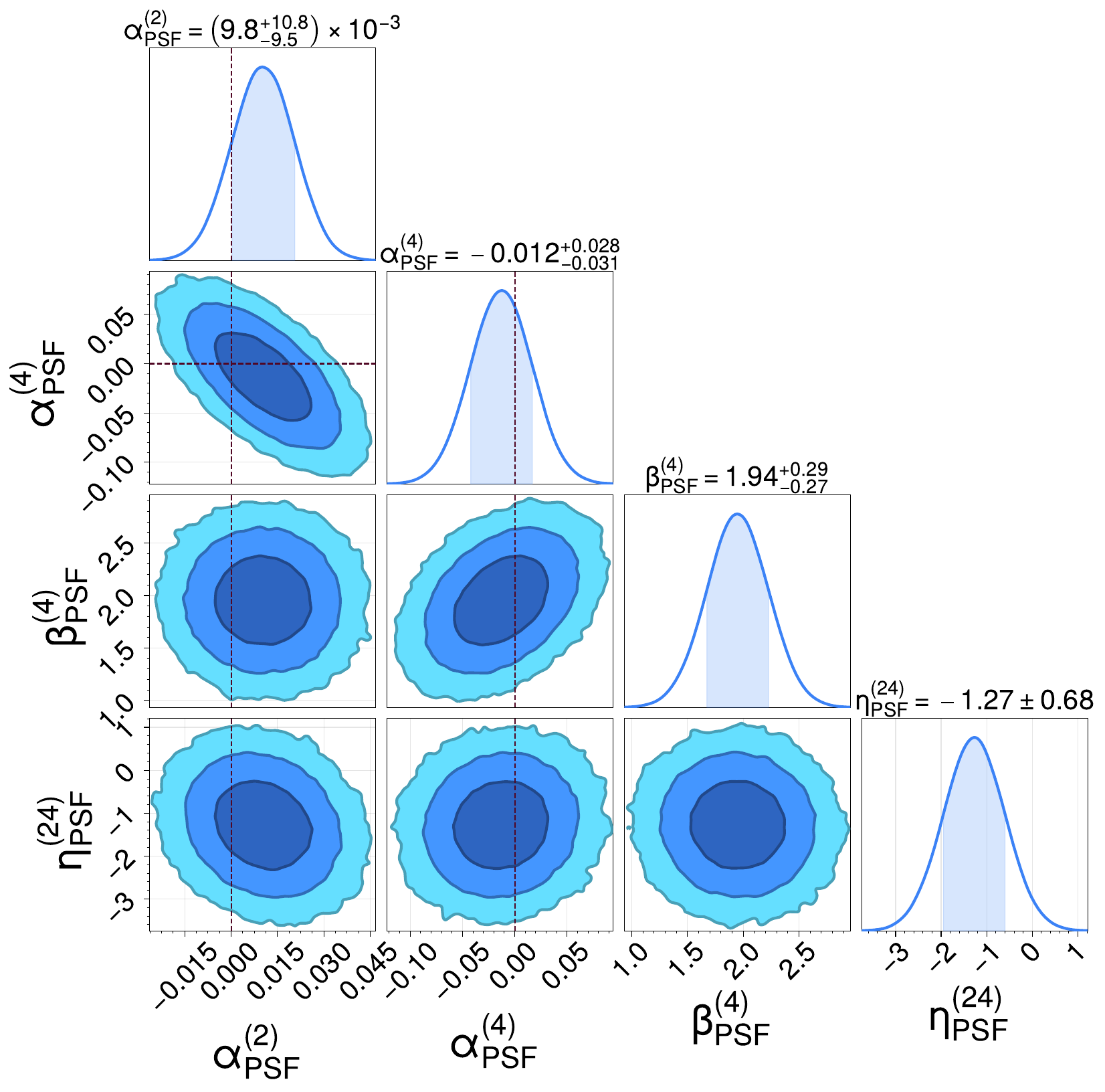}
    \caption{The non-null parameter posterior contours for the PSF contamination model for the fiducial angular scales and binning. The dotted lines show the parameter values that are expected with no PSF leakage, which are $\alpha^{(2)}, \alpha^{(4)}=0$. The $\alpha$ parameters are consistent with zero indicating no PSF leakage. The non-zero $\beta-$ and $\eta$ parameters indicate PSF modeling errors. The full posterior is in Fig.~\ref{fig:mcmc_posterior}.
    }
    \label{fig:abe_mcmc}
\end{figure}

\begin{figure*}
    \includegraphics[width=\textwidth]{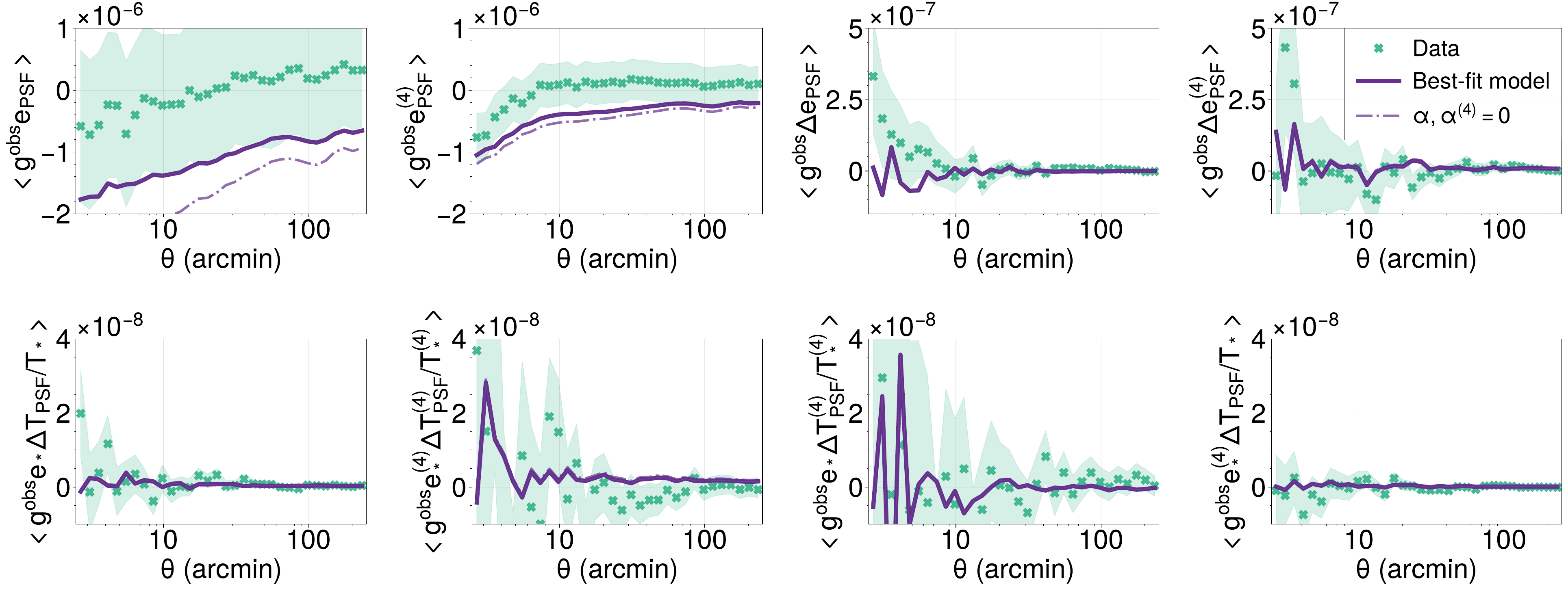}
    \caption{Galaxy-PSF cross-correlation measurements ($\tau$-statistics) for the fiducial angular scales and binning. The measurements are shown as the cross symbols. The 1$\sigma$ error on the data from the diagonal part of the covariance matrix is shown as a shaded region. The best-fit $\tau$-statistic model is shown as the line in each panel. Adjacent angular bins of the $\tau$-statistics are very correlated, so that the goodness-of-fit cannot be judged by visual inspection. The $\chi^2/{\rm dof}$ is 218.62/248 with $p$-value of 0.91.
    }
    \label{fig:taustats}
\end{figure*}

To constrain this model, we measure the full set of PSF shape auto-correlations and the full set of PSF-galaxy shape cross-correlations. Using the model above, we can then make predictions for these signals and constrain the model coefficients. See Appendix~\ref{app:psf}, Eqn.~\ref{eqn:tau0}-\ref{eqn:tau452}, for the full expressions. The galaxy-shear cross-correlation statistics on the left-hand side of Eqn.~\ref{eqn:tau0}-\ref{eqn:tau452} are usually denoted as $\tau$-statistics. The auto-correlations of the PSF quantities are usually denoted as $\rho$-statistics. \cite{y6psf} present the $\rho$-statistics for the Y6 PSF models and demonstrate improvements relative to the DES Y3 PSF models (see their Fig.~20). In our case, we evaluate the PSF models at the locations of the reserve stars using the median galaxy color, producing the reserve star catalogue. We then use this catalog, combined with shape measurements performed on the stars directly and the \mdet\ catalog, to compute the various two-point correlation functions in 32 angular bins ranging between 2.5 - 250 arcmin. We estimate the covariance of our measurements from the 800 mock catalogues (see Sec.~\ref{subsec:mocks}), applying statistical corrections for the finite number of mock realizations \citep{2022MNRAS.510.3207P}.

Given the measurements and their covariance, we find the best-fit parameters to the contamination model (Eqn.~\ref{eqn:model}) by minimizing the $\chi^2$ with the Nelder-Mead algorithm \citep{nelder-mead}. We also sample the posterior using Markov Chain Monte Carlo (MCMC) with the {\sc emcee} package \citep{emcee}. We show a subset of the 1D and 2D marginal posterior contours of the parameters in Fig.~\ref{fig:abe_mcmc}, and 1D marginalized posterior constraints for all parameters are in Table~\ref{tab:abe}. For 2d marginalized plots of the posterior for all parameters, see Fig.~\ref{fig:mcmc_posterior}. Overall, the model fits the data well, with a $\chi^2$/dof of 218.62/248, giving a $p$-value of 0.91. Our best-fit model to the $\tau$-statistics measurement is shown as a violet solid line in Fig.~\ref{fig:taustats} and is in good agreement with the measurements shown as the crosses. The shaded region shows the 1$\sigma$ error from the diagonal part of the covariance matrix. The covariance matrix has significant off-diagonal terms, so the model's goodness-of-fit cannot be judged visually. While most parameters are consistent with zero, we do detect the effects of PSF modeling errors in a few of the $\beta$ and $\eta$ terms. We find no evidence of PSF leakage in our galaxy sample since $\alpha^{(2)}$ and $\alpha^{(4)}$ are consistent with zero. Finally, with the model parameter constraints, we can then use the PSF contamination model to predict the effect of the residual PSF modeling errors on the cosmological shear two-point functions. While we defer a full analysis of this systematic effect to future work \citep{y6_1x2pt}, preliminary results indicate that the current level of contamination would shift the cosmological parameters $S_8$ and $\Omega_m$ by $\lesssim0.1\sigma$.

\subsection{Tangential and Cross-component Shear Around Stars} \label{subsec:tanshear_stars}
In this test, we explore the tangential and cross-component shear around the positions of faint and bright stars. This test is potentially sensitive to PSF leakage and modeling errors. Bright stars, in particular, have large stellar wings and misestimated backgrounds, which may adversely affect shear signals.

Among the stars in our reserved star catalogue, we first select a unique set of stars that made it to the coadd images. We then divide the stars into bright ($m_r$<16.5) and faint ($m_r$>16.5) samples. We then compute the tangential and cross-component shear signals, making sure to subtract the signal around random points. We note that in order to guarantee the uniform distribution of galaxies and stars in our footprint, we apply weights to the star sample, which is the ratio of the number of galaxies and stars in each {\sc HealPIX} grid. These weights correct for a subtle correlation between the stellar density and galaxy density due to detection that can result in a non-null signal (see GS20 for more details).

\begin{figure}
	\includegraphics[width=0.8\columnwidth]{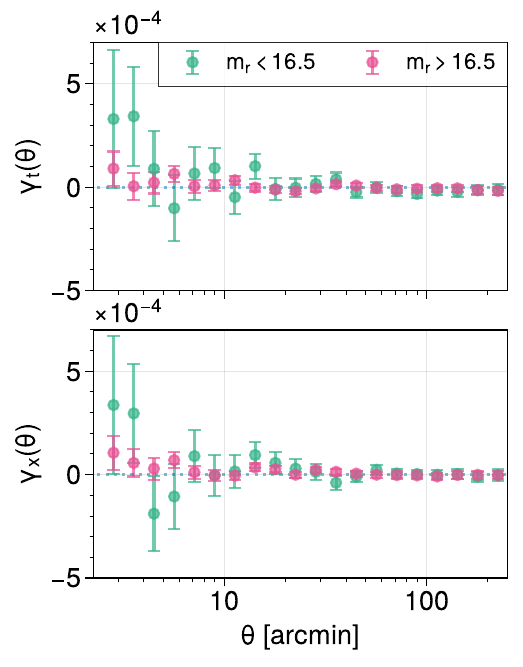}
    \caption{Tangential and cross-component shear around bright and faint stars. \textbf{\textit{Top}}: tangential shear around bright and faint stars with $r$-band magnitude $\lessgtr$ 16.5. The Reduced $\chi^2$ statistic computed using the full covariance matrix for each $\gamma_{\rm t}$ is $\chi^2_{\rm reduced}$=12.2/20 and $\chi^2_{\rm reduced}$=21.3/20, respectively. \textbf{\textit{Bottom}}: cross-component shear around bright and faint stars with $r$-band magnitude $\lessgtr$ 16.5. The reduced $\chi^2$ statistic computed using the diagonal components of the covariance matrix for each $\gamma_{\rm x}$ is $\chi^2_{\rm reduced}$=16.7/20 and $\chi^2_{\rm reduced}$=17.3/20, respectively. }
    \label{fig:tan_star}
\end{figure}

Figure~\ref{fig:tan_star} shows the result of this test. For the shear around bright star sample, the reduced $\chi^2$ statistic with respect to a null signal is $\chi^2_{\rm reduced}$=12.2/20 for $\gamma_t$ and $\chi^2_{\rm reduced}$=16.7/20 for $\gamma_x$. We see no evidence of the impact around bright stars. For the shear around faint star sample, the reduced $\chi^2$ statistic with respect to a null signal is $\chi^2_{\rm reduced}$=21.3/20 for $\gamma_t$ and $\chi^2_{\rm reduced}$=17.3/20 for $\gamma_x$. We again see no significant deviation from a null signal for the faint sample.

\section{Shape Catalogue Systematics Tests} \label{sec:shearsyst}
Now we perform null tests of the shape catalogue against various survey properties and coordinate systems in an effort to search for unknown residual systematic effects. These tests are largely sensitive to additive errors in the catalogue, and their overall statistical power is limited by the shape noise in the catalogue itself. In this section, we performed the following tests,
\begin{itemize}
    \item shear variations in the focal plane, cell-based coadd, coadd coordinates (Sec.~\ref{subsec:meanshear_focalcoord});
    \item tangential and cross-component shear around field/exposure, cell-based coadd, coadd centers (Sec.~\ref{subsec:tanshear_field});
    \item B-modes (Sec.~\ref{subsec:bmode});
    \item dependencies of mean shear on survey properties (Sec.~\ref{subsec:survey_property}).
\end{itemize}

\subsection{Mean Shear Variations in Focal Plane, Cell-based Coadd, Coadd Tile Coordinates} \label{subsec:meanshear_focalcoord}
In this test, we explore the mean shear variations across different image coordinate systems. These variations, especially in CCD coordinates, can help to detect residual effects like persistently mismasked bad CCD columns, charge-transfer issues, glowing edges, etc. Shear variations in focal plane coordinates can point to both systematic effects from the CCDs and systematic effects from the optics. Shear variations in the coadd tile and cell-based coadd coordinates could be sensitive to systematics from our new cell-based coadding algorithms.

We compute the response-corrected mean shear in each of these coordinate systems by projecting each object in the \mdet\ catalogue into the respective coordinate system. In the case of projecting to CCD coordinates, we project each object to its location in each CCD in which it was measured, and so objects are duplicated. We were also careful to align the read directions of the CCDs as needed, which differ in different parts of the focal plane. We then build coarse grids composed of patches of $128\times128$ pixels for CCD/focal plane coordinates, 10 $\times$ 10 pixels grid for cell-based coadd coordinates, and 250 $\times$ 250 pixels grid for coadd coordinates. Finally, we compute the response-corrected mean shear for these coarse grids, removing the overall mean shear from the catalogue.

\begin{figure*}
	\includegraphics[width=\textwidth]{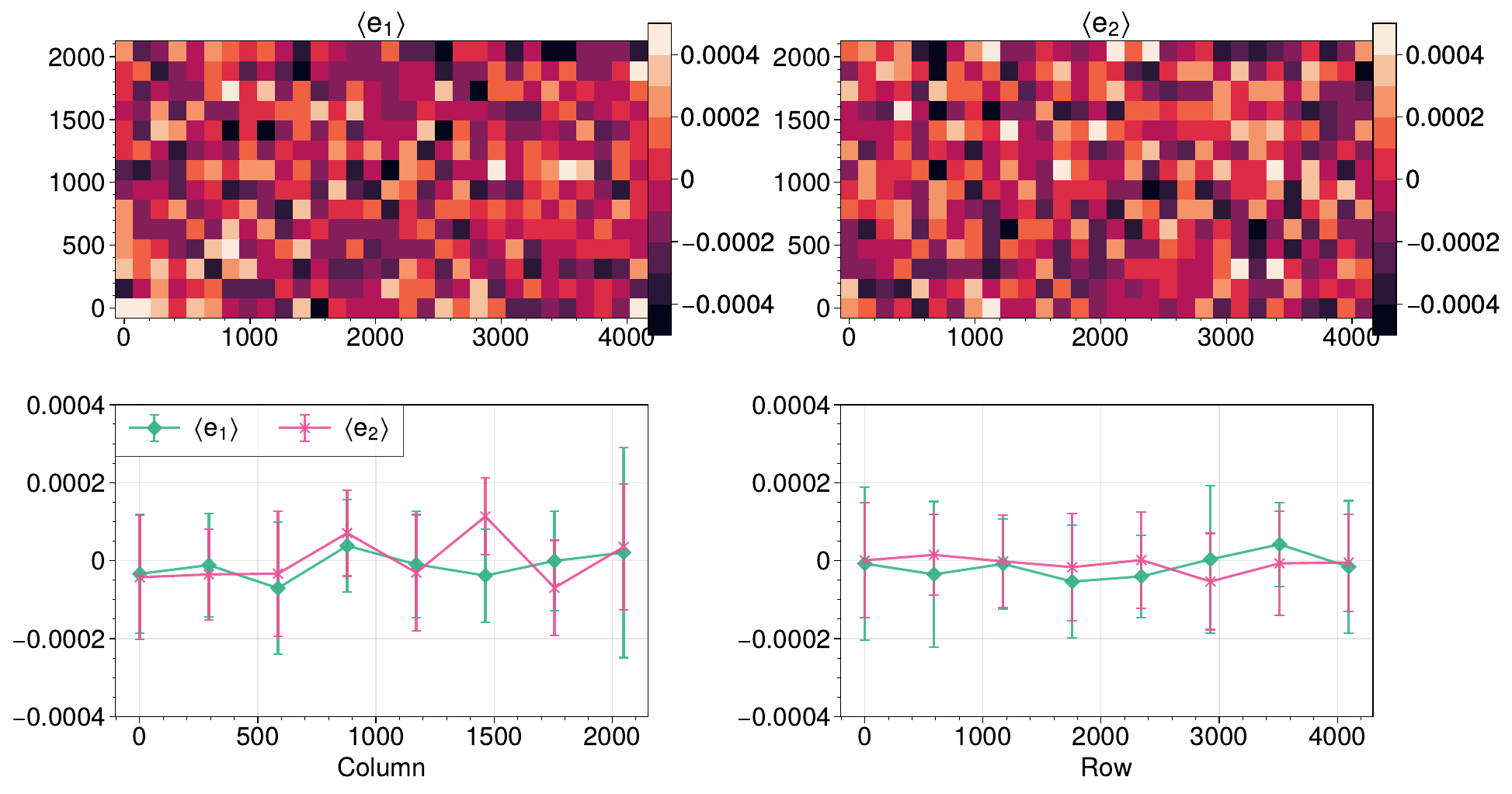}
    \caption{\textbf{\emph{Top Row}}: mean shear ($e_1$ and $e_2$) in the CCD coordinate system. We use cells of $128\times128$ pixels and compute the response corrected mean shear in each cell. All CCDs from all exposures that contribute to a shear measurement are used. Each object detected in the \mdet\ catalogue is binned into the CCD according to its position projected into the CCD coordinate system. \textbf{\emph{Bottom Left}}: The response-corrected mean shear as a function of the CCD column. \textbf{\emph{Bottom Right}}: The response-corrected mean shear as a function of the CCD row. Note that global mean shear residuals in the catalogue are subtracted in all panels. }
    \label{fig:meanshear_focal_xy}
\end{figure*}

\begin{figure*}
    \centering
    \includegraphics[width=0.92\columnwidth]{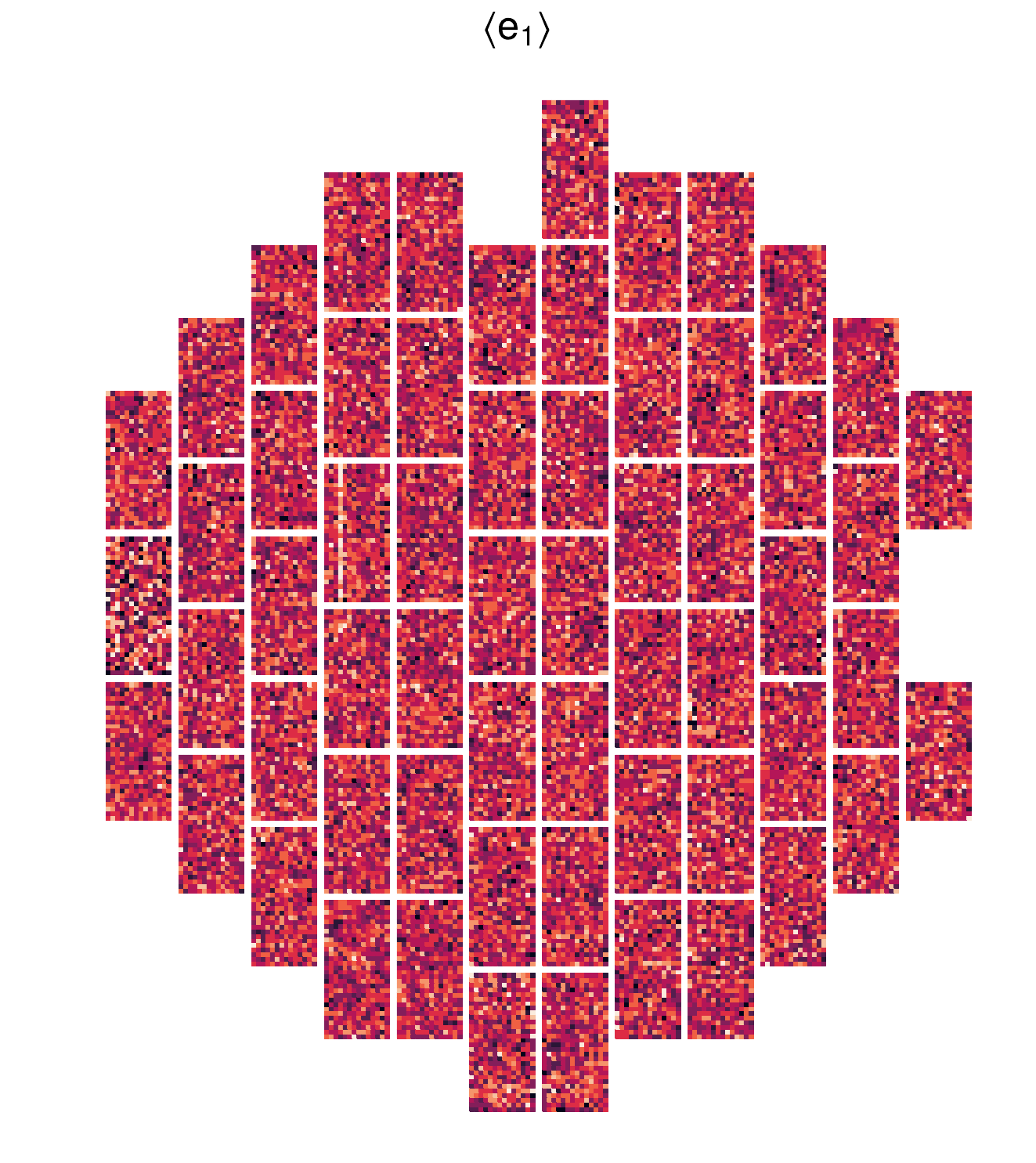}~
    \includegraphics[width=0.92\columnwidth]{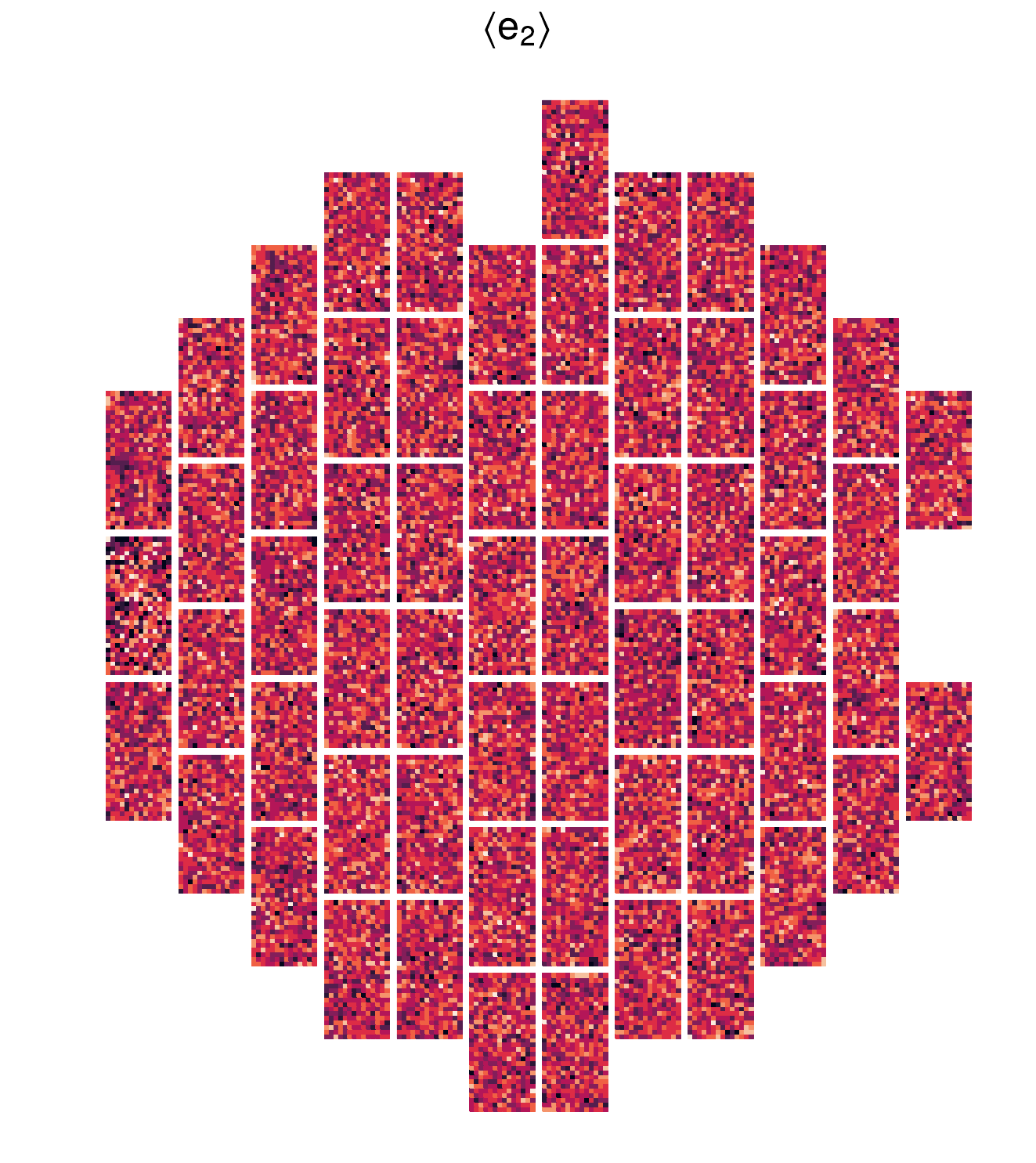}
    \includegraphics[width=0.2\columnwidth]{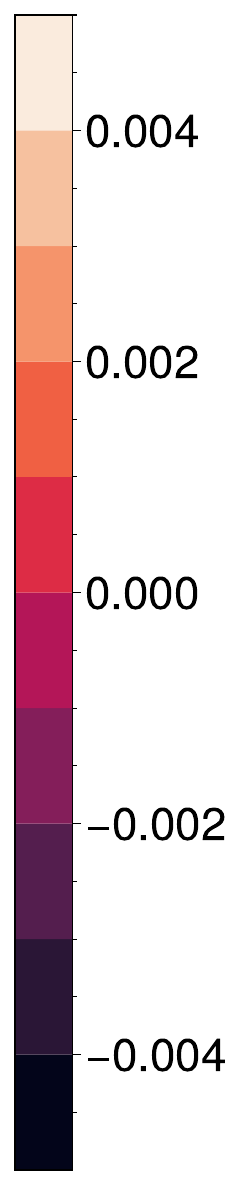}~
    \caption{\label{fig:meanshear_focal_coarse} Mean shear (\textbf{\emph{left}}: $e_1$ and \textbf{\emph{right}}: $e_2$) in focal plane coordinates, which covers the field of view of 2.2 deg$^2$.  Galaxy ellipticities are binned into a grid of $128\times128$ pixel cells covering each CCD. We use all $r$-band exposures and compute the response corrected mean shear in each grid cell. We find a $\chi^2/{\rm dof}$ = 26387/25200 for $e_1$ and $\chi^2/{\rm dof}$ = 25802/25200 for $e_2$. Note that global mean shear residuals in the catalogue are subtracted in all panels.}
\end{figure*}

Figure~\ref{fig:meanshear_focal_xy} shows the mean shear variations where all CCDs are stacked together so that the signal in each CCD can be enhanced if it exists. In the bottom panel, the rows and columns of the grid are summed together to enhance any signals here. In both cases, we find no significant trend of measured shear as a function of focal plane coordinates. We also present the mean shear variations in all CCDs in Fig.~\ref{fig:meanshear_focal_coarse}. Known CCD failures in the focal plane can be seen missing in the two panels of the figure, and a systematically missed bad column is visible in the pixel data in mean $e_1$ on the left half of the focal plane. The change in shear due to this bad column is $\approx+0.005$, but this contribution is down-weighted by the fraction of the data that comes from this CCD, $\approx1/60$, times the fraction of objects that intersect with the bad column, $\lesssim1/10$. Combining these numbers, we get a contamination level of $\lesssim10^{-5}$. This level is much smaller than any mean shear signals we detect in the catalogue and can be ignored. The reduced $\chi^2$ with respect to null detection is $\chi^2/{\rm dof}$ = 26621/25200 for $e_1$ and $\chi^2/{\rm dof}$ = 25844/25200 for $e_2$.

\begin{figure}
	\includegraphics[width=\columnwidth]{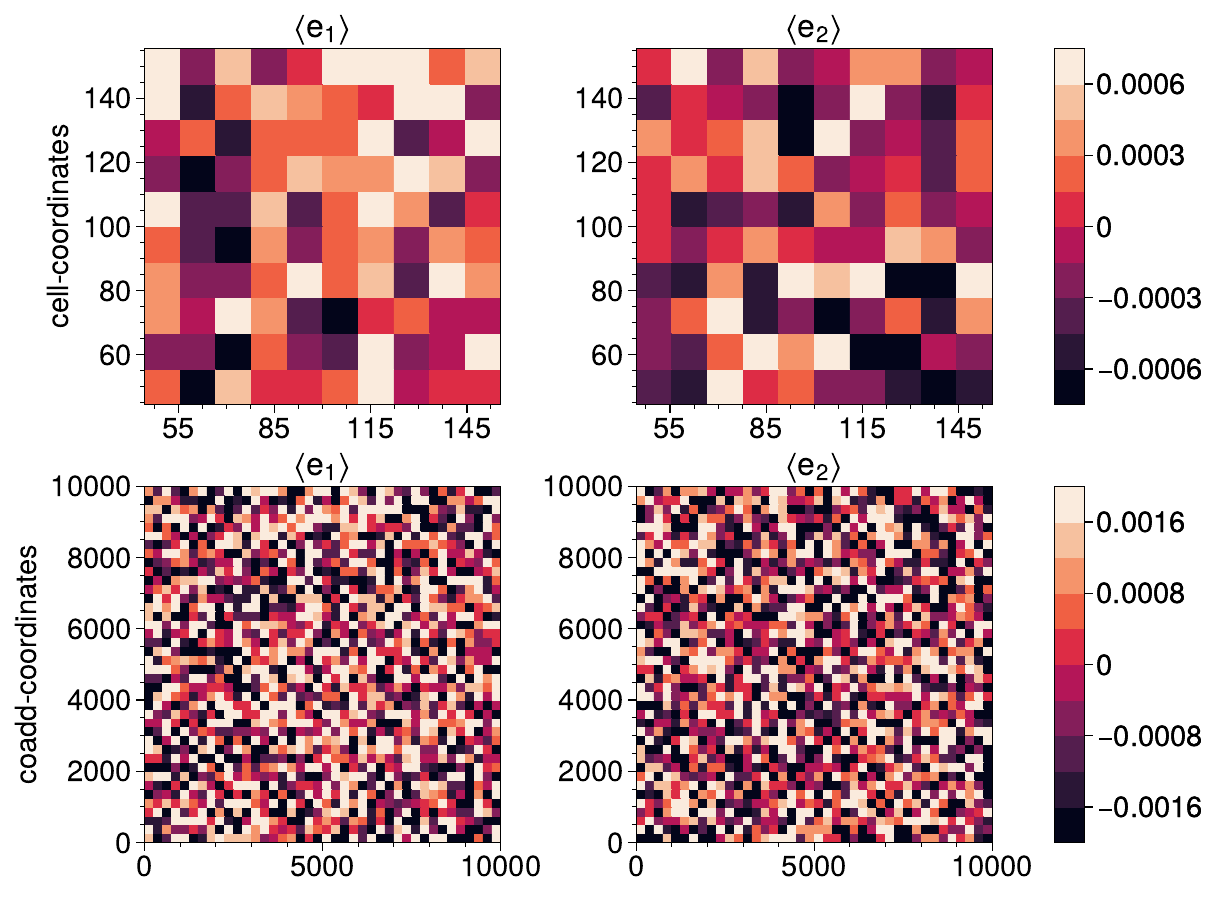}
    \caption{\textbf{\emph{Top Row}}: mean shear ($e_1$ and $e_2$) in cell-based coadd coordinates. Due to a buffer region around each cell-based coadd where object detections are removed since they overlap with adjacent cell-basec coadds, the figure covers the central $\sim100\times100$ pixel region of a cell-based coadd. The $\chi^2$ with respect to zero mean shear is $\chi^2_{\rm reduced}$=66.5/64 for $e_1$ and $\chi^2_{\rm reduced}$=70.3/64 for $e_2$. \textbf{\emph{Bottom Row}}: mean shear ($e_1$ and $e_2$) in coadd tile coordinates. A coadd tile is 10,000 $\times$ 10,000 pixels. The $\chi^2$ with respect to zero mean shear is, $\chi^2_{\rm reduced}$=1501/1444 for $e_1$ and $\chi^2_{\rm reduced}$=1388/1444 for $e_2$. In both coordinate systems, objects are binned into grid cells consisting of multiple pixels and then mean response-corrected shear is measured for each cell. for this test, the global mean shear in the catalogue is subtracted
    }
    \label{fig:meanshear_cell+coadd}
\end{figure}

As for shear variations in cell-based coadd and coadd coordinates, Fig.~\ref{fig:meanshear_cell+coadd} shows no significant pattern appearing in both coordinate systems. The reported $\chi^2$ with respect to zero mean shear is, $\chi^2_{\rm reduced}$=66.5/64 in $e_1$ and $\chi^2_{\rm reduced}$=70.3/64 in $e_2$ for cell-based coadd coordinates and $\chi^2_{\rm reduced}$=1501/1444 in $e_1$ and $\chi^2_{\rm reduced}$=1388/1444 in $e_2$ for coadd coordinates.

\subsection{Tangential Shear Around Cell-based Coadd, Coadd Tile, and Field/Exposure Centers} \label{subsec:tanshear_field}

\begin{figure*}
	\includegraphics[width=\textwidth]{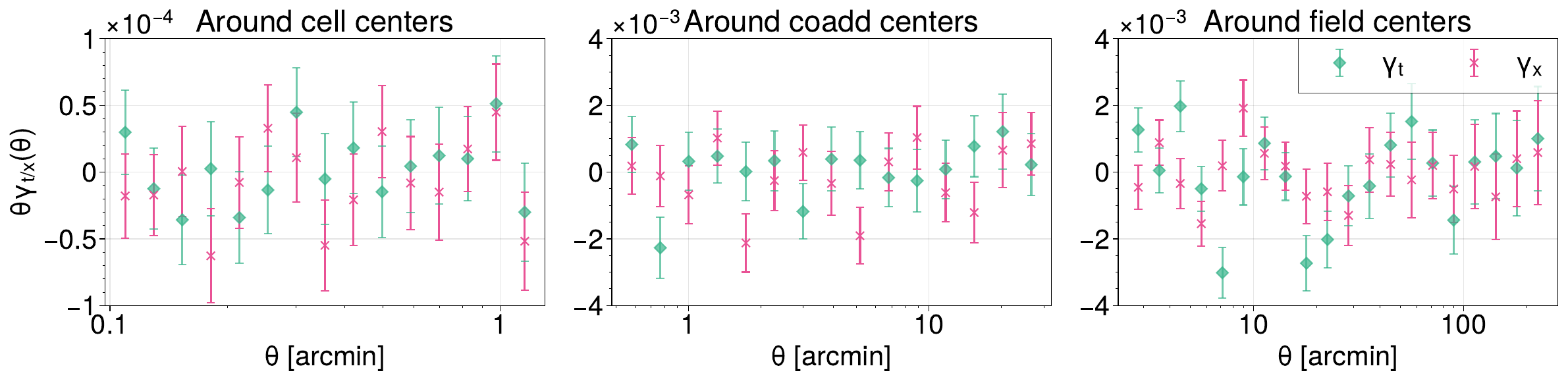}
    \caption{ \textbf{\emph{Left}}: tangential and cross-component shear around the centers of cell-based coadds, which have a dimension of 200 x 200 pixels. The reduced $\chi^2$ statistics computed using the full covariance matrix for each $\gamma_{\rm t}$ and $\gamma_{\rm x}$ is $\chi^2_{\rm reduced}$=6.37/15 and $\chi^2_{\rm reduced}$=16.6/15, respectively. \textbf{\emph{Middle}}: tangential and cross-component shear around the centers of coadd tiles, which have a dimension of 10,000 x 10,000 pixels. The reduced $\chi^2$ statistics computed using the full covariance matrix for each $\gamma_{\rm t}$ and $\gamma_{\rm x}$ is $\chi^2_{\rm reduced}$=12.9/15 and $\chi^2_{\rm reduced}$=19.4/15, respectively. \textbf{\emph{Right}}: tangential and cross-component shear around the centers of fields/exposures. The reduced $\chi^2$ statistics computed using the full covariance matrix for each $\gamma_{\rm t}$ and $\gamma_{\rm x}$ is $\chi^2_{\rm reduced}$=61.1/20 and $\chi^2_{\rm reduced}$=19.4/20, respectively. The number of field/exposure centers used for this test was 47610. }
    \label{fig:tan_field}
\end{figure*}

Shear measurements should not correlate with any special points in the various intermediate coordinate systems used for key measurements. Of particular interest is the correlation of shear signals with centers of DES fields/exposures. Residual PSF leakage and modeling errors driven by the telescope optics could potentially be more easily detected in a coordinate system where the signals have a chance to constructively interfere. We use three different sets of special points. The centers of individual cell-coadds, the centers of DES coadd tiles, and the centers of fields/exposures. The number of field/exposure centers used in this test is 47610 (only $r$-band). We present the result in Fig.~\ref{fig:tan_field}. From the left to right panels, it shows both the tangential and cross-component shear around cell-based coadd, coadd, and field/exposure centers. The reduced $\chi^2$ with respect to a null signal is $\chi^2_{\rm reduced}$=6.37/15 for $\gamma_t$ and $\chi^2_{\rm reduced}$=16.6/15 for $\gamma_x$ around cell-based coadd centres. $\chi^2_{\rm reduced}$=12.9/15 for $\gamma_t$ and $\chi^2_{\rm reduced}$=19.4/15 for $\gamma_x$ around coadd centres. $\chi^2_{\rm reduced}$=61.1/20 for $\gamma_t$ and $\chi^2_{\rm reduced}$=19.4/20 for $\gamma_x$ around field/exposure centres. These $\chi^2$ are measured with full covariance matrices associated with each measurement and we have subtracted the same signals measured around random points.

As seen in GS20, we see a non-null tangential shear signal around field/exposure centers. In order to verify that we are not dominated by cosmic variance, we run the same test on 800 mock catalogues (see Sec.~\ref{subsec:mocks}) and re-estimate the covariance matrix. Nonetheless, we do not see any improvement in our measurement and report $\chi^2_{\rm reduced}$=67.7/20 for $\gamma_t$.

\begin{figure*}
    \includegraphics[width=\textwidth]{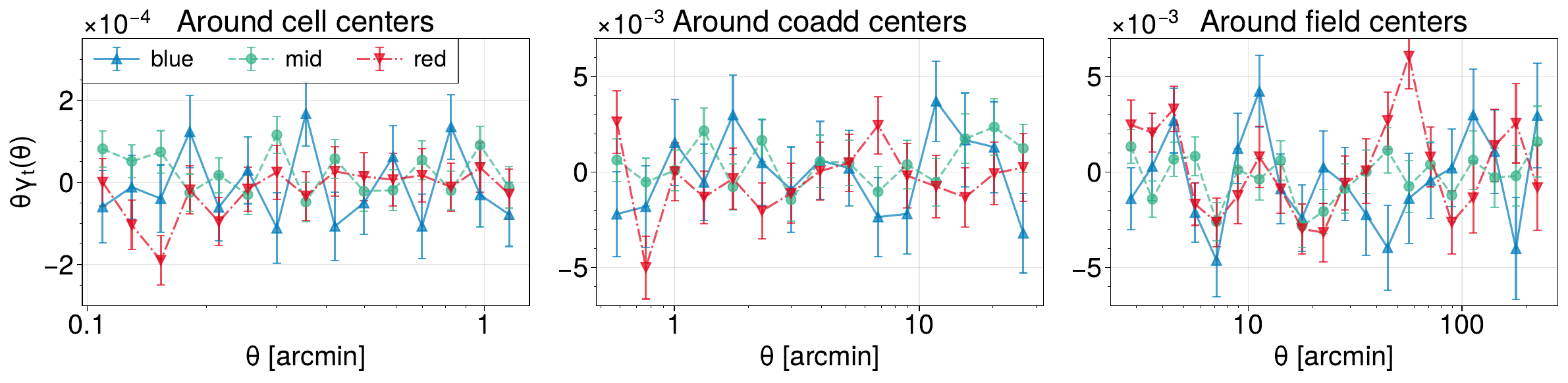}
    \caption{Tangential shear around cell-based coadd, coadd, and field/exposure centers where the galaxy sample is split by their measured $g-i$ color (blue: [-2.00, 0.76], mid: [0.76, 1.49], red: [1.49, 4.00]). }
    \label{fig:tan_cell+coadd+field_color}
\end{figure*}

\begin{figure}
    \includegraphics[width=\columnwidth]{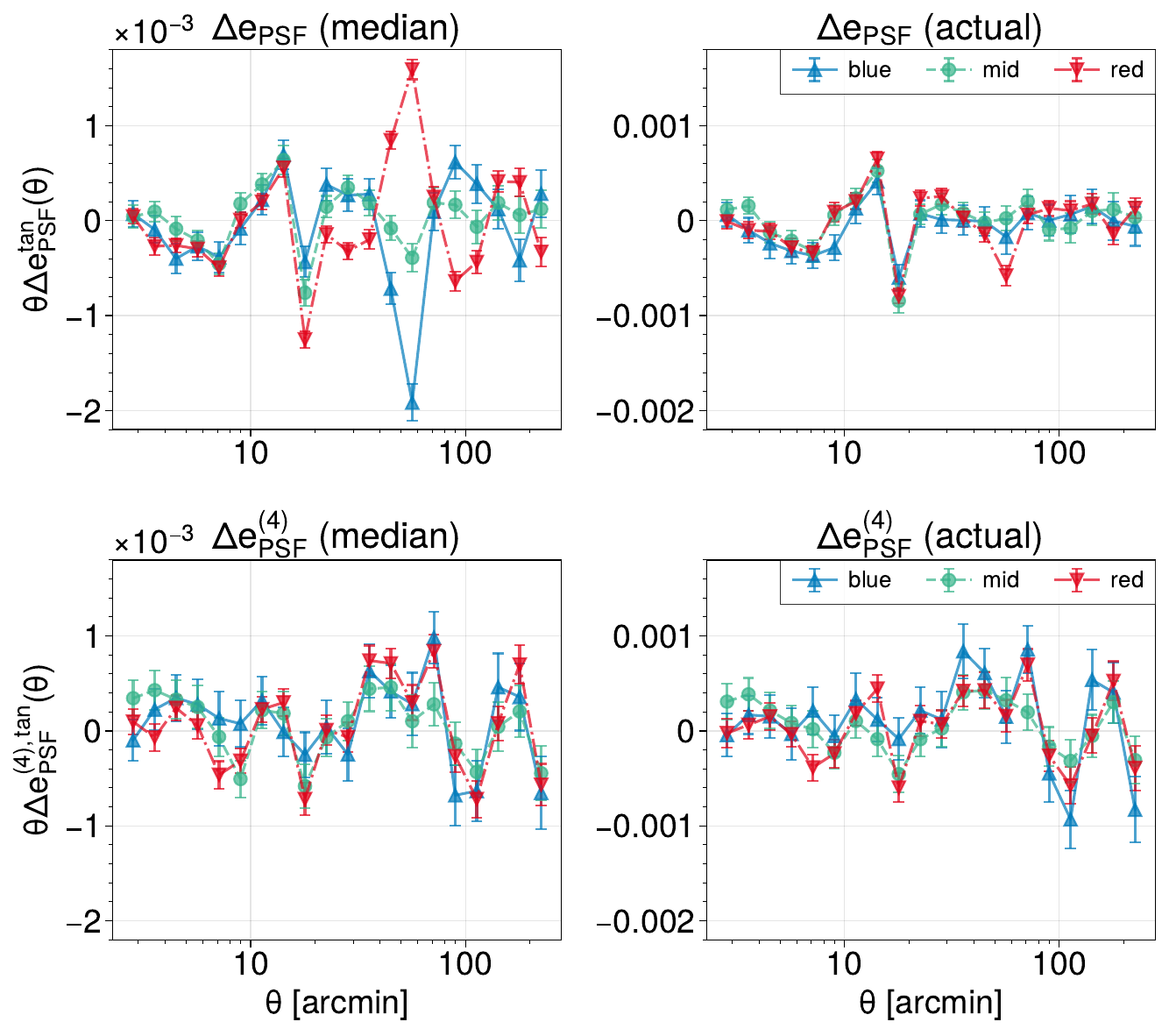}
    \caption{Tangential projection of residual PSF 2nd order (\textbf{\emph{top}}) and 4th order (\textbf{\emph{bottom}}) shapes around field/exposure centers where the star sample is split by their measured $g-i$ color (blue: [-2.00, 0.76], mid: [0.76, 1.49], red: [1.49, 4.00]). \textbf{\emph{Left}} panel shows the case when the PSFs were drawn at their median color, which is the case in our \mdet\ shape catalogue. \textbf{\emph{Right}} panel shows the case when the PSFs were drawn at their actual star colors.}
    \label{fig:tan_field_PSF_color}
\end{figure}

In order to better understand the source of this signal, we split our galaxy sample by their $g-i$ colors. Figure~\ref{fig:tan_cell+coadd+field_color} shows the same measurement of tangential shear in different coordinate systems. We again find no trend around cell-based coadd and coadd centers, but find a significant deviation from the null signal for the field/exposure centers around 7-8 arcmin, 10-20 arcmin, and 40-60 arcmin scale in the right panel of Fig~\ref{fig:tan_cell+coadd+field_color}. This deviation corresponds to the deviations seen in the right panel of Fig~\ref{fig:tan_field}. Some of these effects can be understood as PSF modeling errors due to \mdet\ using the median galaxy color for the PSF. Figure~\ref{fig:tan_field_PSF_color} shows the same statistics as we measured for the galaxies, but for the PSF modeling residuals ($e_* - e_{\rm PSF}$) when using PSF models evaluated at the median galaxy color versus the stars actual color. Even when the PSFs are evaluated at their actual star color, the right panel of the figure suggests the non-zero patterns at scales around 10-20 arcmin, which corresponds to the scales of each CCD.

As for the cosmological relevance of this signal, we note that it is included already in the analysis of the PSF-shear cross-correlations above. Thus no additional correction specific to this signal is required. Further, we find effects that are smaller than those found in GS20 where they concluded that the overall level of contamination in the cosmic shear correlation functions is at the $\sim0.01\%$ level. Thus we conclude that we can safely ignore this systematic effect.

\subsection{B-mode Measurements} \label{subsec:bmode}
In this section, we measure the B-mode signals in our catalogue. To first order, gravitational lensing produces shear fields with only E-modes (i.e. tangential or radial patterns). At higher order, lensing and intrinsic alignment can produce small amounts of B-modes (i.e., shear patterns that appear to be spirals) as well. However, at the current sensitivity of data, measuring and detecting B-mode signal would more likely point to a systematic bias in the shear catalogue, such as PSF modeling.

We measured the B-mode signal in our catalogue with a pseudo-$C_{\ell}$ estimator \citep{hikage_pseudocell}. We first created two HEALPix shear maps with \texttt{nside}=1024 of weighted average of response-corrected galaxy ellipticities $e_1$ and $e_2$. With the pseudo-$C_{\ell}$ estimator in \texttt{NaMaster} \citep{alonso_namaster}, we can measure the E-mode and B-mode power spectra from the shear maps in the multipole of range $\ell$ = [2, 3071] in 30 bins. The pseudo-$C_{\ell}$ estimator accounts for the effects of the survey mask when computing the power spectra. To properly account for the contribution due to shape noise and E/B-mode mixing, we measured the B-mode signal on 800 mock catalogues (see Sec.~\ref{subsec:mocks}) that are generated from a pure E-mode cosmic shear field and took an average of the measured B-mode power spectra. This power spectrum is subtracted from the measured B-mode in the data. The covariance matrix of the pseudo-$C_{\ell}$ B-modes was obtained from the measurement in the simulations as well.

\begin{figure}
	\includegraphics[width=\columnwidth]{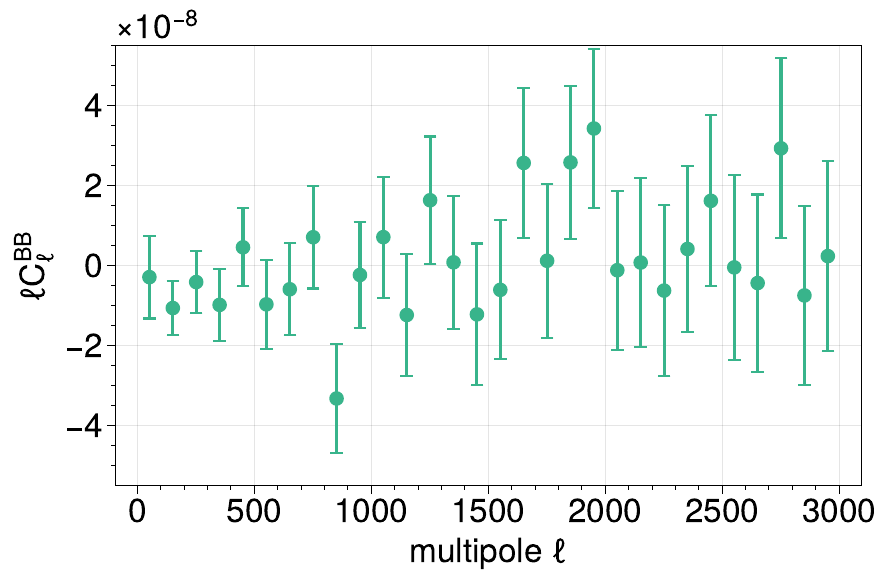}
    \caption{B-mode statistics measured with \texttt{NaMaster} \citep{alonso_namaster}. We find $\chi_2/{\rm dof}=24.2/30$ with $p$-value of 0.76, consistent with a null detection.}
    \label{fig:bmode}
\end{figure}

Figure~\ref{fig:bmode} shows our B-mode statistics for this non-tomographic measurement. We find they are consistent with zero, getting $\chi_2/{\rm dof}=24/30$ with $p$-value of 0.76. The final results, including the tomographic measurement of B-mode statistics, will be presented in future work \citep*{y6_1x2pt}.

\subsection{Null Tests with Survey Property Maps} \label{subsec:survey_property}

Shear measurements should not correlate with unrelated observing conditions at the level of a full survey. Any correlations, if found, would indicate some potential systematic contamination in the catalogue. We use a subset of the survey property maps from \citep{y6gold}. Each map comes as a HEALPix grid with the resolutions specified below. We evaluate the survey property at the object's position and then compute the mean shear in each component in bins of the survey property.

The survey property maps with the number of HEALPix values we consider are,
\begin{itemize}
    \item airmass (weighted mean) -- secant of the zenith angle for the observations. ({\tt nside}=16384)
    \item exposure time (sum) -- the total exposure time in seconds. ({\tt nside}=16384)
    \item FWHM (weighted mean) -- the full width half maximum (in arcseconds) of the PSF (seeing). ({\tt nside}=16384)
    \item sky brightness (weighted mean) -- contributions from airglow, zodiacal light, scattered starlight. ({\tt nside}=16384)
    \item Differential Chromatic Refraction (DCR; weighted mean) -- the effect of DCR on PSF shape $e_1$ and $e_2$ and on the sky coordinates (RA, DEC) ({\tt nside}=16384)
    \item Background Offset -- global and local background over-subtraction estimated from stellar aperture photometry with two different aperture radii. ({\tt nside}=64)
     \item Dust Extinction -- the dust extinction map in \citet{sfd98} from which we used $E(B-V)$ values to correct for the reddening. ({\tt nside}=4096)
    \item Stellar Density -- the density of \emph{\gaia} stars (EDR3; \citealt{gaia_edr3}) found in our footprint. The population is split into blue and red stars by its broad band photometry $G_{\rm BP} - G_{\rm RP}$. ({\tt nside}=16384)
\end{itemize}
A more detailed description of how each survey property is measured can be found in \citealt{y6gold}.

The measurement is conducted as follows. We first divide the range of a survey property signal into several bins and assign galaxy shear to each bin based on their positions in the sky. We then compute the response corrected mean shear for each survey property bin. Finally, we perform a linear fit to the relationship between the shear signals and the survey property map using an ordinary least-squares fitter. The survey property signals are normalized to $\subseteq$ [0,1] prior to performing the fit.

\begin{figure}
    \includegraphics[width=\columnwidth]{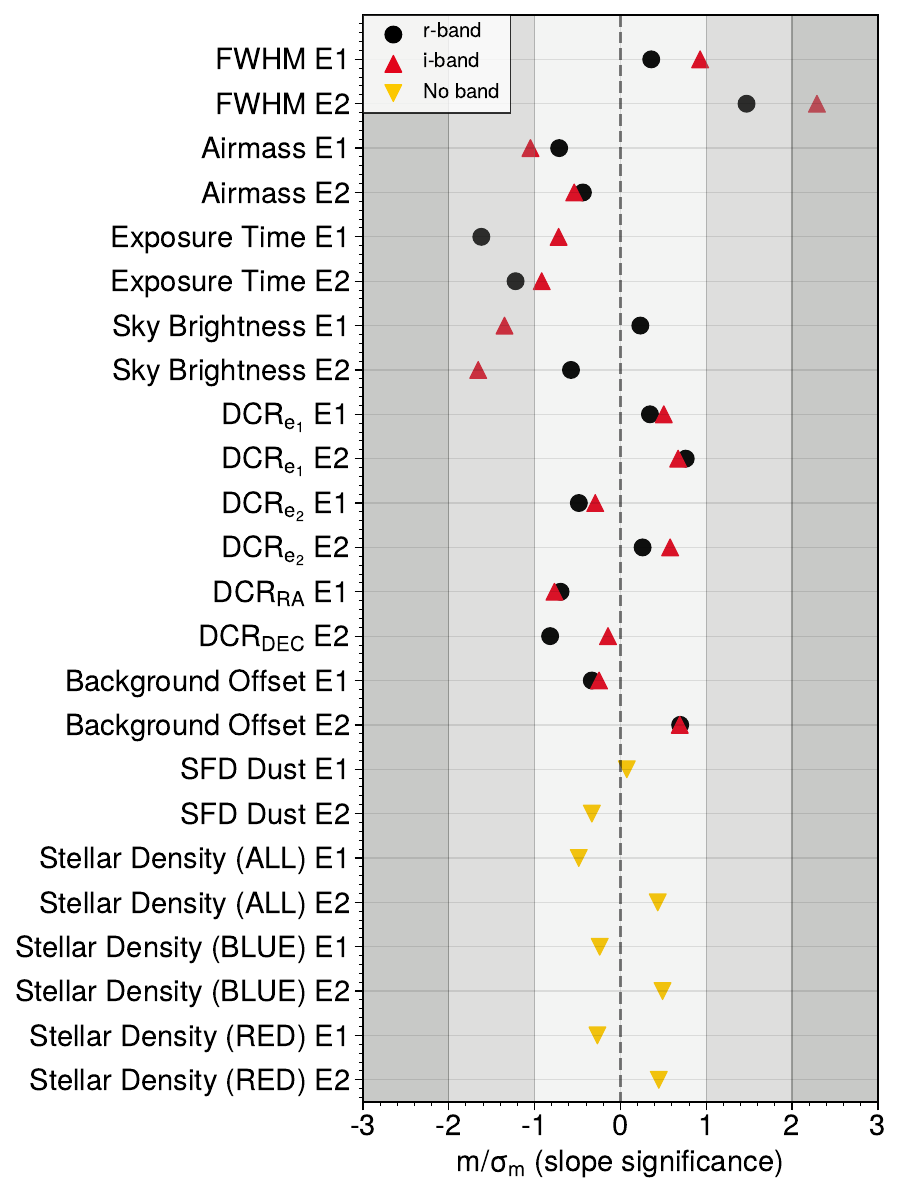}
    \caption{Statistical significance of the dependence of the shear measurements on survey properties. The horizontal axis gives the statistical significance of the slope of the best-fit linear relation between the shear signals and the survey property. $E1$/$E2$ in the y-axis denotes the $e_1$/$e_2$ component of the shear correlated with the survey property. The significance of the observed slope is computed using 800 {\sc Cosmogrid} simulations.
    }
    \label{fig:surveysys}
\end{figure}

To estimate the uncertainties on the fit parameters, we use mock catalogues. These catalogues allow us to estimate the fit parameter uncertainties, including both shape noise and cosmic variance. They also avoid any possible complications of estimating the uncertainties from a contaminated signal. Using the 800 realizations of the {\sc Cosmogrid} simulations, we compute the standard deviation of the slope of the best-fit line $m$. Figure~\ref{fig:surveysys} shows the $m/\sigma_m$, which is the statistical significance of the slope relative to zero. In all survey properties, we find that slopes are statistically consistent with zero, meaning observed shear is not correlated with the survey conditions. We report reduced $\chi^2$ statistics in Table~\ref{tab:survey_prop_chi2}.

\begin{table}
    \centering
    \caption{\label{tab:survey_prop_chi2} The reduced $\chi^2$ with respect to the null detection for each correlation of observed shear with survey property map. All of these results are consistent with null detection at 3$\sigma$.
    }
    \begin{tabular}[width=0.7\columnwidth]{ccc}
\hline\hline
     survey property & $\chi^2/{\rm dof}$ (r-band; $e_1$, $e_2$) & $\chi^2/{\rm dof}$ (i-band) \\ \hline
     Airmass & 0.47, 0.75 & 0.96, 0.38 \\
     Exposure Time & 0.57, 0.51 & 0.71, 1.12\\
     FWHM & 0.53, 0.69 & 0.50, 0.64\\
     Sky Brightness & 0.56, 0.92 & 0.53, 0.61 \\
     DCR (RA) & 0.86, 0.44 & 0.63, 0.81\\
     DCR (DEC) & 0.63, 0.50 & 0.61, 0.52\\
     DCR (PSF $e_1$) & 0.76, 1.32 & 0.50, 1.14\\
     DCR (PSF $e_2$) & 0.67, 0.33 & 0.67, 0.69\\
     Background Offset & 0.95, 1.31 & 1.25, 1.01\\
     \hline
     Dust Extinction & \multicolumn{2}{|c|}{0.89, 0.87}\\
     Stellar Density (ALL) & \multicolumn{2}{|c|}{0.79, 1.38}\\
     Stellar Density (BLUE) & \multicolumn{2}{|c|}{0.40, 0.99}\\
     Stellar Density (RED) & \multicolumn{2}{|c|}{0.43, 1.26}\\
\hline\hline
    \end{tabular}
\end{table}

\section{Image Simulation-based Tests}\label{sec:imagesims}

So far, we have discussed the catalogue-level systematics tests and mostly examined the sources of additive bias, which we can calibrate from data.
The calibration of multiplicative bias, however, needs to be characterized and carried out in realistic image simulations. For this purpose, we use
an image simulation suite generated using an update version of the DES Y3 image simulation code and data \citep{2022MNRAS.509.3371M}. In this work,
we present a short description of the simulations and defer a more detailed presentation to \citealt*{y6imagesims}.

The image simulations are generated independently for each coadd tile. We use a set of $\sim1000$ randomly selected tiles. For each coadd tile, we simulate each input single-epoch (CCD) image using the observed pixel masks, backgrounds, WCS, and PSF solutions.\footnote{For the image simulations, we make the simulation and run the analysis using a simpler WCS from \texttt{SCAMP} \citep{scamp2002} for each CCD image. We made this change so that we could fully integrate the analysis pipeline with that used for synthetic-source injection \citep{y6balrog}.} The input single-epoch images are then put through the identical cell-based coadding and \mdet\ pipelines as the actual DES data. The input catalogue of galaxies for the simulations is derived from the same catalogue used in \cite{2022MNRAS.509.3371M}, but with updates to include realistic galaxy clustering from the Cardinal simulation \citep{2024ApJ...961...59T} and corrections to the bright end of the magnitude function for the deficit of bright objects in the original catalogue. We again used \texttt{fitvd}\footnote{\url{https://github.com/esheldon/fitvd}} models fit to the data as the input sources \citep{2022MNRAS.509.3371M}. For the stars, we used a combination of the LSST \texttt{trilegal} simulations \citep{lsst_trilegal} and \gaia\ EDR3 \citep{gaia_edr3} catalogues. This combination allows us to place very bright stars in their correct locations. Our \mdet\ pipeline uses the locations of \gaia\ stars for special processing steps and so this detail is important. When simulating objects, we use color and position of the object to evaluate the PSF model, allowing our simulations to test both the effects of using the median galaxy color in the \mdet\ code and the effects of ignoring the PSF spatial dependence within single cell-based coadds.

We simulated a series of different image configurations to test both the image simulations themselves as well as the effects of various approximations we use in the \mdet\ pipeline. These configurations are as follows:
\begin{enumerate}
    \item \texttt{grid-exp}: galaxies are modeled as circular exponentials with fixed flux and size and placed in a hexagonal grid; no stars are simulated.
    \item \texttt{grid}: galaxies are sampled from the COSMOS catalogue and placed in a hexagonal grid; no stars are simulated.
	\item \texttt{grid-median\_color}: same as \texttt{grid}, but PSFs are always evaluated at the median color of the galaxy catalog; no stars are simulated.
    \item \texttt{fiducial}: galaxies and their positions are sampled from the COSMOS catalogue.
    \item \texttt{median\_color}: same as fiducial, but PSFs are always evaluated at the median color of the galaxy catalog,
    \item \texttt{no\_stars}: same as fiducial, but no stars are simulated.
\end{enumerate}
For each of these configurations, we simulated $\sim$400 DES tiles per configuration, with the exceptions of \texttt{grid-exp} for which we only simulate 100 tiles and \texttt{fiducial} for which we simulate $\sim$1000 tiles. The random seeds per coadd tile are consistent between the configurations, allowing us to measure the differences in results between the configurations at high precision by canceling most of the shared shape and pixel noise.

In each case, we test the shear recovery by creating pairs of simulations with true shears of ($g_1$, $g_2$)=[(+0.02, 0), (-0.02, 0)] for which we run \mdet\ and combine the mean measured shears according to the estimators from \citet{2019A&A...621A...2P}
\begin{align}\label{eqn:bias}
	\hat{m} &= \frac{\langle \hat{g}^+ \rangle - \langle \hat{g}^- \rangle}{2 |g|} - 1\, , \\
	\hat{c} &= \frac{\langle \hat{g}^+ \rangle + \langle \hat{g}^- \rangle}{2}\, .
\end{align}
We perform jackknife resampling over simulated tiles to estimate the measurement uncertainty. We apply all of the same cuts as in the data, except that objects are not cut to the unique tile boundaries. Instead, we leave a buffer region of unsimulated area near the edge of the coadd tile where no objects are simulated. This choice allows us to avoid small selection effects where objects move in and out of the unique coadd tile region. These would be accounted for by adjacent tiles in the DES data, but our simulations do not have adjacent tiles.

The measurements of $m_1$ and $c_2$ from each configuration are reported in Table.~\ref{tab:imagesims}. We highlight key findings from these simulations.
\begin{enumerate}
    \item The benchmark tests with simple simulations (\texttt{grid-exp}, \texttt{grid}, \texttt{grid-median-color}) show that for isolated objects, our image processing pipeline (PSF models, coaddition, and \mdet) displays no significant multiplicative and additive bias and are at the sub-percent level. Additionally, the difference between \texttt{grid} and \texttt{grid-median-color} configurations demonstrates that the use of median color for evaluating PSF models does not affect our capability to recover shear for isolated objects. The difference in multiplicative and additive bias, obtained by jacknifing over pairs of the same simulated tiles in each configuration to cancel most of the noise, is $\Delta m = (0.44\pm1.70)\times 10^{-3}$, $\Delta c_1 = (-2.2\pm2.2)\times 10^{-5}$ and $\Delta c_2 = (-6.3\pm2.2)\times 10^{-5}$.

    \item The difference of \texttt{no\_stars} and \texttt{fiducial} configurations, again obtained by a paired jackknife, demonstrates that the stellar contamination in the catalogue is small, in terms of both the response of star-like objects and their density. The difference in multiplicative and additive bias is $\Delta m = (0.063\pm 3.675)\times 10^{-3}$, $\Delta c_1 = (0.34\pm6.55)\times 10^{-5}$ and $\Delta c_2 = (-2.8\pm6.6)\times 10^{-5}$.

    \item The difference between \texttt{median\_color} and \texttt{fiducial} configurations validates that our ability to recover shear in settings with realistic levels of blending is robust even with the approach of evaluating PSF models at median galaxy color. Using the paired jackknife technique, we find $\Delta m = (0.035\pm2.017)\times 10^{-3}$, $\Delta c_1 = (-3.5\pm2.5)\times 10^{-5}$ and $\Delta c_2 = (-7.2\pm2.5)\times 10^{-5}$.

    \item The measurements for the \texttt{fiducial} simulations yield no detectable additive effects and constrain the net multiplicative bias to be $(+3.4\pm 6.1)\times 10^{-3}$, where we have reported a $3\sigma$ error bar. This estimate of the multiplicative bias $m$ includes any residual effects due to neglecting PSF color, spatial variation when running \mdet\ on the cell-based coadds, and star-galaxy separation issues. 
\end{enumerate}

A final calibration of $m$ and prior/uncertainty for the cosmological analyses, including the effects of blending due to projected sources between different redshifts, will be derived and reported in \citealt*{y6imagesims}.

\begin{table}
    \centering
    \caption{\label{tab:imagesims} Measurements of multiplicative and additive shear biases via jackknifing Eqn.~\ref{eqn:bias} over measurements from simulated images of Y6 tiles. The reported uncertainties correspond to 3 standard errors of the mean.}
    \begin{tabular}[width=\columnwidth]{lccc}
        \hline\hline
        Configuration & $m_1$ [$10^{-3}$] & $c_2$ [$10^{-4}$] & $\#$ tiles \\
        \hline
        \texttt{grid-exp} & $-0.27 \pm 0.10$ & $+0.002 \pm 0.018$ & 97 \\
        \texttt{grid} & $-3.8 \pm 10$ & $-0.5 \pm 3.7$ & 395 \\
        \texttt{grid-median\_color} & $-4.3 \pm 9.9$ & $+0.1 \pm 3.7$ & 395 \\
        \texttt{no\_stars} & $+3.5 \pm 9.5$ & $+0.4 \pm 4.1$ & 395 \\
        \texttt{median\_color} & $+3.4 \pm 9.8$ & $+0.8 \pm 4.1$ & 395 \\
        \texttt{fiducial} & $+3.4 \pm 6.1$ & $+1.0 \pm 2.6$ & 981 \\
        \hline\hline
    \end{tabular}
\end{table}

\section{Conclusion} \label{sec:conclusion}
In this paper, we presented the cell-based coadding and \mdet\ pipelines, applied them to DES Y6 data to build the DES Y6 \mdet\ shape catalog, and performed extensive testing of the catalogue. The final catalogue comprises 151,922,791 source galaxies from the DES Y6 footprint covering 4422 deg$^2$. The effective number density of galaxies is 8.22 galaxies per arcmin$^2$ with shape noise of $\sigma_{\rm e}=0.29$. Our tests of the shape measurement methodology using image simulations \citet*{y6imagesims} indicate that the residual multiplicative bias in the catalogue is $0.34\%\pm0.61\%$ at 3$\sigma$ uncertainty, which is within 1.3\% requirement for LSST Y1 \citep{lsst_srd}. The remaining systematic errors in the catalogue are largely additive. While we significantly reduced the trend of the mean shear as a function of galaxy color, as shown in Table~\ref{tab:meanshear_color}, there is still a residual mean shear signal whose origin is not fully explained, but which can be empirically corrected. We further found detectable PSF modeling errors in the shear catalogue. However, given the statistical precision of the catalog, preliminary estimates indicate that the residual PSF modeling errors are not a significant contaminant to cosmic shear measurements from the \mdet\ catalogue. Should this conclusion change for a particular cosmological analysis, the PSF contamination model in this work can be used to directly marginalize over the residual systematic effects. This work is accompanied by \citet{y6psf}, which describes the Y6 \textsc{Piff} PSF modeling, and \citet*{y6imagesims}, which describes the state-of-the-art Y6 image simulations in full detail. Further, our \mdet\ catalogue is complemented by another shear catalogue that employs the \bfd\ technique and will be described in \citet*{y6bfd}.

Over the past decade, as the DES collaboration has worked diligently to improve our weak lensing analysis algorithms, our understanding of how to make weak lensing shear measurements has evolved significantly. During the SV analysis, one of the primary concerns in weak lensing measurements at the time was noise bias (e.g., \citealt{2003MNRAS.343..459H, 2012MNRAS.427.2711K, 2012MNRAS.424.2757M, 2012MNRAS.425.1951R}). However, that analysis helped to reveal the importance of shear-dependent catalogue selection effects (e.g., \citealt{2003MNRAS.343..459H, bernstein_jarvis_2002}). In the analysis of Y1 and Y3 data, the collaboration moved to use weak lensing shear measurement techniques that correctly handled these noise bias and selection effects \citep{2017arXiv170202600H, 2017ApJ...841...24S}. The DES Y3 analysis introduced several additional innovations, including improved PSF modeling with \textsc{Piff} and extensive use of image simulations to calibrate cross-redshift, blending-induced shear calibration effects (\citealt*{2021MNRAS.501.1282J, y3-shapecatalog, 2022MNRAS.509.3371M}). The DES Y3 analysis further helped to reveal the importance of shear-dependent detection effects and motivated the introduction of \mdet\ itself \citep{2020ApJ...902..138S}. Finally, we have broadened our understanding of what a complete catalogue testing strategy entails and have made significant improvements to our catalogue testing strategies, including the introduction of more extensive statistical tests with PSF modeling residuals \citep{2016MNRAS.460.2245J} and direct null tests on the cosmic shear two-point correlation functions \citep[see, e.g.,][]{2016PhRvD..94b2002B,2022PhRvD.105b3514A}.

Looking forward, the cell-based coadding with \mdet\ pipeline presented in this work is the precursor to a similar pipeline being developed by the LSST Dark Energy Science Collaboration for use with the Vera C. Rubin Observatory LSST \citep{sheldon_mdet_rubin}. The DES Y6 \mdet\ catalogue already meets key targets for the control of systematic errors needed for the analysis of the first year of LSST data \citep{lsst_srd}, including controlling multiplicative biases. Demonstrating this level of systematics control in the analysis of actual survey data is a significant milestone and bodes well for the longevity of the techniques employed in this work.

\section*{Acknowledgements}

\textbf{Author Contributions}: All authors contributed to
this paper and/or carried out infrastructure work that
made this analysis possible.
MY performed almost all analysis and manuscript preparation. MRB and ESS contributed to analysis and manuscript preparation, and developed cell-based coadd and \mdet\ software. MJ contributed to the analysis and development of \mdet\ and the cell-based coadds. RG and FM ran the cell-based coadding code and \mdet\ on the full survey dataset and provided technical support for the analysis. ESR contributed to generating masks. SM contributed to developing the image simulation suite, ran the simulations, and analyzed the results. TS, MG and MAT contributed to the analysis. DA contributed to running the image simulations. AA, DA, GB, MT, AT, BY contributed to analysis interpretation as members of the DES Y6 shear analysis team.
AA, DG, EH contributed to manuscript preparation as collaboration internal reviewers.
The remaining authors have made contributions to this paper that include, but are not limited to, the construction of DECam and other aspects of collecting the data; data processing and calibration; developing broadly used methods, codes, and simulations; running the pipelines and validation tests; and promoting the science analysis.

This work was based in part on observations at Cerro Tololo Inter-American Observatory, National Optical Astronomy Observatory, which is operated by the Association of Universities for Research in Astronomy (AURA) under a cooperative agreement with the National
Science Foundation. This work was also based in part on data obtained from the ESO Science Archive Facility with DOI: \url{https://doi.org/10.18727/archive/57}. The astronomical community is honored to have the opportunity to conduct astronomical research on I’oligam Du’ag (Kitt Peak) in Arizona, on Maunakea in Hawai’i, and on Cerro Tololo and Cerro Pachón in Chile. We recognize and acknowledge the very significant cultural role and reverence that these sites have to the Tohono O’odham Nation, to the Native Hawaiian community, and to the local communities in Chile, respectively.

Funding for the DES Projects has been provided by the U.S. Department of Energy, the U.S. National Science Foundation, the Ministry of Science and Education of Spain, 
the Science and Technology Facilities Council of the United Kingdom, the Higher Education Funding Council for England, the National Center for Supercomputing 
Applications at the University of Illinois at Urbana-Champaign, the Kavli Institute of Cosmological Physics at the University of Chicago, 
the Center for Cosmology and Astro-Particle Physics at the Ohio State University,
the Mitchell Institute for Fundamental Physics and Astronomy at Texas A\&M University, Financiadora de Estudos e Projetos, 
Funda{\c c}{\~a}o Carlos Chagas Filho de Amparo {\`a} Pesquisa do Estado do Rio de Janeiro, Conselho Nacional de Desenvolvimento Cient{\'i}fico e Tecnol{\'o}gico and 
the Minist{\'e}rio da Ci{\^e}ncia, Tecnologia e Inova{\c c}{\~a}o, the Deutsche Forschungsgemeinschaft and the Collaborating Institutions in the Dark Energy Survey. 

The Collaborating Institutions are Argonne National Laboratory, the University of California at Santa Cruz, the University of Cambridge, Centro de Investigaciones Energ{\'e}ticas, 
Medioambientales y Tecnol{\'o}gicas-Madrid, the University of Chicago, University College London, the DES-Brazil Consortium, the University of Edinburgh, 
the Eidgen{\"o}ssische Technische Hochschule (ETH) Z{\"u}rich, 
Fermi National Accelerator Laboratory, the University of Illinois at Urbana-Champaign, the Institut de Ci{\`e}ncies de l'Espai (IEEC/CSIC), 
the Institut de F{\'i}sica d'Altes Energies, Lawrence Berkeley National Laboratory, the Ludwig-Maximilians Universit{\"a}t M{\"u}nchen and the associated Excellence Cluster Universe, 
the University of Michigan, NSF NOIRLab, the University of Nottingham, The Ohio State University, the University of Pennsylvania, the University of Portsmouth, 
SLAC National Accelerator Laboratory, Stanford University, the University of Sussex, Texas A\&M University, and the OzDES Membership Consortium.

Based in part on observations at NSF Cerro Tololo Inter-American Observatory at NSF NOIRLab (NOIRLab Prop. ID 2012B-0001; PI: J. Frieman), which is managed by the Association of Universities for Research in Astronomy (AURA) under a cooperative agreement with the National Science Foundation.

The DES data management system is supported by the National Science Foundation under Grant Numbers AST-1138766 and AST-1536171.
The DES participants from Spanish institutions are partially supported by MICINN under grants PID2021-123012, PID2021-128989 PID2022-141079, SEV-2016-0588, CEX2020-001058-M and CEX2020-001007-S, some of which include ERDF funds from the European Union. IFAE is partially funded by the CERCA program of the Generalitat de Catalunya.

We  acknowledge support from the Brazilian Instituto Nacional de Ci\^encia
e Tecnologia (INCT) do e-Universo (CNPq grant 465376/2014-2).

Argonne National Laboratory's work was supported under the U.S. Department of Energy contract DE-AC02-06CH11357.

This document was prepared by the DES Collaboration using the resources of the Fermi National Accelerator Laboratory (Fermilab), a U.S. Department of Energy, Office of Science, Office of High Energy Physics HEP User Facility. Fermilab is managed by Fermi Forward Discovery Group, LLC, acting under Contract No. 89243024CSC000002.

%%%%%%%%%%%%%%%%%%%%%%%%%%%%%%%%%%%%%%%%%%%%%%%%%%
\section*{Data Availability}

The full \mdet\ catalogue will be made publicly available following publication of the final DES results, at the URL https://des.ncsa.illinois.edu/releases. The code used to perform the tests in this manuscript will be made available upon reasonable request to the authors.



\bibliographystyle{mnras}
\bibliography{main} 

@string{june = {June}}

@article{gaap,
	adsurl = {https://ui.adsabs.harvard.edu/abs/2008A&A...482.1053K},
	archiveprefix = {arXiv},
	author = {{Kuijken}, K.},
	doi = {10.1051/0004-6361:20066601},
	eprint = {astro-ph/0610606},
	journal = {\aap},
	month = may,
	number = {3},
	pages = {1053-1067},
	primaryclass = {astro-ph},
	title = {{GaaP: PSF- and aperture-matched photometry using shapelets}},
	volume = {482},
	year = 2008}

@article{sep,
	author = {Kyle Barbary},
	doi = {10.21105/joss.00058},
	journal = {Journal of Open Source Software},
	number = {6},
	pages = {58},
	publisher = {The Open Journal},
	title = {SEP: Source Extractor as a library},
	url = {https://doi.org/10.21105/joss.00058},
	volume = {1},
	year = {2016}}

@article{numpy,
	author = {Charles R. Harris and K. Jarrod Millman and St{\'{e}}fan J. van der Walt and Ralf Gommers and Pauli Virtanen and David Cournapeau and Eric Wieser and Julian Taylor and Sebastian Berg and Nathaniel J. Smith and Robert Kern and Matti Picus and Stephan Hoyer and Marten H. van Kerkwijk and Matthew Brett and Allan Haldane and Jaime Fern{\'{a}}ndez del R{\'{i}}o and Mark Wiebe and Pearu Peterson and Pierre G{\'{e}}rard-Marchant and Kevin Sheppard and Tyler Reddy and Warren Weckesser and Hameer Abbasi and Christoph Gohlke and Travis E. Oliphant},
	doi = {10.1038/s41586-020-2649-2},
	journal = {Nature},
	month = sep,
	number = {7825},
	pages = {357--362},
	publisher = {Springer Science and Business Media {LLC}},
	title = {Array programming with {NumPy}},
	url = {https://doi.org/10.1038/s41586-020-2649-2},
	volume = {585},
	year = {2020}}

@misc{condaforge,
	author = {conda-forge community},
	doi = {10.5281/zenodo.4774216},
	month = jul,
	publisher = {Zenodo},
	title = {{The conda-forge Project: Community-based Software Distribution Built on the conda Package Format and Ecosystem}},
	url = {https://doi.org/10.5281/zenodo.4774216},
	year = 2015}

@inproceedings{numba,
	author = {Lam, Siu Kwan and Pitrou, Antoine and Seibert, Stanley},
	booktitle = {Proceedings of the Second Workshop on the LLVM Compiler Infrastructure in HPC},
	pages = {1--6},
	title = {Numba: A llvm-based python jit compiler},
	year = {2015}}

@article{plancherel1910,
	author = {Plancherel, Michel and Leffler, Mittag},
	date = {1910/12/01},
	doi = {10.1007/BF03014877},
	id = {Plancherel1910},
	isbn = {0009-725X},
	journal = {Rendiconti del Circolo Matematico di Palermo (1884-1940)},
	number = {1},
	pages = {289--335},
	title = {Contribution {\`A}{\v L}{\'E}tude de la repr{\'E}sentation D'une fonction arbitraire par des int{\'E}grales d{\'E}finies},
	url = {https://doi.org/10.1007/BF03014877},
	volume = {30},
	year = {1910}}

@article{ct2,
	author = {{Farin}, G.},
	journal = {Computer Aided Geometric Design},
	pages = {83},
	title = {Triangular Bernstein-Bezier patches},
	volume = {3},
	year = {1986}}

@article{ct1,
	author = {{Alfeld}, P.},
	journal = {Computer Aided Geometric Design},
	pages = {169},
	title = {A trivariate Clough-Tocher scheme for tetrahedral data},
	volume = {1},
	year = {1984}}

@article{2003A&A...412...45P,
	adsurl = {https://ui.adsabs.harvard.edu/abs/2003A&A...412...45P},
	author = {{Paturel}, G. and {Petit}, C. and {Prugniel}, Ph. and {Theureau}, G. and {Rousseau}, J. and {Brouty}, M. and {Dubois}, P. and {Cambr{\'e}sy}, L.},
	journal = {\aap},
	month = {December},
	pages = {45-55},
	title = {{HYPERLEDA. I. Identification and designation of galaxies}},
	volume = {412},
	year = {2003}}

@article{2005ApJ...622..759G,
	adsurl = {https://ui.adsabs.harvard.edu/abs/2005ApJ...622..759G},
	archiveprefix = {arXiv},
	author = {{G{\'o}rski}, K.~M. and {Hivon}, E. and {Banday}, A.~J. and {Wandelt}, B.~D. and {Hansen}, F.~K. and {Reinecke}, M. and {Bartelmann}, M.},
	eprint = {astro-ph/0409513},
	journal = {\apj},
	month = {April},
	number = {2},
	pages = {759-771},
	primaryclass = {astro-ph},
	title = {{HEALPix: A Framework for High-Resolution Discretization and Fast Analysis of Data Distributed on the Sphere}},
	volume = {622},
	year = {2005}}

@article{2005astro.ph.10346T,
	adsurl = {https://ui.adsabs.harvard.edu/abs/2005astro.ph.10346T},
	archiveprefix = {arXiv},
	author = {{The Dark Energy Survey Collaboration}},
	eid = {astro-ph/0510346},
	eprint = {astro-ph/0510346},
	journal = {arXiv e-prints},
	month = {October},
	pages = {astro-ph/0510346},
	primaryclass = {astro-ph},
	title = {{The Dark Energy Survey}},
	year = {2005}}

@article{2006MNRAS.368.1323H,
	adsurl = {https://ui.adsabs.harvard.edu/abs/2006MNRAS.368.1323H},
	archiveprefix = {arXiv},
	author = {{Heymans}, Catherine and {Van Waerbeke}, Ludovic and {Bacon}, David and {Berge}, Joel and {Bernstein}, Gary and {Bertin}, Emmanuel and {Bridle}, Sarah and {Brown}, Michael L. and {Clowe}, Douglas and {Dahle}, H{\r{a}}kon and {Erben}, Thomas and {Gray}, Meghan and {Hetterscheidt}, Marco and {Hoekstra}, Henk and {Hudelot}, Patrick and {Jarvis}, Mike and {Kuijken}, Konrad and {Margoniner}, Vera and {Massey}, Richard and {Mellier}, Yannick and {Nakajima}, Reiko and {Refregier}, Alexandre and {Rhodes}, Jason and {Schrabback}, Tim and {Wittman}, David},
	eprint = {astro-ph/0506112},
	journal = {\mnras},
	month = {May},
	number = {3},
	pages = {1323-1339},
	primaryclass = {astro-ph},
	title = {{The Shear Testing Programme - I. Weak lensing analysis of simulated ground-based observations}},
	volume = {368},
	year = {2006}}

@article{2007MNRAS.376...13M,
	adsurl = {https://ui.adsabs.harvard.edu/abs/2007MNRAS.376...13M},
	archiveprefix = {arXiv},
	author = {{Massey}, Richard and {Heymans}, Catherine and {Berg{\'e}}, Joel and {Bernstein}, Gary and {Bridle}, Sarah and {Clowe}, Douglas and {Dahle}, H{\r{a}}kon and {Ellis}, Richard and {Erben}, Thomas and {Hetterscheidt}, Marco and {High}, F. William and {Hirata}, Christopher and {Hoekstra}, Henk and {Hudelot}, Patrick and {Jarvis}, Mike and {Johnston}, David and {Kuijken}, Konrad and {Margoniner}, Vera and {Mandelbaum}, Rachel and {Mellier}, Yannick and {Nakajima}, Reiko and {Paulin-Henriksson}, Stephane and {Peeples}, Molly and {Roat}, Chris and {Refregier}, Alexandre and {Rhodes}, Jason and {Schrabback}, Tim and {Schirmer}, Mischa and {Seljak}, Uro{\v{s}} and {Semboloni}, Elisabetta and {van Waerbeke}, Ludovic},
	eprint = {astro-ph/0608643},
	journal = {\mnras},
	month = {March},
	number = {1},
	pages = {13-38},
	primaryclass = {astro-ph},
	title = {{The Shear Testing Programme 2: Factors affecting high-precision weak-lensing analyses}},
	volume = {376},
	year = {2007}}

@article{2008A&A...484...67P,
	adsurl = {https://ui.adsabs.harvard.edu/abs/2008A&A...484...67P},
	archiveprefix = {arXiv},
	author = {{Paulin-Henriksson}, S. and {Amara}, A. and {Voigt}, L. and {Refregier}, A. and {Bridle}, S.~L.},
	eprint = {0711.4886},
	journal = {\aap},
	month = {June},
	number = {1},
	pages = {67-77},
	primaryclass = {astro-ph},
	title = {{Point spread function calibration requirements for dark energy from cosmic shear}},
	volume = {484},
	year = {2008}}

@article{2009arXiv0912.0201L,
	adsurl = {https://ui.adsabs.harvard.edu/abs/2009arXiv0912.0201L},
	archiveprefix = {arXiv},
	author = {{LSST Science Collaboration} and {Abell}, Paul A. and {Allison}, Julius and {Anderson}, Scott F. and {Andrew}, John R. and {Angel}, J. Roger P. and {Armus}, Lee and {Arnett}, David and {Asztalos}, S.~J. and {Axelrod}, Tim S. and {Bailey}, Stephen and {Ballantyne}, D.~R. and {Bankert}, Justin R. and {Barkhouse}, Wayne A. and {Barr}, Jeffrey D. and {Barrientos}, L. Felipe and {Barth}, Aaron J. and {Bartlett}, James G. and {Becker}, Andrew C. and {Becla}, Jacek and {Beers}, Timothy C. and {Bernstein}, Joseph P. and {Biswas}, Rahul and {Blanton}, Michael R. and {Bloom}, Joshua S. and {Bochanski}, John J. and {Boeshaar}, Pat and {Borne}, Kirk D. and {Bradac}, Marusa and {Brandt}, W.~N. and {Bridge}, Carrie R. and {Brown}, Michael E. and {Brunner}, Robert J. and {Bullock}, James S. and {Burgasser}, Adam J. and {Burge}, James H. and {Burke}, David L. and {Cargile}, Phillip A. and {Chandrasekharan}, Srinivasan and {Chartas}, George and {Chesley}, Steven R. and {Chu}, You-Hua and {Cinabro}, David and {Claire}, Mark W. and {Claver}, Charles F. and {Clowe}, Douglas and {Connolly}, A.~J. and {Cook}, Kem H. and {Cooke}, Jeff and {Cooray}, Asantha and {Covey}, Kevin R. and {Culliton}, Christopher S. and {de Jong}, Roelof and {de Vries}, Willem H. and {Debattista}, Victor P. and {Delgado}, Francisco and {Dell'Antonio}, Ian P. and {Dhital}, Saurav and {Di Stefano}, Rosanne and {Dickinson}, Mark and {Dilday}, Benjamin and {Djorgovski}, S.~G. and {Dobler}, Gregory and {Donalek}, Ciro and {Dubois-Felsmann}, Gregory and {Durech}, Josef and {Eliasdottir}, Ardis and {Eracleous}, Michael and {Eyer}, Laurent and {Falco}, Emilio E. and {Fan}, Xiaohui and {Fassnacht}, Christopher D. and {Ferguson}, Harry C. and {Fernandez}, Yanga R. and {Fields}, Brian D. and {Finkbeiner}, Douglas and {Figueroa}, Eduardo E. and {Fox}, Derek B. and {Francke}, Harold and {Frank}, James S. and {Frieman}, Josh and {Fromenteau}, Sebastien and {Furqan}, Muhammad and {Galaz}, Gaspar and {Gal-Yam}, A. and {Garnavich}, Peter and {Gawiser}, Eric and {Geary}, John and {Gee}, Perry and {Gibson}, Robert R. and {Gilmore}, Kirk and {Grace}, Emily A. and {Green}, Richard F. and {Gressler}, William J. and {Grillmair}, Carl J. and {Habib}, Salman and {Haggerty}, J.~S. and {Hamuy}, Mario and {Harris}, Alan W. and {Hawley}, Suzanne L. and {Heavens}, Alan F. and {Hebb}, Leslie and {Henry}, Todd J. and {Hileman}, Edward and {Hilton}, Eric J. and {Hoadley}, Keri and {Holberg}, J.~B. and {Holman}, Matt J. and {Howell}, Steve B. and {Infante}, Leopoldo and {Ivezic}, Zeljko and {Jacoby}, Suzanne H. and {Jain}, Bhuvnesh and {R} and {Jedicke} and {Jee}, M. James and {Garrett Jernigan}, J. and {Jha}, Saurabh W. and {Johnston}, Kathryn V. and {Jones}, R. Lynne and {Juric}, Mario and {Kaasalainen}, Mikko and {Styliani} and {Kafka} and {Kahn}, Steven M. and {Kaib}, Nathan A. and {Kalirai}, Jason and {Kantor}, Jeff and {Kasliwal}, Mansi M. and {Keeton}, Charles R. and {Kessler}, Richard and {Knezevic}, Zoran and {Kowalski}, Adam and {Krabbendam}, Victor L. and {Krughoff}, K. Simon and {Kulkarni}, Shrinivas and {Kuhlman}, Stephen and {Lacy}, Mark and {Lepine}, Sebastien and {Liang}, Ming and {Lien}, Amy and {Lira}, Paulina and {Long}, Knox S. and {Lorenz}, Suzanne and {Lotz}, Jennifer M. and {Lupton}, R.~H. and {Lutz}, Julie and {Macri}, Lucas M. and {Mahabal}, Ashish A. and {Mandelbaum}, Rachel and {Marshall}, Phil and {May}, Morgan and {McGehee}, Peregrine M. and {Meadows}, Brian T. and {Meert}, Alan and {Milani}, Andrea and {Miller}, Christopher J. and {Miller}, Michelle and {Mills}, David and {Minniti}, Dante and {Monet}, David and {Mukadam}, Anjum S. and {Nakar}, Ehud and {Neill}, Douglas R. and {Newman}, Jeffrey A. and {Nikolaev}, Sergei and {Nordby}, Martin and {O'Connor}, Paul and {Oguri}, Masamune and {Oliver}, John and {Olivier}, Scot S. and {Olsen}, Julia K. and {Olsen}, Knut and {Olszewski}, Edward W. and {Oluseyi}, Hakeem and {Padilla}, Nelson D. and {Parker}, Alex and {Pepper}, Joshua and {Peterson}, John R. and {Petry}, Catherine and {Pinto}, Philip A. and {Pizagno}, James L. and {Popescu}, Bogdan and {Prsa}, Andrej and {Radcka}, Veljko and {Raddick}, M. Jordan and {Rasmussen}, Andrew and {Rau}, Arne and {Rho}, Jeonghee and {Rhoads}, James E. and {Richards}, Gordon T. and {Ridgway}, Stephen T. and {Robertson}, Brant E. and {Roskar}, Rok and {Saha}, Abhijit and {Sarajedini}, Ata and {Scannapieco}, Evan and {Schalk}, Terry and {Schindler}, Rafe and {Schmidt}, Samuel and {Schmidt}, Sarah and {Schneider}, Donald P. and {Schumacher}, German and {Scranton}, Ryan and {Sebag}, Jacques and {Seppala}, Lynn G. and {Shemmer}, Ohad and {Simon}, Joshua D. and {Sivertz}, M. and {Smith}, Howard A. and {Allyn Smith}, J. and {Smith}, Nathan and {Spitz}, Anna H. and {Stanford}, Adam and {Stassun}, Keivan G. and {Strader}, Jay and {Strauss}, Michael A. and {Stubbs}, Christopher W. and {Sweeney}, Donald W. and {Szalay}, Alex and {Szkody}, Paula and {Takada}, Masahiro and {Thorman}, Paul and {Trilling}, David E. and {Trimble}, Virginia and {Tyson}, Anthony and {Van Berg}, Richard and {Vanden Berk}, Daniel and {VanderPlas}, Jake and {Verde}, Licia and {Vrsnak}, Bojan and {Walkowicz}, Lucianne M. and {Wandelt}, Benjamin D. and {Wang}, Sheng and {Wang}, Yun and {Warner}, Michael and {Wechsler}, Risa H. and {West}, Andrew A. and {Wiecha}, Oliver and {Williams}, Benjamin F. and {Willman}, Beth and {Wittman}, David and {Wolff}, Sidney C. and {Wood-Vasey}, W. Michael and {Wozniak}, Przemek and {Young}, Patrick and {Zentner}, Andrew and {Zhan}, Hu},
	eid = {arXiv:0912.0201},
	eprint = {0912.0201},
	journal = {arXiv e-prints},
	month = {December},
	pages = {arXiv:0912.0201},
	primaryclass = {astro-ph.IM},
	title = {{LSST Science Book, Version 2.0}},
	year = {2009}}

@article{hscy3_dalal,
   title={Hyper Suprime-Cam Year 3 results: Cosmology from cosmic shear power spectra},
   volume={108},
   ISSN={2470-0029},
   url={http://dx.doi.org/10.1103/PhysRevD.108.123519},
   DOI={10.1103/physrevd.108.123519},
   number={12},
   journal={Physical Review D},
   publisher={American Physical Society (APS)},
   author={Dalal, Roohi and Li, Xiangchong and Nicola, Andrina and Zuntz, Joe and Strauss, Michael A. and Sugiyama, Sunao and Zhang, Tianqing and Rau, Markus M. and Mandelbaum, Rachel and Takada, Masahiro and More, Surhud and Miyatake, Hironao and Kannawadi, Arun and Shirasaki, Masato and Taniguchi, Takanori and Takahashi, Ryuichi and Osato, Ken and Hamana, Takashi and Oguri, Masamune and Nishizawa, Atsushi J. and Malagón, Andrés A. Plazas and Sunayama, Tomomi and Alonso, David and Slosar, Anže and Luo, Wentao and Armstrong, Robert and Bosch, James and Hsieh, Bau-Ching and Komiyama, Yutaka and Lupton, Robert H. and Lust, Nate B. and MacArthur, Lauren A. and Miyazaki, Satoshi and Murayama, Hitoshi and Nishimichi, Takahiro and Okura, Yuki and Price, Paul A. and Tait, Philip J. and Tanaka, Masayuki and Wang, Shiang-Yu},
   year={2023},
   month=dec }

@article{2010MNRAS.404..350R,
	adsurl = {https://ui.adsabs.harvard.edu/abs/2010MNRAS.404..350R},
	archiveprefix = {arXiv},
	author = {{Rowe}, Barnaby},
	eprint = {0904.3056},
	journal = {\mnras},
	month = {May},
	number = {1},
	pages = {350-366},
	primaryclass = {astro-ph.CO},
	title = {{Improving PSF modelling for weak gravitational lensing using new methods in model selection}},
	volume = {404},
	year = {2010}}

@article{2010MNRAS.405.2044B,
	adsurl = {https://ui.adsabs.harvard.edu/abs/2010MNRAS.405.2044B},
	archiveprefix = {arXiv},
	author = {{Bridle}, Sarah and {Balan}, Sreekumar T. and {Bethge}, Matthias and {Gentile}, Marc and {Harmeling}, Stefan and {Heymans}, Catherine and {Hirsch}, Michael and {Hosseini}, Reshad and {Jarvis}, Mike and {Kirk}, Donnacha and {Kitching}, Thomas and {Kuijken}, Konrad and {Lewis}, Antony and {Paulin-Henriksson}, Stephane and {Sch{\"o}lkopf}, Bernhard and {Velander}, Malin and {Voigt}, Lisa and {Witherick}, Dugan and {Amara}, Adam and {Bernstein}, Gary and {Courbin}, Fr{\'e}d{\'e}ric and {Gill}, Mandeep and {Heavens}, Alan and {Mandelbaum}, Rachel and {Massey}, Richard and {Moghaddam}, Baback and {Rassat}, Anais and {R{\'e}fr{\'e}gier}, Alexandre and {Rhodes}, Jason and {Schrabback}, Tim and {Shawe-Taylor}, John and {Shmakova}, Marina and {van Waerbeke}, Ludovic and {Wittman}, David},
	eprint = {0908.0945},
	journal = {\mnras},
	month = {July},
	number = {3},
	pages = {2044-2061},
	primaryclass = {astro-ph.CO},
	title = {{Results of the GREAT08 Challenge: an image analysis competition for cosmological lensing}},
	volume = {405},
	year = {2010}}

@article{2011arXiv1110.3193L,
	adsurl = {https://ui.adsabs.harvard.edu/abs/2011arXiv1110.3193L},
	archiveprefix = {arXiv},
	author = {{Laureijs}, R. and {Amiaux}, J. and {Arduini}, S. and {Augu{\`e}res}, J. -L. and {Brinchmann}, J. and {Cole}, R. and {Cropper}, M. and {Dabin}, C. and {Duvet}, L. and {Ealet}, A. and {Garilli}, B. and {Gondoin}, P. and {Guzzo}, L. and {Hoar}, J. and {Hoekstra}, H. and {Holmes}, R. and {Kitching}, T. and {Maciaszek}, T. and {Mellier}, Y. and {Pasian}, F. and {Percival}, W. and {Rhodes}, J. and {Saavedra Criado}, G. and {Sauvage}, M. and {Scaramella}, R. and {Valenziano}, L. and {Warren}, S. and {Bender}, R. and {Castander}, F. and {Cimatti}, A. and {Le F{\`e}vre}, O. and {Kurki-Suonio}, H. and {Levi}, M. and {Lilje}, P. and {Meylan}, G. and {Nichol}, R. and {Pedersen}, K. and {Popa}, V. and {Rebolo Lopez}, R. and {Rix}, H. -W. and {Rottgering}, H. and {Zeilinger}, W. and {Grupp}, F. and {Hudelot}, P. and {Massey}, R. and {Meneghetti}, M. and {Miller}, L. and {Paltani}, S. and {Paulin-Henriksson}, S. and {Pires}, S. and {Saxton}, C. and {Schrabback}, T. and {Seidel}, G. and {Walsh}, J. and {Aghanim}, N. and {Amendola}, L. and {Bartlett}, J. and {Baccigalupi}, C. and {Beaulieu}, J. -P. and {Benabed}, K. and {Cuby}, J. -G. and {Elbaz}, D. and {Fosalba}, P. and {Gavazzi}, G. and {Helmi}, A. and {Hook}, I. and {Irwin}, M. and {Kneib}, J. -P. and {Kunz}, M. and {Mannucci}, F. and {Moscardini}, L. and {Tao}, C. and {Teyssier}, R. and {Weller}, J. and {Zamorani}, G. and {Zapatero Osorio}, M.~R. and {Boulade}, O. and {Foumond}, J.~J. and {Di Giorgio}, A. and {Guttridge}, P. and {James}, A. and {Kemp}, M. and {Martignac}, J. and {Spencer}, A. and {Walton}, D. and {Bl{\"u}mchen}, T. and {Bonoli}, C. and {Bortoletto}, F. and {Cerna}, C. and {Corcione}, L. and {Fabron}, C. and {Jahnke}, K. and {Ligori}, S. and {Madrid}, F. and {Martin}, L. and {Morgante}, G. and {Pamplona}, T. and {Prieto}, E. and {Riva}, M. and {Toledo}, R. and {Trifoglio}, M. and {Zerbi}, F. and {Abdalla}, F. and {Douspis}, M. and {Grenet}, C. and {Borgani}, S. and {Bouwens}, R. and {Courbin}, F. and {Delouis}, J. -M. and {Dubath}, P. and {Fontana}, A. and {Frailis}, M. and {Grazian}, A. and {Koppenh{\"o}fer}, J. and {Mansutti}, O. and {Melchior}, M. and {Mignoli}, M. and {Mohr}, J. and {Neissner}, C. and {Noddle}, K. and {Poncet}, M. and {Scodeggio}, M. and {Serrano}, S. and {Shane}, N. and {Starck}, J. -L. and {Surace}, C. and {Taylor}, A. and {Verdoes-Kleijn}, G. and {Vuerli}, C. and {Williams}, O.~R. and {Zacchei}, A. and {Altieri}, B. and {Escudero Sanz}, I. and {Kohley}, R. and {Oosterbroek}, T. and {Astier}, P. and {Bacon}, D. and {Bardelli}, S. and {Baugh}, C. and {Bellagamba}, F. and {Benoist}, C. and {Bianchi}, D. and {Biviano}, A. and {Branchini}, E. and {Carbone}, C. and {Cardone}, V. and {Clements}, D. and {Colombi}, S. and {Conselice}, C. and {Cresci}, G. and {Deacon}, N. and {Dunlop}, J. and {Fedeli}, C. and {Fontanot}, F. and {Franzetti}, P. and {Giocoli}, C. and {Garcia-Bellido}, J. and {Gow}, J. and {Heavens}, A. and {Hewett}, P. and {Heymans}, C. and {Holland}, A. and {Huang}, Z. and {Ilbert}, O. and {Joachimi}, B. and {Jennins}, E. and {Kerins}, E. and {Kiessling}, A. and {Kirk}, D. and {Kotak}, R. and {Krause}, O. and {Lahav}, O. and {van Leeuwen}, F. and {Lesgourgues}, J. and {Lombardi}, M. and {Magliocchetti}, M. and {Maguire}, K. and {Majerotto}, E. and {Maoli}, R. and {Marulli}, F. and {Maurogordato}, S. and {McCracken}, H. and {McLure}, R. and {Melchiorri}, A. and {Merson}, A. and {Moresco}, M. and {Nonino}, M. and {Norberg}, P. and {Peacock}, J. and {Pello}, R. and {Penny}, M. and {Pettorino}, V. and {Di Porto}, C. and {Pozzetti}, L. and {Quercellini}, C. and {Radovich}, M. and {Rassat}, A. and {Roche}, N. and {Ronayette}, S. and {Rossetti}, E. and {Sartoris}, B. and {Schneider}, P. and {Semboloni}, E. and {Serjeant}, S. and {Simpson}, F. and {Skordis}, C. and {Smadja}, G. and {Smartt}, S. and {Spano}, P. and {Spiro}, S. and {Sullivan}, M. and {Tilquin}, A. and {Trotta}, R. and {Verde}, L. and {Wang}, Y. and {Williger}, G. and {Zhao}, G. and {Zoubian}, J. and {Zucca}, E.},
	eid = {arXiv:1110.3193},
	eprint = {1110.3193},
	journal = {arXiv e-prints},
	month = {October},
	pages = {arXiv:1110.3193},
	primaryclass = {astro-ph.CO},
	title = {{Euclid Definition Study Report}},
	year = {2011}}

@article{2012MNRAS.427..146H,
	adsurl = {https://ui.adsabs.harvard.edu/abs/2012MNRAS.427..146H},
	archiveprefix = {arXiv},
	author = {{Heymans}, Catherine and {Van Waerbeke}, Ludovic and {Miller}, Lance and {Erben}, Thomas and {Hildebrandt}, Hendrik and {Hoekstra}, Henk and {Kitching}, Thomas D. and {Mellier}, Yannick and {Simon}, Patrick and {Bonnett}, Christopher and {Coupon}, Jean and {Fu}, Liping and {Harnois D{\'e}raps}, Joachim and {Hudson}, Michael J. and {Kilbinger}, Martin and {Kuijken}, Koenraad and {Rowe}, Barnaby and {Schrabback}, Tim and {Semboloni}, Elisabetta and {van Uitert}, Edo and {Vafaei}, Sanaz and {Velander}, Malin},
	eprint = {1210.0032},
	journal = {\mnras},
	month = {November},
	number = {1},
	pages = {146-166},
	primaryclass = {astro-ph.CO},
	title = {{CFHTLenS: the Canada-France-Hawaii Telescope Lensing Survey}},
	volume = {427},
	year = {2012}}

@article{2013ApJS..205...12K,
	adsurl = {https://ui.adsabs.harvard.edu/abs/2013ApJS..205...12K},
	archiveprefix = {arXiv},
	author = {{Kitching}, T.~D. and {Rowe}, B. and {Gill}, M. and {Heymans}, C. and {Massey}, R. and {Witherick}, D. and {Courbin}, F. and {Georgatzis}, K. and {Gentile}, M. and {Gruen}, D. and {Kilbinger}, M. and {Li}, G.~L. and {Mariglis}, A.~P. and {Meylan}, G. and {Storkey}, A. and {Xin}, B.},
	eid = {12},
	eprint = {1210.1979},
	journal = {\apjs},
	month = {April},
	number = {2},
	pages = {12},
	primaryclass = {astro-ph.IM},
	title = {{Image Analysis for Cosmology: Results from the GREAT10 Star Challenge}},
	volume = {205},
	year = {2013}}

@article{2013ExA....35...25D,
	adsurl = {https://ui.adsabs.harvard.edu/abs/2013ExA....35...25D},
	archiveprefix = {arXiv},
	author = {{de Jong}, Jelte T.~A. and {Verdoes Kleijn}, Gijs A. and {Kuijken}, Konrad H. and {Valentijn}, Edwin A.},
	eprint = {1206.1254},
	journal = {Experimental Astronomy},
	month = {January},
	number = {1-2},
	pages = {25-44},
	primaryclass = {astro-ph.CO},
	title = {{The Kilo-Degree Survey}},
	volume = {35},
	year = {2013}}

@article{2013MNRAS.434.2121C,
	adsurl = {https://ui.adsabs.harvard.edu/abs/2013MNRAS.434.2121C},
	archiveprefix = {arXiv},
	author = {{Chang}, C. and {Jarvis}, M. and {Jain}, B. and {Kahn}, S.~M. and {Kirkby}, D. and {Connolly}, A. and {Krughoff}, S. and {Peng}, E. -H. and {Peterson}, J.~R.},
	eprint = {1305.0793},
	journal = {\mnras},
	month = {September},
	number = {3},
	pages = {2121-2135},
	primaryclass = {astro-ph.CO},
	title = {{The effective number density of galaxies for weak lensing measurements in the LSST project}},
	volume = {434},
	year = {2013}}

@article{2015A&C....10..121R,
	adsurl = {https://ui.adsabs.harvard.edu/abs/2015A&C....10..121R},
	archiveprefix = {arXiv},
	author = {{Rowe}, B.~T.~P. and {Jarvis}, M. and {Mandelbaum}, R. and {Bernstein}, G.~M. and {Bosch}, J. and {Simet}, M. and {Meyers}, J.~E. and {Kacprzak}, T. and {Nakajima}, R. and {Zuntz}, J. and {Miyatake}, H. and {Dietrich}, J.~P. and {Armstrong}, R. and {Melchior}, P. and {Gill}, M.~S.~S.},
	eprint = {1407.7676},
	journal = {Astronomy and Computing},
	month = {April},
	pages = {121-150},
	primaryclass = {astro-ph.IM},
	title = {{GALSIM: The modular galaxy image simulation toolkit}},
	volume = {10},
	year = {2015}}

@article{2015AJ....150..150F,
	adsurl = {https://ui.adsabs.harvard.edu/abs/2015AJ....150..150F},
	archiveprefix = {arXiv},
	author = {{Flaugher}, B. and {Diehl}, H.~T. and {Honscheid}, K. and {Abbott}, T.~M.~C. and {Alvarez}, O. and {Angstadt}, R. and {Annis}, J.~T. and {Antonik}, M. and {Ballester}, O. and {Beaufore}, L. and {Bernstein}, G.~M. and {Bernstein}, R.~A. and {Bigelow}, B. and {Bonati}, M. and {Boprie}, D. and {Brooks}, D. and {Buckley-Geer}, E.~J. and {Campa}, J. and {Cardiel-Sas}, L. and {Castander}, F.~J. and {Castilla}, J. and {Cease}, H. and {Cela-Ruiz}, J.~M. and {Chappa}, S. and {Chi}, E. and {Cooper}, C. and {da Costa}, L.~N. and {Dede}, E. and {Derylo}, G. and {DePoy}, D.~L. and {de Vicente}, J. and {Doel}, P. and {Drlica-Wagner}, A. and {Eiting}, J. and {Elliott}, A.~E. and {Emes}, J. and {Estrada}, J. and {Fausti Neto}, A. and {Finley}, D.~A. and {Flores}, R. and {Frieman}, J. and {Gerdes}, D. and {Gladders}, M.~D. and {Gregory}, B. and {Gutierrez}, G.~R. and {Hao}, J. and {Holland}, S.~E. and {Holm}, S. and {Huffman}, D. and {Jackson}, C. and {James}, D.~J. and {Jonas}, M. and {Karcher}, A. and {Karliner}, I. and {Kent}, S. and {Kessler}, R. and {Kozlovsky}, M. and {Kron}, R.~G. and {Kubik}, D. and {Kuehn}, K. and {Kuhlmann}, S. and {Kuk}, K. and {Lahav}, O. and {Lathrop}, A. and {Lee}, J. and {Levi}, M.~E. and {Lewis}, P. and {Li}, T.~S. and {Mandrichenko}, I. and {Marshall}, J.~L. and {Martinez}, G. and {Merritt}, K.~W. and {Miquel}, R. and {Mu{\~n}oz}, F. and {Neilsen}, E.~H. and {Nichol}, R.~C. and {Nord}, B. and {Ogando}, R. and {Olsen}, J. and {Palaio}, N. and {Patton}, K. and {Peoples}, J. and {Plazas}, A.~A. and {Rauch}, J. and {Reil}, K. and {Rheault}, J. -P. and {Roe}, N.~A. and {Rogers}, H. and {Roodman}, A. and {Sanchez}, E. and {Scarpine}, V. and {Schindler}, R.~H. and {Schmidt}, R. and {Schmitt}, R. and {Schubnell}, M. and {Schultz}, K. and {Schurter}, P. and {Scott}, L. and {Serrano}, S. and {Shaw}, T.~M. and {Smith}, R.~C. and {Soares-Santos}, M. and {Stefanik}, A. and {Stuermer}, W. and {Suchyta}, E. and {Sypniewski}, A. and {Tarle}, G. and {Thaler}, J. and {Tighe}, R. and {Tran}, C. and {Tucker}, D. and {Walker}, A.~R. and {Wang}, G. and {Watson}, M. and {Weaverdyck}, C. and {Wester}, W. and {Woods}, R. and {Yanny}, B. and {DES Collaboration}},
	eid = {150},
	eprint = {1504.02900},
	journal = {\aj},
	month = {November},
	number = {5},
	pages = {150},
	primaryclass = {astro-ph.IM},
	title = {{The Dark Energy Camera}},
	volume = {150},
	year = {2015}}

@article{2015arXiv150303757S,
	adsurl = {https://ui.adsabs.harvard.edu/abs/2015arXiv150303757S},
	archiveprefix = {arXiv},
	author = {{Spergel}, D. and {Gehrels}, N. and {Baltay}, C. and {Bennett}, D. and {Breckinridge}, J. and {Donahue}, M. and {Dressler}, A. and {Gaudi}, B.~S. and {Greene}, T. and {Guyon}, O. and {Hirata}, C. and {Kalirai}, J. and {Kasdin}, N.~J. and {Macintosh}, B. and {Moos}, W. and {Perlmutter}, S. and {Postman}, M. and {Rauscher}, B. and {Rhodes}, J. and {Wang}, Y. and {Weinberg}, D. and {Benford}, D. and {Hudson}, M. and {Jeong}, W. -S. and {Mellier}, Y. and {Traub}, W. and {Yamada}, T. and {Capak}, P. and {Colbert}, J. and {Masters}, D. and {Penny}, M. and {Savransky}, D. and {Stern}, D. and {Zimmerman}, N. and {Barry}, R. and {Bartusek}, L. and {Carpenter}, K. and {Cheng}, E. and {Content}, D. and {Dekens}, F. and {Demers}, R. and {Grady}, K. and {Jackson}, C. and {Kuan}, G. and {Kruk}, J. and {Melton}, M. and {Nemati}, B. and {Parvin}, B. and {Poberezhskiy}, I. and {Peddie}, C. and {Ruffa}, J. and {Wallace}, J.~K. and {Whipple}, A. and {Wollack}, E. and {Zhao}, F.},
	eid = {arXiv:1503.03757},
	eprint = {1503.03757},
	journal = {arXiv e-prints},
	month = {March},
	pages = {arXiv:1503.03757},
	primaryclass = {astro-ph.IM},
	title = {{Wide-Field InfrarRed Survey Telescope-Astrophysics Focused Telescope Assets WFIRST-AFTA 2015 Report}},
	year = {2015}}

@article{2015MNRAS.450.2963M,
	adsurl = {https://ui.adsabs.harvard.edu/abs/2015MNRAS.450.2963M},
	archiveprefix = {arXiv},
	author = {{Mandelbaum}, Rachel and {Rowe}, Barnaby and {Armstrong}, Robert and {Bard}, Deborah and {Bertin}, Emmanuel and {Bosch}, James and {Boutigny}, Dominique and {Courbin}, Frederic and {Dawson}, William A. and {Donnarumma}, Annamaria and {Fenech Conti}, Ian and {Gavazzi}, Rapha{\"e}l and {Gentile}, Marc and {Gill}, Mandeep S.~S. and {Hogg}, David W. and {Huff}, Eric M. and {Jee}, M. James and {Kacprzak}, Tomasz and {Kilbinger}, Martin and {Kuntzer}, Thibault and {Lang}, Dustin and {Luo}, Wentao and {March}, Marisa C. and {Marshall}, Philip J. and {Meyers}, Joshua E. and {Miller}, Lance and {Miyatake}, Hironao and {Nakajima}, Reiko and {Ngol{\'e} Mboula}, Fred Maurice and {Nurbaeva}, Guldariya and {Okura}, Yuki and {Paulin-Henriksson}, St{\'e}phane and {Rhodes}, Jason and {Schneider}, Michael D. and {Shan}, Huanyuan and {Sheldon}, Erin S. and {Simet}, Melanie and {Starck}, Jean-Luc and {Sureau}, Florent and {Tewes}, Malte and {Zarb Adami}, Kristian and {Zhang}, Jun and {Zuntz}, Joe},
	eprint = {1412.1825},
	journal = {\mnras},
	month = {July},
	number = {3},
	pages = {2963-3007},
	primaryclass = {astro-ph.CO},
	title = {{GREAT3 results - I. Systematic errors in shear estimation and the impact of real galaxy morphology}},
	volume = {450},
	year = {2015}}

@article{2016MNRAS.459.4467B,
	adsurl = {https://ui.adsabs.harvard.edu/abs/2016MNRAS.459.4467B},
	archiveprefix = {arXiv},
	author = {{Bernstein}, Gary M. and {Armstrong}, Robert and {Krawiec}, Christina and {March}, Marisa C.},
	eprint = {1508.05655},
	journal = {\mnras},
	month = {July},
	number = {4},
	pages = {4467-4484},
	primaryclass = {astro-ph.IM},
	title = {{An accurate and practical method for inference of weak gravitational lensing from galaxy images}},
	volume = {459},
	year = {2016}}

@article{2016MNRAS.460.2245J,
	adsurl = {https://ui.adsabs.harvard.edu/abs/2016MNRAS.460.2245J},
	archiveprefix = {arXiv},
	author = {{Jarvis}, M. and {Sheldon}, E. and {Zuntz}, J. and {Kacprzak}, T. and {Bridle}, S.~L. and {Amara}, A. and {Armstrong}, R. and {Becker}, M.~R. and {Bernstein}, G.~M. and {Bonnett}, C. and {Chang}, C. and {Das}, R. and {Dietrich}, J.~P. and {Drlica-Wagner}, A. and {Eifler}, T.~F. and {Gangkofner}, C. and {Gruen}, D. and {Hirsch}, M. and {Huff}, E.~M. and {Jain}, B. and {Kent}, S. and {Kirk}, D. and {MacCrann}, N. and {Melchior}, P. and {Plazas}, A.~A. and {Refregier}, A. and {Rowe}, B. and {Rykoff}, E.~S. and {Samuroff}, S. and {S{\'a}nchez}, C. and {Suchyta}, E. and {Troxel}, M.~A. and {Vikram}, V. and {Abbott}, T. and {Abdalla}, F.~B. and {Allam}, S. and {Annis}, J. and {Benoit-L{\'e}vy}, A. and {Bertin}, E. and {Brooks}, D. and {Buckley-Geer}, E. and {Burke}, D.~L. and {Capozzi}, D. and {Carnero Rosell}, A. and {Carrasco Kind}, M. and {Carretero}, J. and {Castander}, F.~J. and {Clampitt}, J. and {Crocce}, M. and {Cunha}, C.~E. and {D'Andrea}, C.~B. and {da Costa}, L.~N. and {DePoy}, D.~L. and {Desai}, S. and {Diehl}, H.~T. and {Doel}, P. and {Fausti Neto}, A. and {Flaugher}, B. and {Fosalba}, P. and {Frieman}, J. and {Gaztanaga}, E. and {Gerdes}, D.~W. and {Gruendl}, R.~A. and {Gutierrez}, G. and {Honscheid}, K. and {James}, D.~J. and {Kuehn}, K. and {Kuropatkin}, N. and {Lahav}, O. and {Li}, T.~S. and {Lima}, M. and {March}, M. and {Martini}, P. and {Miquel}, R. and {Mohr}, J.~J. and {Neilsen}, E. and {Nord}, B. and {Ogando}, R. and {Reil}, K. and {Romer}, A.~K. and {Roodman}, A. and {Sako}, M. and {Sanchez}, E. and {Scarpine}, V. and {Schubnell}, M. and {Sevilla-Noarbe}, I. and {Smith}, R.~C. and {Soares-Santos}, M. and {Sobreira}, F. and {Swanson}, M.~E.~C. and {Tarle}, G. and {Thaler}, J. and {Thomas}, D. and {Walker}, A.~R. and {Wechsler}, R.~H.},
	eprint = {1507.05603},
	journal = {\mnras},
	month = {August},
	number = {2},
	pages = {2245-2281},
	primaryclass = {astro-ph.IM},
	title = {{The DES Science Verification weak lensing shear catalogues}},
	volume = {460},
	year = {2016}}

@article{2017ApJ...841...24S,
	adsurl = {https://ui.adsabs.harvard.edu/abs/2017ApJ...841...24S},
	archiveprefix = {arXiv},
	author = {{Sheldon}, Erin S. and {Huff}, Eric M.},
	eid = {24},
	eprint = {1702.02601},
	journal = {\apj},
	month = {May},
	number = {1},
	pages = {24},
	primaryclass = {astro-ph.CO},
	title = {{Practical Weak-lensing Shear Measurement with Metacalibration}},
	volume = {841},
	year = {2017}}

@article{2017arXiv170202600H,
	adsurl = {https://ui.adsabs.harvard.edu/abs/2017arXiv170202600H},
	archiveprefix = {arXiv},
	author = {{Huff}, Eric and {Mandelbaum}, Rachel},
	eid = {arXiv:1702.02600},
	eprint = {1702.02600},
	journal = {arXiv e-prints},
	month = {February},
	pages = {arXiv:1702.02600},
	primaryclass = {astro-ph.CO},
	title = {{Metacalibration: Direct Self-Calibration of Biases in Shear Measurement}},
	year = {2017}}

@article{2018ARA&A..56..393M,
	adsurl = {https://ui.adsabs.harvard.edu/abs/2018ARA&A..56..393M},
	archiveprefix = {arXiv},
	author = {{Mandelbaum}, Rachel},
	eprint = {1710.03235},
	journal = {\araa},
	month = {September},
	pages = {393-433},
	primaryclass = {astro-ph.CO},
	title = {{Weak Lensing for Precision Cosmology}},
	volume = {56},
	year = {2018}}

@article{2018MNRAS.481.1149Z,
	archiveprefix = {arXiv},
        author = "Zuntz, J. and Sheldon, E. and others",
	eprint = {1708.01533},
	journal = {\mnras},
	month = {November},
	number = {1},
	pages = {1149-1182},
	primaryclass = {astro-ph.CO},
	title = {{Dark Energy Survey Year 1 results: weak lensing shape catalogues}},
	volume = {481},
	year = {2018}}

@article{2018PASJ...70S...4A,
	adsurl = {https://ui.adsabs.harvard.edu/abs/2018PASJ...70S...4A},
	archiveprefix = {arXiv},
	author = {{Aihara}, Hiroaki and {Arimoto}, Nobuo and {Armstrong}, Robert and {Arnouts}, St{\'e}phane and {Bahcall}, Neta A. and {Bickerton}, Steven and {Bosch}, James and {Bundy}, Kevin and {Capak}, Peter L. and {Chan}, James H.~H. and {Chiba}, Masashi and {Coupon}, Jean and {Egami}, Eiichi and {Enoki}, Motohiro and {Finet}, Francois and {Fujimori}, Hiroki and {Fujimoto}, Seiji and {Furusawa}, Hisanori and {Furusawa}, Junko and {Goto}, Tomotsugu and {Goulding}, Andy and {Greco}, Johnny P. and {Greene}, Jenny E. and {Gunn}, James E. and {Hamana}, Takashi and {Harikane}, Yuichi and {Hashimoto}, Yasuhiro and {Hattori}, Takashi and {Hayashi}, Masao and {Hayashi}, Yusuke and {He{\l}miniak}, Krzysztof G. and {Higuchi}, Ryo and {Hikage}, Chiaki and {Ho}, Paul T.~P. and {Hsieh}, Bau-Ching and {Huang}, Kuiyun and {Huang}, Song and {Ikeda}, Hiroyuki and {Imanishi}, Masatoshi and {Inoue}, Akio K. and {Iwasawa}, Kazushi and {Iwata}, Ikuru and {Jaelani}, Anton T. and {Jian}, Hung-Yu and {Kamata}, Yukiko and {Karoji}, Hiroshi and {Kashikawa}, Nobunari and {Katayama}, Nobuhiko and {Kawanomoto}, Satoshi and {Kayo}, Issha and {Koda}, Jin and {Koike}, Michitaro and {Kojima}, Takashi and {Komiyama}, Yutaka and {Konno}, Akira and {Koshida}, Shintaro and {Koyama}, Yusei and {Kusakabe}, Haruka and {Leauthaud}, Alexie and {Lee}, Chien-Hsiu and {Lin}, Lihwai and {Lin}, Yen-Ting and {Lupton}, Robert H. and {Mandelbaum}, Rachel and {Matsuoka}, Yoshiki and {Medezinski}, Elinor and {Mineo}, Sogo and {Miyama}, Shoken and {Miyatake}, Hironao and {Miyazaki}, Satoshi and {Momose}, Rieko and {More}, Anupreeta and {More}, Surhud and {Moritani}, Yuki and {Moriya}, Takashi J. and {Morokuma}, Tomoki and {Mukae}, Shiro and {Murata}, Ryoma and {Murayama}, Hitoshi and {Nagao}, Tohru and {Nakata}, Fumiaki and {Niida}, Mana and {Niikura}, Hiroko and {Nishizawa}, Atsushi J. and {Obuchi}, Yoshiyuki and {Oguri}, Masamune and {Oishi}, Yukie and {Okabe}, Nobuhiro and {Okamoto}, Sakurako and {Okura}, Yuki and {Ono}, Yoshiaki and {Onodera}, Masato and {Onoue}, Masafusa and {Osato}, Ken and {Ouchi}, Masami and {Price}, Paul A. and {Pyo}, Tae-Soo and {Sako}, Masao and {Sawicki}, Marcin and {Shibuya}, Takatoshi and {Shimasaku}, Kazuhiro and {Shimono}, Atsushi and {Shirasaki}, Masato and {Silverman}, John D. and {Simet}, Melanie and {Speagle}, Joshua and {Spergel}, David N. and {Strauss}, Michael A. and {Sugahara}, Yuma and {Sugiyama}, Naoshi and {Suto}, Yasushi and {Suyu}, Sherry H. and {Suzuki}, Nao and {Tait}, Philip J. and {Takada}, Masahiro and {Takata}, Tadafumi and {Tamura}, Naoyuki and {Tanaka}, Manobu M. and {Tanaka}, Masaomi and {Tanaka}, Masayuki and {Tanaka}, Yoko and {Terai}, Tsuyoshi and {Terashima}, Yuichi and {Toba}, Yoshiki and {Tominaga}, Nozomu and {Toshikawa}, Jun and {Turner}, Edwin L. and {Uchida}, Tomohisa and {Uchiyama}, Hisakazu and {Umetsu}, Keiichi and {Uraguchi}, Fumihiro and {Urata}, Yuji and {Usuda}, Tomonori and {Utsumi}, Yousuke and {Wang}, Shiang-Yu and {Wang}, Wei-Hao and {Wong}, Kenneth C. and {Yabe}, Kiyoto and {Yamada}, Yoshihiko and {Yamanoi}, Hitomi and {Yasuda}, Naoki and {Yeh}, Sherry and {Yonehara}, Atsunori and {Yuma}, Suraphong},
	eid = {S4},
	eprint = {1704.05858},
	journal = {\pasj},
	month = {January},
	pages = {S4},
	primaryclass = {astro-ph.IM},
	title = {{The Hyper Suprime-Cam SSP Survey: Overview and survey design}},
	volume = {70},
	year = {2018}}

@article{2019ApJ...873..111I,
	adsurl = {https://ui.adsabs.harvard.edu/abs/2019ApJ...873..111I},
	archiveprefix = {arXiv},
	author = {{Ivezi{\'c}}, {\v{Z}}eljko and {Kahn}, Steven M. and {Tyson}, J. Anthony and {Abel}, Bob and {Acosta}, Emily and {Allsman}, Robyn and {Alonso}, David and {AlSayyad}, Yusra and {Anderson}, Scott F. and {Andrew}, John and {Angel}, James Roger P. and {Angeli}, George Z. and {Ansari}, Reza and {Antilogus}, Pierre and {Araujo}, Constanza and {Armstrong}, Robert and {Arndt}, Kirk T. and {Astier}, Pierre and {Aubourg}, {\'E}ric and {Auza}, Nicole and {Axelrod}, Tim S. and {Bard}, Deborah J. and {Barr}, Jeff D. and {Barrau}, Aurelian and {Bartlett}, James G. and {Bauer}, Amanda E. and {Bauman}, Brian J. and {Baumont}, Sylvain and {Bechtol}, Ellen and {Bechtol}, Keith and {Becker}, Andrew C. and {Becla}, Jacek and {Beldica}, Cristina and {Bellavia}, Steve and {Bianco}, Federica B. and {Biswas}, Rahul and {Blanc}, Guillaume and {Blazek}, Jonathan and {Blandford}, Roger D. and {Bloom}, Josh S. and {Bogart}, Joanne and {Bond}, Tim W. and {Booth}, Michael T. and {Borgland}, Anders W. and {Borne}, Kirk and {Bosch}, James F. and {Boutigny}, Dominique and {Brackett}, Craig A. and {Bradshaw}, Andrew and {Brandt}, William Nielsen and {Brown}, Michael E. and {Bullock}, James S. and {Burchat}, Patricia and {Burke}, David L. and {Cagnoli}, Gianpietro and {Calabrese}, Daniel and {Callahan}, Shawn and {Callen}, Alice L. and {Carlin}, Jeffrey L. and {Carlson}, Erin L. and {Chandrasekharan}, Srinivasan and {Charles-Emerson}, Glenaver and {Chesley}, Steve and {Cheu}, Elliott C. and {Chiang}, Hsin-Fang and {Chiang}, James and {Chirino}, Carol and {Chow}, Derek and {Ciardi}, David R. and {Claver}, Charles F. and {Cohen-Tanugi}, Johann and {Cockrum}, Joseph J. and {Coles}, Rebecca and {Connolly}, Andrew J. and {Cook}, Kem H. and {Cooray}, Asantha and {Covey}, Kevin R. and {Cribbs}, Chris and {Cui}, Wei and {Cutri}, Roc and {Daly}, Philip N. and {Daniel}, Scott F. and {Daruich}, Felipe and {Daubard}, Guillaume and {Daues}, Greg and {Dawson}, William and {Delgado}, Francisco and {Dellapenna}, Alfred and {de Peyster}, Robert and {de Val-Borro}, Miguel and {Digel}, Seth W. and {Doherty}, Peter and {Dubois}, Richard and {Dubois-Felsmann}, Gregory P. and {Durech}, Josef and {Economou}, Frossie and {Eifler}, Tim and {Eracleous}, Michael and {Emmons}, Benjamin L. and {Fausti Neto}, Angelo and {Ferguson}, Henry and {Figueroa}, Enrique and {Fisher-Levine}, Merlin and {Focke}, Warren and {Foss}, Michael D. and {Frank}, James and {Freemon}, Michael D. and {Gangler}, Emmanuel and {Gawiser}, Eric and {Geary}, John C. and {Gee}, Perry and {Geha}, Marla and {Gessner}, Charles J.~B. and {Gibson}, Robert R. and {Gilmore}, D. Kirk and {Glanzman}, Thomas and {Glick}, William and {Goldina}, Tatiana and {Goldstein}, Daniel A. and {Goodenow}, Iain and {Graham}, Melissa L. and {Gressler}, William J. and {Gris}, Philippe and {Guy}, Leanne P. and {Guyonnet}, Augustin and {Haller}, Gunther and {Harris}, Ron and {Hascall}, Patrick A. and {Haupt}, Justine and {Hernandez}, Fabio and {Herrmann}, Sven and {Hileman}, Edward and {Hoblitt}, Joshua and {Hodgson}, John A. and {Hogan}, Craig and {Howard}, James D. and {Huang}, Dajun and {Huffer}, Michael E. and {Ingraham}, Patrick and {Innes}, Walter R. and {Jacoby}, Suzanne H. and {Jain}, Bhuvnesh and {Jammes}, Fabrice and {Jee}, M. James and {Jenness}, Tim and {Jernigan}, Garrett and {Jevremovi{\'c}}, Darko and {Johns}, Kenneth and {Johnson}, Anthony S. and {Johnson}, Margaret W.~G. and {Jones}, R. Lynne and {Juramy-Gilles}, Claire and {Juri{\'c}}, Mario and {Kalirai}, Jason S. and {Kallivayalil}, Nitya J. and {Kalmbach}, Bryce and {Kantor}, Jeffrey P. and {Karst}, Pierre and {Kasliwal}, Mansi M. and {Kelly}, Heather and {Kessler}, Richard and {Kinnison}, Veronica and {Kirkby}, David and {Knox}, Lloyd and {Kotov}, Ivan V. and {Krabbendam}, Victor L. and {Krughoff}, K. Simon and {Kub{\'a}nek}, Petr and {Kuczewski}, John and {Kulkarni}, Shri and {Ku}, John and {Kurita}, Nadine R. and {Lage}, Craig S. and {Lambert}, Ron and {Lange}, Travis and {Langton}, J. Brian and {Le Guillou}, Laurent and {Levine}, Deborah and {Liang}, Ming and {Lim}, Kian-Tat and {Lintott}, Chris J. and {Long}, Kevin E. and {Lopez}, Margaux and {Lotz}, Paul J. and {Lupton}, Robert H. and {Lust}, Nate B. and {MacArthur}, Lauren A. and {Mahabal}, Ashish and {Mandelbaum}, Rachel and {Markiewicz}, Thomas W. and {Marsh}, Darren S. and {Marshall}, Philip J. and {Marshall}, Stuart and {May}, Morgan and {McKercher}, Robert and {McQueen}, Michelle and {Meyers}, Joshua and {Migliore}, Myriam and {Miller}, Michelle and {Mills}, David J. and {Miraval}, Connor and {Moeyens}, Joachim and {Moolekamp}, Fred E. and {Monet}, David G. and {Moniez}, Marc and {Monkewitz}, Serge and {Montgomery}, Christopher and {Morrison}, Christopher B. and {Mueller}, Fritz and {Muller}, Gary P. and {Mu{\~n}oz Arancibia}, Freddy and {Neill}, Douglas R. and {Newbry}, Scott P. and {Nief}, Jean-Yves and {Nomerotski}, Andrei and {Nordby}, Martin and {O'Connor}, Paul and {Oliver}, John and {Olivier}, Scot S. and {Olsen}, Knut and {O'Mullane}, William and {Ortiz}, Sandra and {Osier}, Shawn and {Owen}, Russell E. and {Pain}, Reynald and {Palecek}, Paul E. and {Parejko}, John K. and {Parsons}, James B. and {Pease}, Nathan M. and {Peterson}, J. Matt and {Peterson}, John R. and {Petravick}, Donald L. and {Libby Petrick}, M.~E. and {Petry}, Cathy E. and {Pierfederici}, Francesco and {Pietrowicz}, Stephen and {Pike}, Rob and {Pinto}, Philip A. and {Plante}, Raymond and {Plate}, Stephen and {Plutchak}, Joel P. and {Price}, Paul A. and {Prouza}, Michael and {Radeka}, Veljko and {Rajagopal}, Jayadev and {Rasmussen}, Andrew P. and {Regnault}, Nicolas and {Reil}, Kevin A. and {Reiss}, David J. and {Reuter}, Michael A. and {Ridgway}, Stephen T. and {Riot}, Vincent J. and {Ritz}, Steve and {Robinson}, Sean and {Roby}, William and {Roodman}, Aaron and {Rosing}, Wayne and {Roucelle}, Cecille and {Rumore}, Matthew R. and {Russo}, Stefano and {Saha}, Abhijit and {Sassolas}, Benoit and {Schalk}, Terry L. and {Schellart}, Pim and {Schindler}, Rafe H. and {Schmidt}, Samuel and {Schneider}, Donald P. and {Schneider}, Michael D. and {Schoening}, William and {Schumacher}, German and {Schwamb}, Megan E. and {Sebag}, Jacques and {Selvy}, Brian and {Sembroski}, Glenn H. and {Seppala}, Lynn G. and {Serio}, Andrew and {Serrano}, Eduardo and {Shaw}, Richard A. and {Shipsey}, Ian and {Sick}, Jonathan and {Silvestri}, Nicole and {Slater}, Colin T. and {Smith}, J. Allyn and {Smith}, R. Chris and {Sobhani}, Shahram and {Soldahl}, Christine and {Storrie-Lombardi}, Lisa and {Stover}, Edward and {Strauss}, Michael A. and {Street}, Rachel A. and {Stubbs}, Christopher W. and {Sullivan}, Ian S. and {Sweeney}, Donald and {Swinbank}, John D. and {Szalay}, Alexander and {Takacs}, Peter and {Tether}, Stephen A. and {Thaler}, Jon J. and {Thayer}, John Gregg and {Thomas}, Sandrine and {Thornton}, Adam J. and {Thukral}, Vaikunth and {Tice}, Jeffrey and {Trilling}, David E. and {Turri}, Max and {Van Berg}, Richard and {Vanden Berk}, Daniel and {Vetter}, Kurt and {Virieux}, Francoise and {Vucina}, Tomislav and {Wahl}, William and {Walkowicz}, Lucianne and {Walsh}, Brian and {Walter}, Christopher W. and {Wang}, Daniel L. and {Wang}, Shin-Yawn and {Warner}, Michael and {Wiecha}, Oliver and {Willman}, Beth and {Winters}, Scott E. and {Wittman}, David and {Wolff}, Sidney C. and {Wood-Vasey}, W. Michael and {Wu}, Xiuqin and {Xin}, Bo and {Yoachim}, Peter and {Zhan}, Hu},
	eid = {111},
	eprint = {0805.2366},
	journal = {\apj},
	month = {March},
	number = {2},
	pages = {111},
	primaryclass = {astro-ph},
	title = {{LSST: From Science Drivers to Reference Design and Anticipated Data Products}},
	volume = {873},
	year = {2019}}

@article{2020ApJ...902..138S,
	adsurl = {https://ui.adsabs.harvard.edu/abs/2020ApJ...902..138S},
	archiveprefix = {arXiv},
	author = {{Sheldon}, Erin S. and {Becker}, Matthew R. and {MacCrann}, Niall and {Jarvis}, Michael},
	eid = {138},
	eprint = {1911.02505},
	journal = {\apj},
	month = {October},
	number = {2},
	pages = {138},
	primaryclass = {astro-ph.CO},
	title = {{Mitigating Shear-dependent Object Detection Biases with Metacalibration}},
	volume = {902},
	year = {2020}}

@article{2022MNRAS.509.3371M,
	archiveprefix = "arXiv",
	author = {{MacCrann, Becker et al.}},
	doi = "10.1093/mnras/stab2870",
	eprint = "2012.08567",
	journal = "MNRAS",
	number = "3",
	pages = "3371-3394",
	primaryclass = "astro-ph.CO",
	title = "{Dark Energy Survey Y3 results: blending shear and redshift biases in image simulations}",
	volume = "509",
	year = "2022",
}

@ARTICLE{2024ApJ...961...59T,
       author = {{To}, Chun-Hao and {DeRose}, Joseph and {Wechsler}, Risa H. and {Rykoff}, Eli and {Wu}, Hao-Yi and {Adhikari}, Susmita and {Krause}, Elisabeth and {Rozo}, Eduardo and {Weinberg}, David H.},
        title = "{Buzzard to Cardinal: Improved Mock Catalogs for Large Galaxy Surveys}",
      journal = {\apj},
         year = 2024,
        month = jan,
       volume = {961},
       number = {1},
          eid = {59},
        pages = {59},
          doi = {10.3847/1538-4357/ad0e61},
archivePrefix = {arXiv},
       eprint = {2303.12104},
 primaryClass = {astro-ph.CO},
       adsurl = {https://ui.adsabs.harvard.edu/abs/2024ApJ...961...59T}
}

@ARTICLE{2016PhRvD..94b2002B,
       author = {{Becker}, M.~R. and {Troxel}, M.~A. and {MacCrann}, N. and {Krause}, E. and {Eifler}, T.~F. and {Friedrich}, O. and {Nicola}, A. and {Refregier}, A. and {Amara}, A. and {Bacon}, D. and {Bernstein}, G.~M. and {Bonnett}, C. and {Bridle}, S.~L. and {Busha}, M.~T. and {Chang}, C. and {Dodelson}, S. and {Erickson}, B. and {Evrard}, A.~E. and {Frieman}, J. and {Gaztanaga}, E. and {Gruen}, D. and {Hartley}, W. and {Jain}, B. and {Jarvis}, M. and {Kacprzak}, T. and {Kirk}, D. and {Kravtsov}, A. and {Leistedt}, B. and {Peiris}, H.~V. and {Rykoff}, E.~S. and {Sabiu}, C. and {S{\'a}nchez}, C. and {Seo}, H. and {Sheldon}, E. and {Wechsler}, R.~H. and {Zuntz}, J. and {Abbott}, T. and {Abdalla}, F.~B. and {Allam}, S. and {Armstrong}, R. and {Banerji}, M. and {Bauer}, A.~H. and {Benoit-L{\'e}vy}, A. and {Bertin}, E. and {Brooks}, D. and {Buckley-Geer}, E. and {Burke}, D.~L. and {Capozzi}, D. and {Carnero Rosell}, A. and {Carrasco Kind}, M. and {Carretero}, J. and {Castander}, F.~J. and {Crocce}, M. and {Cunha}, C.~E. and {D'Andrea}, C.~B. and {da Costa}, L.~N. and {DePoy}, D.~L. and {Desai}, S. and {Diehl}, H.~T. and {Dietrich}, J.~P. and {Doel}, P. and {Fausti Neto}, A. and {Fernandez}, E. and {Finley}, D.~A. and {Flaugher}, B. and {Fosalba}, P. and {Gerdes}, D.~W. and {Gruendl}, R.~A. and {Gutierrez}, G. and {Honscheid}, K. and {James}, D.~J. and {Kuehn}, K. and {Kuropatkin}, N. and {Lahav}, O. and {Li}, T.~S. and {Lima}, M. and {Maia}, M.~A.~G. and {March}, M. and {Martini}, P. and {Melchior}, P. and {Miller}, C.~J. and {Miquel}, R. and {Mohr}, J.~J. and {Nichol}, R.~C. and {Nord}, B. and {Ogando}, R. and {Plazas}, A.~A. and {Reil}, K. and {Romer}, A.~K. and {Roodman}, A. and {Sako}, M. and {Sanchez}, E. and {Scarpine}, V. and {Schubnell}, M. and {Sevilla-Noarbe}, I. and {Smith}, R.~C. and {Soares-Santos}, M. and {Sobreira}, F. and {Suchyta}, E. and {Swanson}, M.~E.~C. and {Tarle}, G. and {Thaler}, J. and {Thomas}, D. and {Vikram}, V. and {Walker}, A.~R. and {Dark Energy Survey Collaboration}},
        title = "{Cosmic shear measurements with Dark Energy Survey Science Verification data}",
      journal = {\prd},
         year = 2016,
        month = jul,
       volume = {94},
       number = {2},
          eid = {022002},
        pages = {022002},
          doi = {10.1103/PhysRevD.94.022002},
archivePrefix = {arXiv},
       eprint = {1507.05598},
 primaryClass = {astro-ph.CO},
       adsurl = {https://ui.adsabs.harvard.edu/abs/2016PhRvD..94b2002B}
}

@article{2021ApJS..255...20A,
	adsurl = {https://ui.adsabs.harvard.edu/abs/2021ApJS..255...20A},
	archiveprefix = {arXiv},
	author = {DES collaboration, {Abbott}, T.~M.~C. and {Adam{\'o}w}, M. and {Aguena}, M. and {Allam}, S. and {Amon}, A. and {Annis}, J. and {Avila}, S. and {Bacon}, D. and {Banerji}, M. and {Bechtol}, K. and {Becker}, M.~R. and {Bernstein}, G.~M. and {Bertin}, E. and {Bhargava}, S. and {Bridle}, S.~L. and {Brooks}, D. and {Burke}, D.~L. and {Carnero Rosell}, A. and {Carrasco Kind}, M. and {Carretero}, J. and {Castander}, F.~J. and {Cawthon}, R. and {Chang}, C. and {Choi}, A. and {Conselice}, C. and {Costanzi}, M. and {Crocce}, M. and {da Costa}, L.~N. and {Davis}, T.~M. and {De Vicente}, J. and {DeRose}, J. and {Desai}, S. and {Diehl}, H.~T. and {Dietrich}, J.~P. and {Drlica-Wagner}, A. and {Eckert}, K. and {Elvin-Poole}, J. and {Everett}, S. and {Evrard}, A.~E. and {Ferrero}, I. and {Fert{\'e}}, A. and {Flaugher}, B. and {Fosalba}, P. and {Friedel}, D. and {Frieman}, J. and {Garc{\'\i}a-Bellido}, J. and {Gaztanaga}, E. and {Gelman}, L. and {Gerdes}, D.~W. and {Giannantonio}, T. and {Gill}, M.~S.~S. and {Gruen}, D. and {Gruendl}, R.~A. and {Gschwend}, J. and {Gutierrez}, G. and {Hartley}, W.~G. and {Hinton}, S.~R. and {Hollowood}, D.~L. and {Honscheid}, K. and {Huterer}, D. and {James}, D.~J. and {Jeltema}, T. and {Johnson}, M.~D. and {Kent}, S. and {Kron}, R. and {Kuehn}, K. and {Kuropatkin}, N. and {Lahav}, O. and {Li}, T.~S. and {Lidman}, C. and {Lin}, H. and {MacCrann}, N. and {Maia}, M.~A.~G. and {Manning}, T.~A. and {Maloney}, J.~D. and {March}, M. and {Marshall}, J.~L. and {Martini}, P. and {Melchior}, P. and {Menanteau}, F. and {Miquel}, R. and {Morgan}, R. and {Myles}, J. and {Neilsen}, E. and {Ogando}, R.~L.~C. and {Palmese}, A. and {Paz-Chinch{\'o}n}, F. and {Petravick}, D. and {Pieres}, A. and {Plazas}, A.~A. and {Pond}, C. and {Rodriguez-Monroy}, M. and {Romer}, A.~K. and {Roodman}, A. and {Rykoff}, E.~S. and {Sako}, M. and {Sanchez}, E. and {Santiago}, B. and {Scarpine}, V. and {Serrano}, S. and {Sevilla-Noarbe}, I. and {Smith}, J. Allyn and {Smith}, M. and {Soares-Santos}, M. and {Suchyta}, E. and {Swanson}, M.~E.~C. and {Tarle}, G. and {Thomas}, D. and {To}, C. and {Tremblay}, P.~E. and {Troxel}, M.~A. and {Tucker}, D.~L. and {Turner}, D.~J. and {Varga}, T.~N. and {Walker}, A.~R. and {Wechsler}, R.~H. and {Weller}, J. and {Wester}, W. and {Wilkinson}, R.~D. and {Yanny}, B. and {Zhang}, Y. and {Nikutta}, R. and {Fitzpatrick}, M. and {Jacques}, A. and {Scott}, A. and {Olsen}, K. and {Huang}, L. and {Herrera}, D. and {Juneau}, S. and {Nidever}, D. and {Weaver}, B.~A. and {Adean}, C. and {Correia}, V. and {de Freitas}, M. and {Freitas}, F.~N. and {Singulani}, C. and {Vila-Verde}, G. and {Linea Science Server}},
	eid = {20},
	eprint = {2101.05765},
	journal = {\apjs},
	month = {August},
	number = {2},
	pages = {20},
	primaryclass = {astro-ph.IM},
	title = {{The Dark Energy Survey Data Release 2}},
	volume = {255},
	year = {2021}}

@article{2021MNRAS.501.1282J,
	adsurl = {https://ui.adsabs.harvard.edu/abs/2021MNRAS.501.1282J},
	archiveprefix = {arXiv},
	author = {{Jarvis}, M. and {Bernstein}, G.~M. and {Amon}, A. and {Davis}, C. and {L{\'e}get}, P.~F. and {Bechtol}, K. and {Harrison}, I. and {Gatti}, M. and {Roodman}, A. and {Chang}, C. and {Chen}, R. and {Choi}, A. and {Desai}, S. and {Drlica-Wagner}, A. and {Gruen}, D. and {Gruendl}, R.~A. and {Hernandez}, A. and {MacCrann}, N. and {Meyers}, J. and {Navarro-Alsina}, A. and {Pandey}, S. and {Plazas}, A.~A. and {Secco}, L.~F. and {Sheldon}, E. and {Troxel}, M.~A. and {Vorperian}, S. and {Wei}, K. and {Zuntz}, J. and {Abbott}, T.~M.~C. and {Aguena}, M. and {Allam}, S. and {Avila}, S. and {Bhargava}, S. and {Bridle}, S.~L. and {Brooks}, D. and {Carnero Rosell}, A. and {Carrasco Kind}, M. and {Carretero}, J. and {Costanzi}, M. and {da Costa}, L.~N. and {De Vicente}, J. and {Diehl}, H.~T. and {Doel}, P. and {Everett}, S. and {Flaugher}, B. and {Fosalba}, P. and {Frieman}, J. and {Garc{\'\i}a-Bellido}, J. and {Gaztanaga}, E. and {Gerdes}, D.~W. and {Gutierrez}, G. and {Hinton}, S.~R. and {Hollowood}, D.~L. and {Honscheid}, K. and {James}, D.~J. and {Kent}, S. and {Kuehn}, K. and {Kuropatkin}, N. and {Lahav}, O. and {Maia}, M.~A.~G. and {March}, M. and {Marshall}, J.~L. and {Melchior}, P. and {Menanteau}, F. and {Miquel}, R. and {Ogando}, R.~L.~C. and {Paz-Chinch{\'o}n}, F. and {Rykoff}, E.~S. and {Sanchez}, E. and {Scarpine}, V. and {Schubnell}, M. and {Serrano}, S. and {Sevilla-Noarbe}, I. and {Smith}, M. and {Suchyta}, E. and {Swanson}, M.~E.~C. and {Tarle}, G. and {Varga}, T.~N. and {Walker}, A.~R. and {Wester}, W. and {Wilkinson}, R.~D. and {Wilkinson}, R.~D. and {DES Collaboration}},
	eprint = {2011.03409},
	journal = {\mnras},
	month = {February},
	number = {1},
	pages = {1282-1299},
	primaryclass = {astro-ph.IM},
	title = {{Dark Energy Survey year 3 results: point spread function modelling}},
	volume = {501},
	year = {2021}}

@article{y3-shapecatalog,
    author = "Gatti, M. and Sheldon, E. and others",
    collaboration = "DES",
    title = "{Dark Energy Survey Year 3 Results: Weak Lensing Shape Catalogue}",
    eprint = "2011.03408",
    archivePrefix = "arXiv",
    primaryClass = "astro-ph.CO",
    reportNumber = "FERMILAB-PUB-20-545-AE, DES-2015-0048",
    doi = "10.1093/mnras/stab918",
    journal = "MNRAS",
    volume = "504",
    number = "3",
    pages = "4312-4336",
    year = "2021"
}

@article{2022arXiv221203257Z,
	adsurl = {https://ui.adsabs.harvard.edu/abs/2023MNRAS.525.2441Z},
	archiveprefix = {arXiv},
	author = {{Zhang}, Tianqing and {Li}, Xiangchong and {Dalal}, Roohi and {Mandelbaum}, Rachel and {Strauss}, Michael A. and {Kannawadi}, Arun and {Miyatake}, Hironao and {Nicola}, Andrina and {Malag{\'o}n}, Andr{\'e}s A. Plazas and {Shirasaki}, Masato and {Sugiyama}, Sunao and {Takada}, Masahiro and {More}, Surhud},
	doi = {10.1093/mnras/stad1801},
	eprint = {2212.03257},
	journal = {\mnras},
	month = {October},
	number = {2},
	pages = {2441-2471},
	primaryclass = {astro-ph.CO},
	title = {{A general framework for removing point-spread function additive systematics in cosmological weak lensing analysis}},
	volume = {525},
	year = {2023}}

@article{2022MNRAS.510.3207P,
	adsurl = {https://ui.adsabs.harvard.edu/abs/2022MNRAS.510.3207P},
	archiveprefix = {arXiv},
	author = {{Percival}, Will J. and {Friedrich}, Oliver and {Sellentin}, Elena and {Heavens}, Alan},
	eprint = {2108.10402},
	journal = {\mnras},
	month = {March},
	number = {3},
	pages = {3207-3221},
	primaryclass = {astro-ph.IM},
	title = {{Matching Bayesian and frequentist coverage probabilities when using an approximate data covariance matrix}},
	volume = {510},
	year = {2022}}

@article{2022PhRvD.105b3514A,
	adsurl = {https://ui.adsabs.harvard.edu/abs/2022PhRvD.105b3514A},
	archiveprefix = {arXiv},
	author = {{Amon}, A. and {Gruen}, D. and {Troxel}, M.~A. and {MacCrann}, N. and {Dodelson}, S. and {Choi}, A. and {Doux}, C. and {Secco}, L.~F. and {Samuroff}, S. and {Krause}, E. and {Cordero}, J. and {Myles}, J. and {DeRose}, J. and {Wechsler}, R.~H. and {Gatti}, M. and {Navarro-Alsina}, A. and {Bernstein}, G.~M. and {Jain}, B. and {Blazek}, J. and {Alarcon}, A. and {Fert{\'e}}, A. and {Lemos}, P. and {Raveri}, M. and {Campos}, A. and {Prat}, J. and {S{\'a}nchez}, C. and {Jarvis}, M. and {Alves}, O. and {Andrade-Oliveira}, F. and {Baxter}, E. and {Bechtol}, K. and {Becker}, M.~R. and {Bridle}, S.~L. and {Camacho}, H. and {Carnero Rosell}, A. and {Carrasco Kind}, M. and {Cawthon}, R. and {Chang}, C. and {Chen}, R. and {Chintalapati}, P. and {Crocce}, M. and {Davis}, C. and {Diehl}, H.~T. and {Drlica-Wagner}, A. and {Eckert}, K. and {Eifler}, T.~F. and {Elvin-Poole}, J. and {Everett}, S. and {Fang}, X. and {Fosalba}, P. and {Friedrich}, O. and {Gaztanaga}, E. and {Giannini}, G. and {Gruendl}, R.~A. and {Harrison}, I. and {Hartley}, W.~G. and {Herner}, K. and {Huang}, H. and {Huff}, E.~M. and {Huterer}, D. and {Kuropatkin}, N. and {Leget}, P. and {Liddle}, A.~R. and {McCullough}, J. and {Muir}, J. and {Pandey}, S. and {Park}, Y. and {Porredon}, A. and {Refregier}, A. and {Rollins}, R.~P. and {Roodman}, A. and {Rosenfeld}, R. and {Ross}, A.~J. and {Rykoff}, E.~S. and {Sanchez}, J. and {Sevilla-Noarbe}, I. and {Sheldon}, E. and {Shin}, T. and {Troja}, A. and {Tutusaus}, I. and {Tutusaus}, I. and {Varga}, T.~N. and {Weaverdyck}, N. and {Yanny}, B. and {Yin}, B. and {Zhang}, Y. and {Zuntz}, J. and {Aguena}, M. and {Allam}, S. and {Annis}, J. and {Bacon}, D. and {Bertin}, E. and {Bhargava}, S. and {Brooks}, D. and {Buckley-Geer}, E. and {Burke}, D.~L. and {Carretero}, J. and {Costanzi}, M. and {da Costa}, L.~N. and {Pereira}, M.~E.~S. and {De Vicente}, J. and {Desai}, S. and {Dietrich}, J.~P. and {Doel}, P. and {Ferrero}, I. and {Flaugher}, B. and {Frieman}, J. and {Garc{\'\i}a-Bellido}, J. and {Gaztanaga}, E. and {Gerdes}, D.~W. and {Giannantonio}, T. and {Gschwend}, J. and {Gutierrez}, G. and {Hinton}, S.~R. and {Hollowood}, D.~L. and {Honscheid}, K. and {Hoyle}, B. and {James}, D.~J. and {Kron}, R. and {Kuehn}, K. and {Lahav}, O. and {Lima}, M. and {Lin}, H. and {Maia}, M.~A.~G. and {Marshall}, J.~L. and {Martini}, P. and {Melchior}, P. and {Menanteau}, F. and {Miquel}, R. and {Mohr}, J.~J. and {Morgan}, R. and {Ogando}, R.~L.~C. and {Palmese}, A. and {Paz-Chinch{\'o}n}, F. and {Petravick}, D. and {Pieres}, A. and {Romer}, A.~K. and {Sanchez}, E. and {Scarpine}, V. and {Schubnell}, M. and {Serrano}, S. and {Smith}, M. and {Soares-Santos}, M. and {Tarle}, G. and {Thomas}, D. and {To}, C. and {Weller}, J. and {DES Collaboration}},
	eid = {023514},
	eprint = {2105.13543},
	journal = {\prd},
	month = {January},
	number = {2},
	pages = {023514},
	primaryclass = {astro-ph.CO},
	title = {{Dark Energy Survey Year 3 results: Cosmology from cosmic shear and robustness to data calibration}},
	volume = {105},
	year = {2022}}

@article{2022PhRvD.105h3518F,
	adsurl = {https://ui.adsabs.harvard.edu/abs/2022PhRvD.105h3518F},
	archiveprefix = {arXiv},
	author = {{Fluri}, Janis and {Kacprzak}, Tomasz and {Lucchi}, Aurelien and {Schneider}, Aurel and {Refregier}, Alexandre and {Hofmann}, Thomas},
	eid = {083518},
	eprint = {2201.07771},
	journal = {\prd},
	month = {April},
	number = {8},
	pages = {083518},
	primaryclass = {astro-ph.CO},
	title = {{Full w CDM analysis of KiDS-1000 weak lensing maps using deep learning}},
	volume = {105},
	year = {2022}}

@article{2023JCAP...02..050K,
	adsurl = {https://ui.adsabs.harvard.edu/abs/2023JCAP...02..050K},
	archiveprefix = {arXiv},
	author = {{Kacprzak}, Tomasz and {Fluri}, Janis and {Schneider}, Aurel and {Refregier}, Alexandre and {Stadel}, Joachim},
	eid = {050},
	eprint = {2209.04662},
	journal = {\jcap},
	month = {February},
	number = {2},
	pages = {050},
	primaryclass = {astro-ph.CO},
	title = {{CosmoGridV1: a simulated ��CDM theory prediction for map-level cosmological inference}},
	volume = {2023},
	year = {2023}}

@article{2023MNRAS.520.2328Z,
	adsurl = {https://ui.adsabs.harvard.edu/abs/2023MNRAS.520.2328Z},
	author = {{Zhang}, Tianqing and {Almoubayyed}, Husni and {Mandelbaum}, Rachel and {Meyers}, Joshua E. and {Jarvis}, Mike and {Kannawadi}, Arun and {Schmitz}, Morgan A. and {Guinot}, Axel and {LSST Dark Energy Science Collaboration}},
	journal = {\mnras},
	month = {April},
	number = {2},
	pages = {2328-2350},
	title = {{Impact of point spread function higher moments error on weak gravitational lensing - II. A comprehensive study}},
	volume = {520},
	year = {2023}}

@article{2023OJAp....6E...5M,
	adsurl = {https://ui.adsabs.harvard.edu/abs/2023OJAp....6E...5M},
	archiveprefix = {arXiv},
	author = {{Mandelbaum}, Rachel and {Jarvis}, Mike and {Lupton}, Robert H. and {Bosch}, James and {Kannawadi}, Arun and {Murphy}, Michael D. and {Zhang}, Tianqing and {LSST Dark Energy Science Collaboration}},
	eid = {5},
	eprint = {2209.09253},
	journal = {The Open Journal of Astrophysics},
	month = {February},
	pages = {5},
	primaryclass = {astro-ph.IM},
	title = {{PSFs of coadded images}},
	volume = {6},
	year = {2023}}

@article{desdm,
	adsurl = {https://ui.adsabs.harvard.edu/abs/2018PASP..130g4501M},
	archiveprefix = {arXiv},
	author = {{Morganson}, E. and {Gruendl}, R.~A. and {Menanteau}, F. and {Carrasco Kind}, M. and {Chen}, Y. -C. and {Daues}, G. and {Drlica-Wagner}, A. and {Friedel}, D.~N. and {Gower}, M. and {Johnson}, M.~W.~G. and {Johnson}, M.~D. and {Kessler}, R. and {Paz-Chinch{\'o}n}, F. and {Petravick}, D. and {Pond}, C. and {Yanny}, B. and {Allam}, S. and {Armstrong}, R. and {Barkhouse}, W. and {Bechtol}, K. and {Benoit-L{\'e}vy}, A. and {Bernstein}, G.~M. and {Bertin}, E. and {Buckley-Geer}, E. and {Covarrubias}, R. and {Desai}, S. and {Diehl}, H.~T. and {Goldstein}, D.~A. and {Gruen}, D. and {Li}, T.~S. and {Lin}, H. and {Marriner}, J. and {Mohr}, J.~J. and {Neilsen}, E. and {Ngeow}, C. -C. and {Paech}, K. and {Rykoff}, E.~S. and {Sako}, M. and {Sevilla-Noarbe}, I. and {Sheldon}, E. and {Sobreira}, F. and {Tucker}, D.~L. and {Wester}, W. and {DES Collaboration}},
	doi = {10.1088/1538-3873/aab4ef},
	eprint = {1801.03177},
	journal = {\pasp},
	month = {July},
	number = {989},
	pages = {074501},
	primaryclass = {astro-ph.IM},
	title = {{The Dark Energy Survey Image Processing Pipeline}},
	volume = {130},
	year = {2018}}

@article{fgcm,
	adsurl = {https://ui.adsabs.harvard.edu/abs/2018AJ....155...41B},
	archiveprefix = {arXiv},
	author = {{Burke}, D.~L. and {Rykoff}, E.~S. and {Allam}, S. and {Annis}, J. and {Bechtol}, K. and {Bernstein}, G.~M. and {Drlica-Wagner}, A. and {Finley}, D.~A. and {Gruendl}, R.~A. and {James}, D.~J. and {Kent}, S. and {Kessler}, R. and {Kuhlmann}, S. and {Lasker}, J. and {Li}, T.~S. and {Scolnic}, D. and {Smith}, J. and {Tucker}, D.~L. and {Wester}, W. and {Yanny}, B. and {Abbott}, T.~M.~C. and {Abdalla}, F.~B. and {Benoit-L{\'e}vy}, A. and {Bertin}, E. and {Carnero Rosell}, A. and {Carrasco Kind}, M. and {Carretero}, J. and {Cunha}, C.~E. and {D'Andrea}, C.~B. and {da Costa}, L.~N. and {Desai}, S. and {Diehl}, H.~T. and {Doel}, P. and {Estrada}, J. and {Garc{\'\i}a-Bellido}, J. and {Gruen}, D. and {Gutierrez}, G. and {Honscheid}, K. and {Kuehn}, K. and {Kuropatkin}, N. and {Maia}, M.~A.~G. and {March}, M. and {Marshall}, J.~L. and {Melchior}, P. and {Menanteau}, F. and {Miquel}, R. and {Plazas}, A.~A. and {Sako}, M. and {Sanchez}, E. and {Scarpine}, V. and {Schindler}, R. and {Sevilla-Noarbe}, I. and {Smith}, M. and {Smith}, R.~C. and {Soares-Santos}, M. and {Sobreira}, F. and {Suchyta}, E. and {Tarle}, G. and {Walker}, A.~R. and {DES Collaboration}},
	doi = {10.3847/1538-3881/aa9f22},
	eid = {41},
	eprint = {1706.01542},
	journal = {\aj},
	month = {January},
	number = {1},
	pages = {41},
	primaryclass = {astro-ph.IM},
	title = {{Forward Global Photometric Calibration of the Dark Energy Survey}},
	volume = {155},
	year = {2018}}

@article{Fosalba2015,
	adsurl = {https://ui.adsabs.harvard.edu/abs/2015MNRAS.447.1319F},
	archiveprefix = {arXiv},
	author = {{Fosalba}, P. and {Gazta{\~n}aga}, E. and {Castander}, F.~J. and {Crocce}, M.},
	eprint = {1312.2947},
	journal = {\mnras},
	month = {February},
	number = {2},
	pages = {1319-1332},
	primaryclass = {astro-ph.CO},
	title = {{The MICE Grand Challenge light-cone simulation - III. Galaxy lensing mocks from all-sky lensing maps}},
	volume = {447},
	year = {2015}}

@article{KaiserSquires,
	adsurl = {https://ui.adsabs.harvard.edu/abs/1993ApJ...404..441K},
	author = {{Kaiser}, Nick and {Squires}, Gordon},
	journal = {\apj},
	month = {Feb},
	pages = {441},
	title = {{Mapping the Dark Matter with Weak Gravitational Lensing}},
	volume = {404},
	year = {1993}}

@article{sheldon_mdet_rubin,
	adsurl = {https://ui.adsabs.harvard.edu/abs/2023OJAp....6E..17S},
	archiveprefix = {arXiv},
	author = {{Sheldon}, Erin S. and {Becker}, Matthew R. and {Jarvis}, Michael and {Armstrong}, Robert and {LSST Dark Energy Science Collaboration}},
	doi = {10.21105/astro.2303.03947},
	eid = {17},
	eprint = {2303.03947},
	journal = {The Open Journal of Astrophysics},
	month = {May},
	pages = {17},
	primaryclass = {astro-ph.IM},
	title = {{Metadetection Weak Lensing for the Vera C. Rubin Observatory}},
	volume = {6},
	year = {2023}}

@article{treerings,
	adsurl = {https://ui.adsabs.harvard.edu/abs/2014JInst...9C4001P},
	archiveprefix = {arXiv},
	author = {{Plazas}, A.~A. and {Bernstein}, G.~M. and {Sheldon}, E.~S.},
	doi = {10.1088/1748-0221/9/04/C04001},
	eid = {C04001},
	eprint = {1403.6127},
	journal = {Journal of Instrumentation},
	month = {April},
	number = {4},
	pages = {C04001},
	primaryclass = {astro-ph.IM},
	title = {{Transverse electric fields' effects in the Dark Energy Camera CCDs}},
	volume = {9},
	year = {2014}}

@article{y3-massmapping,
	adsurl = {https://ui.adsabs.harvard.edu/abs/2021MNRAS.505.4626J},
	author = {{Jeffrey}, N. and {Gatti}, M. and {Chang}, C. and {Whiteway}, L. and {Demirbozan}, U. and {Kovacs}, A. and {Pollina}, G. and {Bacon}, D. and {Hamaus}, N. and {Kacprzak}, T. and {Lahav}, O. and {Lanusse}, F. and {Mawdsley}, B. and {Nadathur}, S. and {Starck}, J.~L. and {Vielzeuf}, P. and {Zeurcher}, D. and {Alarcon}, A. and {Amon}, A. and {Bechtol}, K. and {Bernstein}, G.~M. and {Campos}, A. and {Rosell}, A. Carnero and {Kind}, M. Carrasco and {Cawthon}, R. and {Chen}, R. and {Choi}, A. and {Cordero}, J. and {Davis}, C. and {DeRose}, J. and {Doux}, C. and {Drlica-Wagner}, A. and {Eckert}, K. and {Elsner}, F. and {Elvin-Poole}, J. and {Everett}, S. and {Fert{\'e}}, A. and {Giannini}, G. and {Gruen}, D. and {Gruendl}, R.~A. and {Harrison}, I. and {Hartley}, W.~G. and {Herner}, K. and {Huff}, E.~M. and {Huterer}, D. and {Kuropatkin}, N. and {Jarvis}, M. and {Leget}, P.~F. and {MacCrann}, N. and {McCullough}, J. and {Muir}, J. and {Myles}, J. and {Navarro-Alsina}, A. and {Pandey}, S. and {Prat}, J. and {Raveri}, M. and {Rollins}, R.~P. and {Ross}, A.~J. and {Rykoff}, E.~S. and {S{\'a}nchez}, C. and {Secco}, L.~F. and {Sevilla-Noarbe}, I. and {Sheldon}, E. and {Shin}, T. and {Troxel}, M.~A. and {Tutusaus}, I. and {Varga}, T.~N. and {Yanny}, B. and {Yin}, B. and {Zhang}, Y. and {Zuntz}, J. and {Abbott}, T.~M.~C. and {Aguena}, M. and {Allam}, S. and {Andrade-Oliveira}, F. and {Becker}, M.~R. and {Bertin}, E. and {Bhargava}, S. and {Brooks}, D. and {Burke}, D.~L. and {Carretero}, J. and {Castander}, F.~J. and {Conselice}, C. and {Costanzi}, M. and {Crocce}, M. and {da Costa}, L.~N. and {Pereira}, M.~E.~S. and {De Vicente}, J. and {Desai}, S. and {Diehl}, H.~T. and {Dietrich}, J.~P. and {Doel}, P. and {Ferrero}, I. and {Flaugher}, B. and {Fosalba}, P. and {Garc{\'\i}a-Bellido}, J. and {Gaztanaga}, E. and {Gerdes}, D.~W. and {Giannantonio}, T. and {Gschwend}, J. and {Gutierrez}, G. and {Hinton}, S.~R. and {Hollowood}, D.~L. and {Hoyle}, B. and {Jain}, B. and {James}, D.~J. and {Lima}, M. and {Maia}, M.~A.~G. and {March}, M. and {Marshall}, J.~L. and {Melchior}, P. and {Menanteau}, F. and {Miquel}, R. and {Mohr}, J.~J. and {Morgan}, R. and {Ogando}, R.~L.~C. and {Palmese}, A. and {Paz-Chinch{\'o}n}, F. and {Plazas}, A.~A. and {Rodriguez-Monroy}, M. and {Roodman}, A. and {Sanchez}, E. and {Scarpine}, V. and {Serrano}, S. and {Smith}, M. and {Soares-Santos}, M. and {Suchyta}, E. and {Tarle}, G. and {Thomas}, D. and {To}, C. and {Weller}, J. and {DES Collaboration}},
	journal = {\mnras},
	month = {August},
	number = {3},
	pages = {4626-4645},
	primaryclass = {astro-ph.CO},
	title = {{Dark Energy Survey Year 3 results: Curved-sky weak lensing mass map reconstruction}},
	volume = {505},
	year = {2021}}

@inproceedings{sextractor,
	adsurl = {https://ui.adsabs.harvard.edu/abs/2011ASPC..442..435B},
	author = {{Bertin}, E.},
	booktitle = {Astronomical Data Analysis Software and Systems XX},
	editor = {{Evans}, I.~N. and {Accomazzi}, A. and {Mink}, D.~J. and {Rots}, A.~H.},
	month = jul,
	pages = {435},
	series = {Astronomical Society of the Pacific Conference Series},
	title = {{Automated Morphometry with SExtractor and PSFEx}},
	volume = {442},
	year = 2011}

@article{2022PASJ...74..421L,
	adsurl = {https://ui.adsabs.harvard.edu/abs/2022PASJ...74..421L},
	archiveprefix = {arXiv},
	author = {{Li}, Xiangchong and {Miyatake}, Hironao and {Luo}, Wentao and {More}, Surhud and {Oguri}, Masamune and {Hamana}, Takashi and {Mandelbaum}, Rachel and {Shirasaki}, Masato and {Takada}, Masahiro and {Armstrong}, Robert and {Kannawadi}, Arun and {Takita}, Satoshi and {Miyazaki}, Satoshi and {Nishizawa}, Atsushi J. and {Plazas Malagon}, Andres A. and {Strauss}, Michael A. and {Tanaka}, Masayuki and {Yoshida}, Naoki},
	doi = {10.1093/pasj/psac006},
	eprint = {2107.00136},
	journal = {\pasj},
	month = apr,
	number = {2},
	pages = {421-459},
	primaryclass = {astro-ph.CO},
	title = {{The three-year shear catalog of the Subaru Hyper Suprime-Cam SSP Survey}},
	volume = {74},
	year = 2022}

@article{2015ApJ...807..182M,
	adsurl = {https://ui.adsabs.harvard.edu/abs/2015ApJ...807..182M},
	archiveprefix = {arXiv},
	author = {{Meyers}, Joshua E. and {Burchat}, Patricia R.},
	doi = {10.1088/0004-637X/807/2/182},
	eid = {182},
	eprint = {1409.6273},
	journal = {\apj},
	month = jul,
	number = {2},
	pages = {182},
	primaryclass = {astro-ph.CO},
	title = {{Impact of Atmospheric Chromatic Effects on Weak Lensing Measurements}},
	volume = {807},
	year = 2015}

@article{bernstein_bfd1,
	adsurl = {https://ui.adsabs.harvard.edu/abs/2014MNRAS.438.1880B},
	archiveprefix = {arXiv},
	author = {{Bernstein}, Gary M. and {Armstrong}, Robert},
	doi = {10.1093/mnras/stt2326},
	eprint = {1304.1843},
	journal = {\mnras},
	month = feb,
	number = {2},
	pages = {1880-1893},
	primaryclass = {astro-ph.CO},
	title = {{Bayesian lensing shear measurement}},
	volume = {438},
	year = 2014}

@article{bernstein_bfd2,
	adsurl = {https://ui.adsabs.harvard.edu/abs/2016MNRAS.459.4467B},
	archiveprefix = {arXiv},
	author = {{Bernstein}, Gary M. and {Armstrong}, Robert and {Krawiec}, Christina and {March}, Marisa C.},
	doi = {10.1093/mnras/stw879},
	eprint = {1508.05655},
	journal = {\mnras},
	month = jul,
	number = {4},
	pages = {4467-4484},
	primaryclass = {astro-ph.IM},
	title = {{An accurate and practical method for inference of weak gravitational lensing from galaxy images}},
	volume = {459},
	year = 2016}

@article{2003MNRAS.343..459H,
	adsurl = {https://ui.adsabs.harvard.edu/abs/2003MNRAS.343..459H},
	archiveprefix = {arXiv},
	author = {{Hirata}, Christopher and {Seljak}, Uro{\v{s}}},
	doi = {10.1046/j.1365-8711.2003.06683.x},
	eprint = {astro-ph/0301054},
	journal = {\mnras},
	month = aug,
	number = {2},
	pages = {459-480},
	primaryclass = {astro-ph},
	title = {{Shear calibration biases in weak-lensing surveys}},
	volume = {343},
	year = 2003}

@article{2012MNRAS.427.2711K,
	adsurl = {https://ui.adsabs.harvard.edu/abs/2012MNRAS.427.2711K},
	archiveprefix = {arXiv},
	author = {{Kacprzak}, Tomasz and {Zuntz}, Joe and {Rowe}, Barnaby and {Bridle}, Sarah and {Refregier}, Alexandre and {Amara}, Adam and {Voigt}, Lisa and {Hirsch}, Michael},
	doi = {10.1111/j.1365-2966.2012.21622.x},
	eprint = {1203.5049},
	journal = {\mnras},
	month = dec,
	number = {4},
	pages = {2711-2722},
	primaryclass = {astro-ph.CO},
	title = {{Measurement and calibration of noise bias in weak lensing galaxy shape estimation}},
	volume = {427},
	year = 2012}

@article{2012MNRAS.424.2757M,
	adsurl = {https://ui.adsabs.harvard.edu/abs/2012MNRAS.424.2757M},
	archiveprefix = {arXiv},
	author = {{Melchior}, P. and {Viola}, M.},
	doi = {10.1111/j.1365-2966.2012.21381.x},
	eprint = {1204.5147},
	journal = {\mnras},
	month = aug,
	number = {4},
	pages = {2757-2769},
	primaryclass = {astro-ph.IM},
	title = {{Means of confusion: how pixel noise affects shear estimates for weak gravitational lensing}},
	volume = {424},
	year = 2012}

@article{2012MNRAS.425.1951R,
	adsurl = {https://ui.adsabs.harvard.edu/abs/2012MNRAS.425.1951R},
	archiveprefix = {arXiv},
	author = {{Refregier}, Alexandre and {Kacprzak}, Tomasz and {Amara}, Adam and {Bridle}, Sarah and {Rowe}, Barnaby},
	doi = {10.1111/j.1365-2966.2012.21483.x},
	eprint = {1203.5050},
	journal = {\mnras},
	month = sep,
	number = {3},
	pages = {1951-1957},
	primaryclass = {astro-ph.CO},
	title = {{Noise bias in weak lensing shape measurements}},
	volume = {425},
	year = 2012}

@article{2014MNRAS.441.2528K,
	adsurl = {https://ui.adsabs.harvard.edu/abs/2014MNRAS.441.2528K},
	archiveprefix = {arXiv},
	author = {{Kacprzak}, Tomasz and {Bridle}, Sarah and {Rowe}, Barnaby and {Voigt}, Lisa and {Zuntz}, Joe and {Hirsch}, Michael and {MacCrann}, Niall},
	doi = {10.1093/mnras/stu588},
	eprint = {1308.4663},
	journal = {\mnras},
	month = jul,
	number = {3},
	pages = {2528-2538},
	primaryclass = {astro-ph.CO},
	title = {{S{\'e}rsic galaxy models in weak lensing shape measurement: model bias, noise bias and their interaction}},
	volume = {441},
	year = 2014}

@article{2010A&A...510A..75M,
	adsurl = {https://ui.adsabs.harvard.edu/abs/2010A&A...510A..75M},
	archiveprefix = {arXiv},
	author = {{Melchior}, P. and {B{\"o}hnert}, A. and {Lombardi}, M. and {Bartelmann}, M.},
	doi = {10.1051/0004-6361/200912785},
	eid = {A75},
	eprint = {0906.5092},
	journal = {\aap},
	month = feb,
	pages = {A75},
	primaryclass = {astro-ph.IM},
	title = {{Limitations on shapelet-based weak-lensing measurements}},
	volume = {510},
	year = 2010}

@article{2010MNRAS.404..458V,
	adsurl = {https://ui.adsabs.harvard.edu/abs/2010MNRAS.404..458V},
	archiveprefix = {arXiv},
	author = {{Voigt}, L.~M. and {Bridle}, S.~L.},
	doi = {10.1111/j.1365-2966.2010.16300.x},
	eprint = {0905.4801},
	journal = {\mnras},
	month = may,
	number = {1},
	pages = {458-467},
	primaryclass = {astro-ph.CO},
	title = {{Limitations of model-fitting methods for lensing shear estimation}},
	volume = {404},
	year = 2010}

@article{bernstein_jarvis_2002,
	adsurl = {https://ui.adsabs.harvard.edu/abs/2002AJ....123..583B},
	archiveprefix = {arXiv},
	author = {{Bernstein}, G.~M. and {Jarvis}, M.},
	doi = {10.1086/338085},
	eprint = {astro-ph/0107431},
	journal = {\aj},
	month = feb,
	number = {2},
	pages = {583-618},
	primaryclass = {astro-ph},
	title = {{Shapes and Shears, Stars and Smears: Optimal Measurements for Weak Lensing}},
	volume = {123},
	year = 2002}

@article{2019A&A...624A..92K,
	adsurl = {https://ui.adsabs.harvard.edu/abs/2019A&A...624A..92K},
	archiveprefix = {arXiv},
	author = {{Kannawadi}, Arun and {Hoekstra}, Henk and {Miller}, Lance and {Viola}, Massimo and {Fenech Conti}, Ian and {Herbonnet}, Ricardo and {Erben}, Thomas and {Heymans}, Catherine and {Hildebrandt}, Hendrik and {Kuijken}, Konrad and {Vakili}, Mohammadjavad and {Wright}, Angus H.},
	doi = {10.1051/0004-6361/201834819},
	eid = {A92},
	eprint = {1812.03983},
	journal = {\aap},
	month = apr,
	pages = {A92},
	primaryclass = {astro-ph.CO},
	title = {{Towards emulating cosmic shear data: revisiting the calibration of the shear measurements for the Kilo-Degree Survey}},
	volume = {624},
	year = 2019}

@article{shearoncoadds,
	adsurl = {https://ui.adsabs.harvard.edu/abs/2024arXiv240701771A},
	archiveprefix = {arXiv},
	author = {{Armstrong}, Robert and {Sheldon}, Erin and {Huff}, Eric and {Bosch}, Jim and {Rykoff}, Eli and {Mandelbaum}, Rachel and {Kannawadi}, Arun and {Melchior}, Peter and {Lupton}, Robert and {Becker}, Matthew R. and {Al-Sayyed}, Yusra and {The LSST Dark Energy Science Collaboration}},
	doi = {10.48550/arXiv.2407.01771},
	eid = {arXiv:2407.01771},
	eprint = {2407.01771},
	journal = {arXiv e-prints},
	month = jul,
	pages = {arXiv:2407.01771},
	primaryclass = {astro-ph.CO},
	title = {{The little coadd that could: Estimating shear from coadded images}},
	year = 2024}

@article{anacal_1,
	adsurl = {https://ui.adsabs.harvard.edu/abs/2023MNRAS.521.4904L},
	archiveprefix = {arXiv},
	author = {{Li}, Xiangchong and {Mandelbaum}, Rachel},
	doi = {10.1093/mnras/stad890},
	eprint = {2208.10522},
	journal = {\mnras},
	month = jun,
	number = {4},
	pages = {4904-4926},
	primaryclass = {astro-ph.CO},
	title = {{Analytical weak-lensing shear responses of galaxy properties and galaxy detection}},
	volume = {521},
	year = 2023}

@ARTICLE{anacal_2,
       author = {{Li}, Xiangchong and {Mandelbaum}, Rachel},
        title = "{Analytical noise bias correction for precise weak lensing shear inference}",
      journal = {\mnras},
         year = 2024,
        month = dec,
          doi = {10.1093/mnras/stae2764},
archivePrefix = {arXiv},
       eprint = {2408.06337},
 primaryClass = {astro-ph.CO},
       adsurl = {https://ui.adsabs.harvard.edu/abs/2024MNRAS.tmp.2638L}
}

@article{y3gold,
	adsurl = {https://ui.adsabs.harvard.edu/abs/2021ApJS..254...24S},
	archiveprefix = {arXiv},
	author = {{Sevilla-Noarbe}, I. and {Bechtol}, K. and {Carrasco Kind}, M. and {Carnero Rosell}, A. and {Becker}, M.~R. and {Drlica-Wagner}, A. and {Gruendl}, R.~A. and {Rykoff}, E.~S. and {Sheldon}, E. and {Yanny}, B. and {Alarcon}, A. and {Allam}, S. and {Amon}, A. and {Benoit-L{\'e}vy}, A. and {Bernstein}, G.~M. and {Bertin}, E. and {Burke}, D.~L. and {Carretero}, J. and {Choi}, A. and {Diehl}, H.~T. and {Everett}, S. and {Flaugher}, B. and {Gaztanaga}, E. and {Gschwend}, J. and {Harrison}, I. and {Hartley}, W.~G. and {Hoyle}, B. and {Jarvis}, M. and {Johnson}, M.~D. and {Kessler}, R. and {Kron}, R. and {Kuropatkin}, N. and {Leistedt}, B. and {Li}, T.~S. and {Menanteau}, F. and {Morganson}, E. and {Ogando}, R.~L.~C. and {Palmese}, A. and {Paz-Chinch{\'o}n}, F. and {Pieres}, A. and {Pond}, C. and {Rodriguez-Monroy}, M. and {Smith}, J. Allyn and {Stringer}, K.~M. and {Troxel}, M.~A. and {Tucker}, D.~L. and {de Vicente}, J. and {Wester}, W. and {Zhang}, Y. and {Abbott}, T.~M.~C. and {Aguena}, M. and {Annis}, J. and {Avila}, S. and {Bhargava}, S. and {Bridle}, S.~L. and {Brooks}, D. and {Brout}, D. and {Castander}, F.~J. and {Cawthon}, R. and {Chang}, C. and {Conselice}, C. and {Costanzi}, M. and {Crocce}, M. and {da Costa}, L.~N. and {Pereira}, M.~E.~S. and {Davis}, T.~M. and {Desai}, S. and {Dietrich}, J.~P. and {Doel}, P. and {Eckert}, K. and {Evrard}, A.~E. and {Ferrero}, I. and {Fosalba}, P. and {Garc{\'\i}a-Bellido}, J. and {Gerdes}, D.~W. and {Giannantonio}, T. and {Gruen}, D. and {Gutierrez}, G. and {Hinton}, S.~R. and {Hollowood}, D.~L. and {Honscheid}, K. and {Huff}, E.~M. and {Huterer}, D. and {James}, D.~J. and {Jeltema}, T. and {Kuehn}, K. and {Lahav}, O. and {Lidman}, C. and {Lima}, M. and {Lin}, H. and {Maia}, M.~A.~G. and {Marshall}, J.~L. and {Martini}, P. and {Melchior}, P. and {Miquel}, R. and {Mohr}, J.~J. and {Morgan}, R. and {Neilsen}, E. and {Plazas}, A.~A. and {Romer}, A.~K. and {Roodman}, A. and {Sanchez}, E. and {Scarpine}, V. and {Schubnell}, M. and {Serrano}, S. and {Smith}, M. and {Suchyta}, E. and {Tarle}, G. and {Thomas}, D. and {To}, C. and {Varga}, T.~N. and {Wechsler}, R.~H. and {Weller}, J. and {Wilkinson}, R.~D. and {DES Collaboration}},
	doi = {10.3847/1538-4365/abeb66},
	eid = {24},
	eprint = {2011.03407},
	journal = {\apjs},
	month = jun,
	number = {2},
	pages = {24},
	primaryclass = {astro-ph.CO},
	title = {{Dark Energy Survey Year 3 Results: Photometric Data Set for Cosmology}},
	volume = {254},
	year = 2021}

@article{asinh_mag,
	adsurl = {https://ui.adsabs.harvard.edu/abs/1999AJ....118.1406L},
	archiveprefix = {arXiv},
	author = {{Lupton}, Robert H. and {Gunn}, James E. and {Szalay}, Alexander S.},
	doi = {10.1086/301004},
	eprint = {astro-ph/9903081},
	journal = {\aj},
	month = sep,
	number = {3},
	pages = {1406-1410},
	primaryclass = {astro-ph},
	title = {{A Modified Magnitude System that Produces Well-Behaved Magnitudes, Colors, and Errors Even for Low Signal-to-Noise Ratio Measurements}},
	volume = {118},
	year = 1999}

@article{cosmos20,
	adsurl = {https://ui.adsabs.harvard.edu/abs/2022ApJS..258...11W},
	archiveprefix = {arXiv},
	author = {{Weaver}, J.~R. and {Kauffmann}, O.~B. and {Ilbert}, O. and {McCracken}, H.~J. and {Moneti}, A. and {Toft}, S. and {Brammer}, G. and {Shuntov}, M. and {Davidzon}, I. and {Hsieh}, B.~C. and {Laigle}, C. and {Anastasiou}, A. and {Jespersen}, C.~K. and {Vinther}, J. and {Capak}, P. and {Casey}, C.~M. and {McPartland}, C.~J.~R. and {Milvang-Jensen}, B. and {Mobasher}, B. and {Sanders}, D.~B. and {Zalesky}, L. and {Arnouts}, S. and {Aussel}, H. and {Dunlop}, J.~S. and {Faisst}, A. and {Franx}, M. and {Furtak}, L.~J. and {Fynbo}, J.~P.~U. and {Gould}, K.~M.~L. and {Greve}, T.~R. and {Gwyn}, S. and {Kartaltepe}, J.~S. and {Kashino}, D. and {Koekemoer}, A.~M. and {Kokorev}, V. and {Le F{\`e}vre}, O. and {Lilly}, S. and {Masters}, D. and {Magdis}, G. and {Mehta}, V. and {Peng}, Y. and {Riechers}, D.~A. and {Salvato}, M. and {Sawicki}, M. and {Scarlata}, C. and {Scoville}, N. and {Shirley}, R. and {Silverman}, J.~D. and {Sneppen}, A. and {Smolc̆i{\'c}}, V. and {Steinhardt}, C. and {Stern}, D. and {Tanaka}, M. and {Taniguchi}, Y. and {Teplitz}, H.~I. and {Vaccari}, M. and {Wang}, W. -H. and {Zamorani}, G.},
	doi = {10.3847/1538-4365/ac3078},
	eid = {11},
	eprint = {2110.13923},
	journal = {\apjs},
	month = jan,
	number = {1},
	pages = {11},
	primaryclass = {astro-ph.GA},
	title = {{COSMOS2020: A Panchromatic View of the Universe to z{\ensuremath{\sim}}10 from Two Complementary Catalogs}},
	volume = {258},
	year = 2022}

@article{2019JOSS....4.1298Z,
	adsurl = {https://ui.adsabs.harvard.edu/abs/2019JOSS....4.1298Z},
	author = {{Zonca}, Andrea and {Singer}, Leo and {Lenz}, Daniel and {Reinecke}, Martin and {Rosset}, Cyrille and {Hivon}, Eric and {Gorski}, Krzysztof},
	doi = {10.21105/joss.01298},
	eid = {1298},
	journal = {The Journal of Open Source Software},
	month = mar,
	number = {35},
	pages = {1298},
	title = {{healpy: equal area pixelization and spherical harmonics transforms for data on the sphere in Python}},
	volume = {4},
	year = 2019}

@article{nelder-mead,
	author = {Nelder, John A. and Mead, R.},
	ee = {https://www.wikidata.org/entity/Q55954356},
	journal = {Comput. J.},
	number = 4,
	pages = {308-313},
	title = {A Simplex Method for Function Minimization.},
	url = {http://dblp.uni-trier.de/db/journals/cj/cj7.html#NelderM65},
	volume = 7,
	year = 1965}

@article{emcee,
	adsurl = {https://ui.adsabs.harvard.edu/abs/2013PASP..125..306F},
	archiveprefix = {arXiv},
	author = {{Foreman-Mackey}, Daniel and {Hogg}, David W. and {Lang}, Dustin and {Goodman}, Jonathan},
	doi = {10.1086/670067},
	eprint = {1202.3665},
	journal = {\pasp},
	month = mar,
	number = {925},
	pages = {306},
	primaryclass = {astro-ph.IM},
	title = {{emcee: The MCMC Hammer}},
	volume = {125},
	year = 2013}

@article{y3balrog,
	adsurl = {https://ui.adsabs.harvard.edu/abs/2022ApJS..258...15E},
	archiveprefix = {arXiv},
	author = {{Everett}, S. and {Yanny}, B. and {Kuropatkin}, N. and {Huff}, E.~M. and {Zhang}, Y. and {Myles}, J. and {Masegian}, A. and {Elvin-Poole}, J. and {Allam}, S. and {Bernstein}, G.~M. and {Sevilla-Noarbe}, I. and {Splettstoesser}, M. and {Sheldon}, E. and {Jarvis}, M. and {Amon}, A. and {Harrison}, I. and {Choi}, A. and {Hartley}, W.~G. and {Alarcon}, A. and {S{\'a}nchez}, C. and {Gruen}, D. and {Eckert}, K. and {Prat}, J. and {Tabbutt}, M. and {Busti}, V. and {Becker}, M.~R. and {MacCrann}, N. and {Diehl}, H.~T. and {Tucker}, D.~L. and {Bertin}, E. and {Jeltema}, T. and {Drlica-Wagner}, A. and {Gruendl}, R.~A. and {Bechtol}, K. and {Carnero Rosell}, A. and {Abbott}, T.~M.~C. and {Aguena}, M. and {Annis}, J. and {Bacon}, D. and {Bhargava}, S. and {Brooks}, D. and {Burke}, D.~L. and {Carrasco Kind}, M. and {Carretero}, J. and {Castander}, F.~J. and {Conselice}, C. and {Costanzi}, M. and {da Costa}, L.~N. and {Pereira}, M.~E.~S. and {De Vicente}, J. and {DeRose}, J. and {Desai}, S. and {Eifler}, T.~F. and {Evrard}, A.~E. and {Ferrero}, I. and {Fosalba}, P. and {Frieman}, J. and {Garc{\'\i}a-Bellido}, J. and {Gaztanaga}, E. and {Gerdes}, D.~W. and {Gutierrez}, G. and {Hinton}, S.~R. and {Hollowood}, D.~L. and {Honscheid}, K. and {Huterer}, D. and {James}, D.~J. and {Kent}, S. and {Krause}, E. and {Kuehn}, K. and {Lahav}, O. and {Lima}, M. and {Lin}, H. and {Maia}, M.~A.~G. and {Marshall}, J.~L. and {Melchior}, P. and {Menanteau}, F. and {Miquel}, R. and {Mohr}, J.~J. and {Morgan}, R. and {Muir}, J. and {Ogando}, R.~L.~C. and {Palmese}, A. and {Paz-Chinch{\'o}n}, F. and {Plazas}, A.~A. and {Rodriguez-Monroy}, M. and {Romer}, A.~K. and {Roodman}, A. and {Sanchez}, E. and {Scarpine}, V. and {Serrano}, S. and {Smith}, M. and {Soares-Santos}, M. and {Suchyta}, E. and {Swanson}, M.~E.~C. and {Tarle}, G. and {To}, C. and {Troxel}, M.~A. and {Varga}, T.~N. and {Weller}, J. and {Wilkinson}, R.~D. and {Wilkinson}, R.~D.},
	doi = {10.3847/1538-4365/ac26c1},
	eid = {15},
	eprint = {2012.12825},
	journal = {\apjs},
	month = jan,
	number = {1},
	pages = {15},
	primaryclass = {astro-ph.CO},
	title = {{Dark Energy Survey Year 3 Results: Measuring the Survey Transfer Function with Balrog}},
	volume = {258},
	year = 2022}

@article{sfd98,
	adsurl = {http://adsabs.harvard.edu/abs/1998ApJ...500..525S},
	author = {{Schlegel}, D.~J. and {Finkbeiner}, D.~P. and {Davis}, M.},
	doi = {10.1086/305772},
	eprint = {astro-ph/9710327},
	journal = {\apj},
	month = jun,
	pages = {525-553},
	title = {{Maps of Dust Infrared Emission for Use in Estimation of Reddening and Cosmic Microwave Background Radiation Foregrounds}},
	volume = 500,
	year = 1998}

@ARTICLE{y6gold,
       author = {{Bechtol}, K. and {Sevilla-Noarbe}, I. and {Drlica-Wagner}, A. and {Yanny}, B. and {Gruendl}, R.~A. and {Sheldon}, E. and {Rykoff}, E.~S. and {De Vicente}, J. and {Adamow}, M. and {Anbajagane}, D. and {Becker}, M.~R. and {Bernstein}, G.~M. and {Carnero Rosell}, A. and {Gschwend}, J. and {Gorsuch}, M. and {Hartley}, W.~G. and {Jarvis}, M. and {Jeltema}, T. and {Kron}, R. and {Manning}, T.~A. and {O'Donnell}, J. and {Pieres}, A. and {Rodr{\'\i}guez-Monroy}, M. and {Sanchez Cid}, D. and {Tabbutt}, M. and {Toribio San Cipriano}, L. and {Tucker}, D.~L. and {Weaverdyck}, N. and {Yamamoto}, M. and {Abbott}, T.~M.~C. and {Aguena}, M. and {Alarc{\'o}n}, A. and {Allam}, S. and {Amon}, A. and {Andrade-Oliveira}, F. and {Avila}, S. and {Bernardinelli}, P.~H. and {Bertin}, E. and {Blazek}, J. and {Brooks}, D. and {Burke}, D.~L. and {Carretero}, J. and {Castander}, F.~J. and {Cawthon}, R. and {Chang}, C. and {Choi}, A. and {Conselice}, C. and {Costanzi}, M. and {Crocce}, M. and {da Costa}, L.~N. and {Davis}, T.~M. and {Desai}, S. and {Diehl}, H.~T. and {Dodelson}, S. and {Doel}, P. and {Doux}, C. and {Fert{\'e}}, A. and {Flaugher}, B. and {Fosalba}, P. and {Frieman}, J. and {Garc{\'\i}a-Bellido}, J. and {Gatti}, M. and {Gaztanaga}, E. and {Giannini}, G. and {Gruen}, D. and {Gutierrez}, G. and {Herner}, K. and {Hinton}, S.~R. and {Hollowood}, D.~L. and {Honscheid}, K. and {Huterer}, D. and {Jeffrey}, N. and {Krause}, E. and {Kuehn}, K. and {Lahav}, O. and {Lee}, S. and {Lidman}, C. and {Lima}, M. and {Lin}, H. and {Marshall}, J.~L. and {Mena-Fern{\'a}ndez}, J. and {Miquel}, R. and {Mohr}, J.~J. and {Muir}, J. and {Myles}, J. and {Ogando}, R.~L.~C. and {Palmese}, A. and {Plazas Malag{\'o}n}, A.~A. and {Porredon}, A. and {Prat}, J. and {Raveri}, M. and {Romer}, A.~K. and {Roodman}, A. and {Samuroff}, S. and {Sanchez}, E. and {Scarpine}, V. and {Smith}, M. and {Soares-Santos}, M. and {Suchyta}, E. and {Tarle}, G. and {Troxel}, M.~A. and {Vikram}, V. and {Walker}, A.~R. and {Weller}, J. and {Wiseman}, P. and {Zhang}, Y.},
        title = "{Dark Energy Survey Year 6 Results: Photometric Data Set for Cosmology}",
      journal = {arXiv e-prints},
         year = 2025,
        month = jan,
          eid = {arXiv:2501.05739},
        pages = {arXiv:2501.05739},
          doi = {10.48550/arXiv.2501.05739},
archivePrefix = {arXiv},
       eprint = {2501.05739},
 primaryClass = {astro-ph.CO},
       adsurl = {https://ui.adsabs.harvard.edu/abs/2025arXiv250105739B}
}

@ARTICLE{y6psf,
       author = {{Schutt}, T. and {Jarvis}, M. and {Roodman}, A. and {Amon}, A. and {Becker}, M.~R. and {Gruendl}, R.~A. and {Yamamoto}, M. and {Bechtol}, K. and {Bernstein}, G.~M. and {Gatti}, M. and {Rykoff}, E.~S. and {Sheldon}, E. and {Troxel}, M.~A. and {Abbott}, T.~M.~C. and {Aguena}, M. and {Andrade-Oliveira}, F. and {Brooks}, D. and {Rosell}, A. Carnero and {Carretero}, J. and {Chang}, C. and {Choi}, A. and {Crocce}, M. and {da Costa}, L.~N. and {Davis}, T.~M. and {De Vicente}, J. and {Desai}, S. and {Diehl}, H.~T. and {Doel}, P. and {Fert{\'e}}, A. and {Frieman}, J. and {Garc{\'\i}a-Bellido}, J. and {Gaztanaga}, E. and {Gruen}, D. and {Gutierrez}, G. and {Hinton}, S.~R. and {Hollowood}, D.~L. and {Honscheid}, K. and {Kuehn}, K. and {Lahav}, O. and {Lee}, S. and {Lima}, M. and {Marshall}, J.~L. and {Mena-Fern{\'a}ndez}, J. and {Miquel}, R. and {Mohr}, J.~J. and {Myles}, J. and {Ogando}, R.~L.~C. and {Pieres}, A. and {Malag{\'o}n}, A.~A. Plazas and {Porredon}, A. and {Samuroff}, S. and {Sanchez}, E. and {Cid}, D. Sanchez and {Sevilla-Noarbe}, I. and {Smith}, M. and {Suchyta}, E. and {Tarle}, G. and {Vikram}, V. and {Walker}, A.~R. and {Weaverdyck}, N.},
        title = "{Dark Energy Survey Year 6 Results: Point-Spread Function Modeling}",
      journal = {The Open Journal of Astrophysics},
         year = 2025,
        month = mar,
       volume = {8},
          eid = {26},
        pages = {26},
          doi = {10.33232/001c.132299},
archivePrefix = {arXiv},
       eprint = {2501.05781},
 primaryClass = {astro-ph.CO},
       adsurl = {https://ui.adsabs.harvard.edu/abs/2025OJAp....8E..26S}
}

@ARTICLE{y6balrog,
       author = {{Anbajagane}, Dhayaa and {Tabbutt}, M. and {Beas-Gonzalez}, J. and {Yanny}, B. and {Everett}, S. and {Becker}, M.~R. and {Yamamoto}, M. and {Legnani}, E. and {De Vicente}, J. and {Bechtol}, K. and {Elvin-Poole}, J. and {Bernstein}, G.~M. and {Choi}, A. and {Gatti}, M. and {Giannini}, G. and {Gruendl}, R.~A. and {Jarvis}, M. and {Lee}, S. and {Mena-Fern{\'a}ndez}, J. and {Porredon}, A. and {Rodriguez-Monroy}, M. and {Rozo}, E. and {Rykoff}, E.~S. and {Schutt}, T. and {Sheldon}, E. and {Troxel}, M.~A. and {Weaverdyck}, N. and {Wetzell}, V. and {Aguena}, M. and {Alarcon}, A. and {Allam}, S. and {Amon}, A. and {Andrade-Oliveira}, F. and {Brooks}, D. and {Rosell}, A. Carnero and {Carretero}, J. and {Chang}, C. and {Crocce}, M. and {da Costa}, L.~N. and {Pereira}, M.~E.~S. and {Davis}, T.~M. and {Desai}, S. and {Diehl}, H.~T. and {Dodelson}, S. and {Doel}, P. and {Drlica-Wagner}, A. and {Fert{\'e}}, A. and {Frieman}, J. and {Garc{\'\i}a-Bellido}, J. and {Gaztanaga}, E. and {Gruen}, D. and {Gutierrez}, G. and {Hartley}, W.~G. and {Herner}, K. and {Hinton}, S.~R. and {Hollowood}, D.~L. and {Honscheid}, K. and {Huterer}, D. and {James}, D.~J. and {Krause}, E. and {Kuehn}, K. and {Lahav}, O. and {Marshall}, J.~L. and {Miquel}, R. and {Muir}, J. and {Myles}, J. and {Pieres}, A. and {Malag{\'o}n}, A.~A. Plazas and {Prat}, J. and {Raveri}, M. and {Samuroff}, S. and {Sanchez}, E. and {Cid}, D. Sanchez and {Sevilla-Noarbe}, I. and {Smith}, M. and {Suchyta}, E. and {Tarle}, G. and {Tucker}, D.~L. and {Walker}, A.~R. and {Wiseman}, P. and {Zhang}, Y.},
        title = "{Dark Energy Survey Year 6 Results: Synthetic-source Injection Across the Full Survey Using Balrog}",
      journal = {The Open Journal of Astrophysics},
         year = 2025,
        month = may,
       volume = {8},
          eid = {65},
        pages = {65},
          doi = {10.33232/001c.138627},
archivePrefix = {arXiv},
       eprint = {2501.05683},
 primaryClass = {astro-ph.CO},
       adsurl = {https://ui.adsabs.harvard.edu/abs/2025OJAp....8E..65A}
}

@article{y6imagesims,
	author = {Mau, S. and Becker, M. R. and others},
        journal = "in preparation.",
	note = {{in preparation}},
	title = {{Y6 Image Simulations}},
	year = "2025"}

@article{y6_source_redshift,
	author = {Yin, B. and others},
	journal = "in preparation.",
	note = {{in preparation}},
	title = {{Y6 Source Redshift}},
	year = "2025"}

@article{y6_1x2pt,
	author = {{DES collaboration}},
	journal = "in preparation.",
	note = {{in preparation}},
	title = {{Y6 Cosmic Shear Cosmology}},
	year = "2025"}

@article{2019A&A...621A...2P,
	adsurl = {https://ui.adsabs.harvard.edu/abs/2019A&A...621A...2P},
	archiveprefix = {arXiv},
	author = {{Pujol}, Arnau and {Kilbinger}, Martin and {Sureau}, Florent and {Bobin}, Jerome},
	doi = {10.1051/0004-6361/201833740},
	eid = {A2},
	eprint = {1806.10537},
	journal = {\aap},
	month = jan,
	pages = {A2},
	primaryclass = {astro-ph.CO},
	title = {{A highly precise shear bias estimator independent of the measured shape noise}},
	volume = {621},
	year = 2019}

@article{y6bfd,
	author = {Gatti, M and Wetzell, V and others},
	journal = "in preparation.",
	note = {{in preparation}},
	title = {{Y6 BFD}},
	year = "2025"}

@inproceedings{2006SPIE.6276E..08D,
	adsurl = {https://ui.adsabs.harvard.edu/abs/2006SPIE.6276E..08D},
	author = {{Derylo}, Gregory and {Diehl}, H. Thomas and {Estrada}, Juan},
	booktitle = {High Energy, Optical, and Infrared Detectors for Astronomy II},
	doi = {10.1117/12.672505},
	editor = {{Dorn}, David A. and {Holland}, Andrew D.},
	eid = {627608},
	month = jun,
	pages = {627608},
	series = {Society of Photo-Optical Instrumentation Engineers (SPIE) Conference Series},
	title = {{0.250mm-thick CCD packaging for the Dark Energy Survey Camera array}},
	volume = {6276},
	year = 2006}

@article{gaia2,
	adsurl = {https://ui.adsabs.harvard.edu/abs/2018A&A...616A...1G},
	archiveprefix = {arXiv},
	author = {{Gaia Collaboration} and {Brown}, A.~G.~A. and {Vallenari}, A. and {Prusti}, T. and {de Bruijne}, J.~H.~J. and {Babusiaux}, C. and {Bailer-Jones}, C.~A.~L. and {Biermann}, M. and {Evans}, D.~W. and {Eyer}, L. and {Jansen}, F. and {Jordi}, C. and {Klioner}, S.~A. and {Lammers}, U. and {Lindegren}, L. and {Luri}, X. and {Mignard}, F. and {Panem}, C. and {Pourbaix}, D. and {Randich}, S. and {Sartoretti}, P. and {Siddiqui}, H.~I. and {Soubiran}, C. and {van Leeuwen}, F. and {Walton}, N.~A. and {Arenou}, F. and {Bastian}, U. and {Cropper}, M. and {Drimmel}, R. and {Katz}, D. and {Lattanzi}, M.~G. and {Bakker}, J. and {Cacciari}, C. and {Casta{\~n}eda}, J. and {Chaoul}, L. and {Cheek}, N. and {De Angeli}, F. and {Fabricius}, C. and {Guerra}, R. and {Holl}, B. and {Masana}, E. and {Messineo}, R. and {Mowlavi}, N. and {Nienartowicz}, K. and {Panuzzo}, P. and {Portell}, J. and {Riello}, M. and {Seabroke}, G.~M. and {Tanga}, P. and {Th{\'e}venin}, F. and {Gracia-Abril}, G. and {Comoretto}, G. and {Garcia-Reinaldos}, M. and {Teyssier}, D. and {Altmann}, M. and {Andrae}, R. and {Audard}, M. and {Bellas-Velidis}, I. and {Benson}, K. and {Berthier}, J. and {Blomme}, R. and {Burgess}, P. and {Busso}, G. and {Carry}, B. and {Cellino}, A. and {Clementini}, G. and {Clotet}, M. and {Creevey}, O. and {Davidson}, M. and {De Ridder}, J. and {Delchambre}, L. and {Dell'Oro}, A. and {Ducourant}, C. and {Fern{\'a}ndez-Hern{\'a}ndez}, J. and {Fouesneau}, M. and {Fr{\'e}mat}, Y. and {Galluccio}, L. and {Garc{\'\i}a-Torres}, M. and {Gonz{\'a}lez-N{\'u}{\~n}ez}, J. and {Gonz{\'a}lez-Vidal}, J.~J. and {Gosset}, E. and {Guy}, L.~P. and {Halbwachs}, J. -L. and {Hambly}, N.~C. and {Harrison}, D.~L. and {Hern{\'a}ndez}, J. and {Hestroffer}, D. and {Hodgkin}, S.~T. and {Hutton}, A. and {Jasniewicz}, G. and {Jean-Antoine-Piccolo}, A. and {Jordan}, S. and {Korn}, A.~J. and {Krone-Martins}, A. and {Lanzafame}, A.~C. and {Lebzelter}, T. and {L{\"o}ffler}, W. and {Manteiga}, M. and {Marrese}, P.~M. and {Mart{\'\i}n-Fleitas}, J.~M. and {Moitinho}, A. and {Mora}, A. and {Muinonen}, K. and {Osinde}, J. and {Pancino}, E. and {Pauwels}, T. and {Petit}, J. -M. and {Recio-Blanco}, A. and {Richards}, P.~J. and {Rimoldini}, L. and {Robin}, A.~C. and {Sarro}, L.~M. and {Siopis}, C. and {Smith}, M. and {Sozzetti}, A. and {S{\"u}veges}, M. and {Torra}, J. and {van Reeven}, W. and {Abbas}, U. and {Abreu Aramburu}, A. and {Accart}, S. and {Aerts}, C. and {Altavilla}, G. and {{\'A}lvarez}, M.~A. and {Alvarez}, R. and {Alves}, J. and {Anderson}, R.~I. and {Andrei}, A.~H. and {Anglada Varela}, E. and {Antiche}, E. and {Antoja}, T. and {Arcay}, B. and {Astraatmadja}, T.~L. and {Bach}, N. and {Baker}, S.~G. and {Balaguer-N{\'u}{\~n}ez}, L. and {Balm}, P. and {Barache}, C. and {Barata}, C. and {Barbato}, D. and {Barblan}, F. and {Barklem}, P.~S. and {Barrado}, D. and {Barros}, M. and {Barstow}, M.~A. and {Bartholom{\'e} Mu{\~n}oz}, S. and {Bassilana}, J. -L. and {Becciani}, U. and {Bellazzini}, M. and {Berihuete}, A. and {Bertone}, S. and {Bianchi}, L. and {Bienaym{\'e}}, O. and {Blanco-Cuaresma}, S. and {Boch}, T. and {Boeche}, C. and {Bombrun}, A. and {Borrachero}, R. and {Bossini}, D. and {Bouquillon}, S. and {Bourda}, G. and {Bragaglia}, A. and {Bramante}, L. and {Breddels}, M.~A. and {Bressan}, A. and {Brouillet}, N. and {Br{\"u}semeister}, T. and {Brugaletta}, E. and {Bucciarelli}, B. and {Burlacu}, A. and {Busonero}, D. and {Butkevich}, A.~G. and {Buzzi}, R. and {Caffau}, E. and {Cancelliere}, R. and {Cannizzaro}, G. and {Cantat-Gaudin}, T. and {Carballo}, R. and {Carlucci}, T. and {Carrasco}, J.~M. and {Casamiquela}, L. and {Castellani}, M. and {Castro-Ginard}, A. and {Charlot}, P. and {Chemin}, L. and {Chiavassa}, A. and {Cocozza}, G. and {Costigan}, G. and {Cowell}, S. and {Crifo}, F. and {Crosta}, M. and {Crowley}, C. and {Cuypers}, J. and {Dafonte}, C. and {Damerdji}, Y. and {Dapergolas}, A. and {David}, P. and {David}, M. and {de Laverny}, P. and {De Luise}, F. and {De March}, R. and {de Martino}, D. and {de Souza}, R. and {de Torres}, A. and {Debosscher}, J. and {del Pozo}, E. and {Delbo}, M. and {Delgado}, A. and {Delgado}, H.~E. and {Di Matteo}, P. and {Diakite}, S. and {Diener}, C. and {Distefano}, E. and {Dolding}, C. and {Drazinos}, P. and {Dur{\'a}n}, J. and {Edvardsson}, B. and {Enke}, H. and {Eriksson}, K. and {Esquej}, P. and {Eynard Bontemps}, G. and {Fabre}, C. and {Fabrizio}, M. and {Faigler}, S. and {Falc{\~a}o}, A.~J. and {Farr{\`a}s Casas}, M. and {Federici}, L. and {Fedorets}, G. and {Fernique}, P. and {Figueras}, F. and {Filippi}, F. and {Findeisen}, K. and {Fonti}, A. and {Fraile}, E. and {Fraser}, M. and {Fr{\'e}zouls}, B. and {Gai}, M. and {Galleti}, S. and {Garabato}, D. and {Garc{\'\i}a-Sedano}, F. and {Garofalo}, A. and {Garralda}, N. and {Gavel}, A. and {Gavras}, P. and {Gerssen}, J. and {Geyer}, R. and {Giacobbe}, P. and {Gilmore}, G. and {Girona}, S. and {Giuffrida}, G. and {Glass}, F. and {Gomes}, M. and {Granvik}, M. and {Gueguen}, A. and {Guerrier}, A. and {Guiraud}, J. and {Guti{\'e}rrez-S{\'a}nchez}, R. and {Haigron}, R. and {Hatzidimitriou}, D. and {Hauser}, M. and {Haywood}, M. and {Heiter}, U. and {Helmi}, A. and {Heu}, J. and {Hilger}, T. and {Hobbs}, D. and {Hofmann}, W. and {Holland}, G. and {Huckle}, H.~E. and {Hypki}, A. and {Icardi}, V. and {Jan{\ss}en}, K. and {Jevardat de Fombelle}, G. and {Jonker}, P.~G. and {Juh{\'a}sz}, {\'A}. L. and {Julbe}, F. and {Karampelas}, A. and {Kewley}, A. and {Klar}, J. and {Kochoska}, A. and {Kohley}, R. and {Kolenberg}, K. and {Kontizas}, M. and {Kontizas}, E. and {Koposov}, S.~E. and {Kordopatis}, G. and {Kostrzewa-Rutkowska}, Z. and {Koubsky}, P. and {Lambert}, S. and {Lanza}, A.~F. and {Lasne}, Y. and {Lavigne}, J. -B. and {Le Fustec}, Y. and {Le Poncin-Lafitte}, C. and {Lebreton}, Y. and {Leccia}, S. and {Leclerc}, N. and {Lecoeur-Taibi}, I. and {Lenhardt}, H. and {Leroux}, F. and {Liao}, S. and {Licata}, E. and {Lindstr{\o}m}, H.~E.~P. and {Lister}, T.~A. and {Livanou}, E. and {Lobel}, A. and {L{\'o}pez}, M. and {Managau}, S. and {Mann}, R.~G. and {Mantelet}, G. and {Marchal}, O. and {Marchant}, J.~M. and {Marconi}, M. and {Marinoni}, S. and {Marschalk{\'o}}, G. and {Marshall}, D.~J. and {Martino}, M. and {Marton}, G. and {Mary}, N. and {Massari}, D. and {Matijevi{\v{c}}}, G. and {Mazeh}, T. and {McMillan}, P.~J. and {Messina}, S. and {Michalik}, D. and {Millar}, N.~R. and {Molina}, D. and {Molinaro}, R. and {Moln{\'a}r}, L. and {Montegriffo}, P. and {Mor}, R. and {Morbidelli}, R. and {Morel}, T. and {Morris}, D. and {Mulone}, A.~F. and {Muraveva}, T. and {Musella}, I. and {Nelemans}, G. and {Nicastro}, L. and {Noval}, L. and {O'Mullane}, W. and {Ord{\'e}novic}, C. and {Ord{\'o}{\~n}ez-Blanco}, D. and {Osborne}, P. and {Pagani}, C. and {Pagano}, I. and {Pailler}, F. and {Palacin}, H. and {Palaversa}, L. and {Panahi}, A. and {Pawlak}, M. and {Piersimoni}, A.~M. and {Pineau}, F. -X. and {Plachy}, E. and {Plum}, G. and {Poggio}, E. and {Poujoulet}, E. and {Pr{\v{s}}a}, A. and {Pulone}, L. and {Racero}, E. and {Ragaini}, S. and {Rambaux}, N. and {Ramos-Lerate}, M. and {Regibo}, S. and {Reyl{\'e}}, C. and {Riclet}, F. and {Ripepi}, V. and {Riva}, A. and {Rivard}, A. and {Rixon}, G. and {Roegiers}, T. and {Roelens}, M. and {Romero-G{\'o}mez}, M. and {Rowell}, N. and {Royer}, F. and {Ruiz-Dern}, L. and {Sadowski}, G. and {Sagrist{\`a} Sell{\'e}s}, T. and {Sahlmann}, J. and {Salgado}, J. and {Salguero}, E. and {Sanna}, N. and {Santana-Ros}, T. and {Sarasso}, M. and {Savietto}, H. and {Schultheis}, M. and {Sciacca}, E. and {Segol}, M. and {Segovia}, J.~C. and {S{\'e}gransan}, D. and {Shih}, I. -C. and {Siltala}, L. and {Silva}, A.~F. and {Smart}, R.~L. and {Smith}, K.~W. and {Solano}, E. and {Solitro}, F. and {Sordo}, R. and {Soria Nieto}, S. and {Souchay}, J. and {Spagna}, A. and {Spoto}, F. and {Stampa}, U. and {Steele}, I.~A. and {Steidelm{\"u}ller}, H. and {Stephenson}, C.~A. and {Stoev}, H. and {Suess}, F.~F. and {Surdej}, J. and {Szabados}, L. and {Szegedi-Elek}, E. and {Tapiador}, D. and {Taris}, F. and {Tauran}, G. and {Taylor}, M.~B. and {Teixeira}, R. and {Terrett}, D. and {Teyssandier}, P. and {Thuillot}, W. and {Titarenko}, A. and {Torra Clotet}, F. and {Turon}, C. and {Ulla}, A. and {Utrilla}, E. and {Uzzi}, S. and {Vaillant}, M. and {Valentini}, G. and {Valette}, V. and {van Elteren}, A. and {Van Hemelryck}, E. and {van Leeuwen}, M. and {Vaschetto}, M. and {Vecchiato}, A. and {Veljanoski}, J. and {Viala}, Y. and {Vicente}, D. and {Vogt}, S. and {von Essen}, C. and {Voss}, H. and {Votruba}, V. and {Voutsinas}, S. and {Walmsley}, G. and {Weiler}, M. and {Wertz}, O. and {Wevers}, T. and {Wyrzykowski}, {\L}. and {Yoldas}, A. and {{\v{Z}}erjal}, M. and {Ziaeepour}, H. and {Zorec}, J. and {Zschocke}, S. and {Zucker}, S. and {Zurbach}, C. and {Zwitter}, T.},
	doi = {10.1051/0004-6361/201833051},
	eid = {A1},
	eprint = {1804.09365},
	journal = {\aap},
	month = aug,
	pages = {A1},
	primaryclass = {astro-ph.GA},
	title = {{Gaia Data Release 2. Summary of the contents and survey properties}},
	volume = {616},
	year = 2018}

@article{lsst_srd,
	adsurl = {https://ui.adsabs.harvard.edu/abs/2018arXiv180901669T},
	archiveprefix = {arXiv},
	author = {{The LSST Dark Energy Science Collaboration} and {Mandelbaum}, Rachel and {Eifler}, Tim and {Hlo{\v{z}}ek}, Ren{\'e}e and {Collett}, Thomas and {Gawiser}, Eric and {Scolnic}, Daniel and {Alonso}, David and {Awan}, Humna and {Biswas}, Rahul and {Blazek}, Jonathan and {Burchat}, Patricia and {Chisari}, Nora Elisa and {Dell'Antonio}, Ian and {Digel}, Seth and {Frieman}, Josh and {Goldstein}, Daniel A. and {Hook}, Isobel and {Ivezi{\'c}}, {\v{Z}}eljko and {Kahn}, Steven M. and {Kamath}, Sowmya and {Kirkby}, David and {Kitching}, Thomas and {Krause}, Elisabeth and {Leget}, Pierre-Fran{\c{c}}ois and {Marshall}, Philip J. and {Meyers}, Joshua and {Miyatake}, Hironao and {Newman}, Jeffrey A. and {Nichol}, Robert and {Rykoff}, Eli and {Sanchez}, F. Javier and {Slosar}, An{\v{z}}e and {Sullivan}, Mark and {Troxel}, M.~A.},
	doi = {10.48550/arXiv.1809.01669},
	eid = {arXiv:1809.01669},
	eprint = {1809.01669},
	journal = {arXiv e-prints},
	month = sep,
	pages = {arXiv:1809.01669},
	primaryclass = {astro-ph.CO},
	title = {{The LSST Dark Energy Science Collaboration (DESC) Science Requirements Document}},
	year = 2018}

@article{scipy,
	adsurl = {https://rdcu.be/b08Wh},
	author = {Virtanen, Pauli and Gommers, Ralf and Oliphant, Travis E. and Haberland, Matt and Reddy, Tyler and Cournapeau, David and Burovski, Evgeni and Peterson, Pearu and Weckesser, Warren and Bright, Jonathan and {van der Walt}, St{\'e}fan J. and Brett, Matthew and Wilson, Joshua and Millman, K. Jarrod and Mayorov, Nikolay and Nelson, Andrew R. J. and Jones, Eric and Kern, Robert and Larson, Eric and Carey, C J and Polat, {\.I}lhan and Feng, Yu and Moore, Eric W. and {VanderPlas}, Jake and Laxalde, Denis and Perktold, Josef and Cimrman, Robert and Henriksen, Ian and Quintero, E. A. and Harris, Charles R. and Archibald, Anne M. and Ribeiro, Ant{\^o}nio H. and Pedregosa, Fabian and {van Mulbregt}, Paul and {SciPy 1.0 Contributors}},
	doi = {10.1038/s41592-019-0686-2},
	journal = {Nature Methods},
	pages = {261--272},
	title = {{{SciPy} 1.0: Fundamental Algorithms for Scientific Computing in Python}},
	volume = {17},
	year = {2020}}

@article{diehl2012dark,
	author = {Diehl, Thomas and Dark Energy Survey Collaboration and others},
	journal = {Physics Procedia},
	pages = {1332--1340},
	publisher = {Elsevier},
	title = {The dark energy survey camera (DECam)},
	volume = {37},
	year = {2012}}

@article{gaia_edr3,
	adsurl = {https://ui.adsabs.harvard.edu/abs/2021A&A...649A...1G},
	archiveprefix = {arXiv},
	author = {{Gaia Collaboration} and {Brown}, A.~G.~A. and {Vallenari}, A. and {Prusti}, T. and {de Bruijne}, J.~H.~J. and {Babusiaux}, C. and {Biermann}, M. and {Creevey}, O.~L. and {Evans}, D.~W. and {Eyer}, L. and {Hutton}, A. and {Jansen}, F. and {Jordi}, C. and {Klioner}, S.~A. and {Lammers}, U. and {Lindegren}, L. and {Luri}, X. and {Mignard}, F. and {Panem}, C. and {Pourbaix}, D. and {Randich}, S. and {Sartoretti}, P. and {Soubiran}, C. and {Walton}, N.~A. and {Arenou}, F. and {Bailer-Jones}, C.~A.~L. and {Bastian}, U. and {Cropper}, M. and {Drimmel}, R. and {Katz}, D. and {Lattanzi}, M.~G. and {van Leeuwen}, F. and {Bakker}, J. and {Cacciari}, C. and {Casta{\~n}eda}, J. and {De Angeli}, F. and {Ducourant}, C. and {Fabricius}, C. and {Fouesneau}, M. and {Fr{\'e}mat}, Y. and {Guerra}, R. and {Guerrier}, A. and {Guiraud}, J. and {Jean-Antoine Piccolo}, A. and {Masana}, E. and {Messineo}, R. and {Mowlavi}, N. and {Nicolas}, C. and {Nienartowicz}, K. and {Pailler}, F. and {Panuzzo}, P. and {Riclet}, F. and {Roux}, W. and {Seabroke}, G.~M. and {Sordo}, R. and {Tanga}, P. and {Th{\'e}venin}, F. and {Gracia-Abril}, G. and {Portell}, J. and {Teyssier}, D. and {Altmann}, M. and {Andrae}, R. and {Bellas-Velidis}, I. and {Benson}, K. and {Berthier}, J. and {Blomme}, R. and {Brugaletta}, E. and {Burgess}, P.~W. and {Busso}, G. and {Carry}, B. and {Cellino}, A. and {Cheek}, N. and {Clementini}, G. and {Damerdji}, Y. and {Davidson}, M. and {Delchambre}, L. and {Dell'Oro}, A. and {Fern{\'a}ndez-Hern{\'a}ndez}, J. and {Galluccio}, L. and {Garc{\'\i}a-Lario}, P. and {Garcia-Reinaldos}, M. and {Gonz{\'a}lez-N{\'u}{\~n}ez}, J. and {Gosset}, E. and {Haigron}, R. and {Halbwachs}, J. -L. and {Hambly}, N.~C. and {Harrison}, D.~L. and {Hatzidimitriou}, D. and {Heiter}, U. and {Hern{\'a}ndez}, J. and {Hestroffer}, D. and {Hodgkin}, S.~T. and {Holl}, B. and {Jan{\ss}en}, K. and {Jevardat de Fombelle}, G. and {Jordan}, S. and {Krone-Martins}, A. and {Lanzafame}, A.~C. and {L{\"o}ffler}, W. and {Lorca}, A. and {Manteiga}, M. and {Marchal}, O. and {Marrese}, P.~M. and {Moitinho}, A. and {Mora}, A. and {Muinonen}, K. and {Osborne}, P. and {Pancino}, E. and {Pauwels}, T. and {Petit}, J. -M. and {Recio-Blanco}, A. and {Richards}, P.~J. and {Riello}, M. and {Rimoldini}, L. and {Robin}, A.~C. and {Roegiers}, T. and {Rybizki}, J. and {Sarro}, L.~M. and {Siopis}, C. and {Smith}, M. and {Sozzetti}, A. and {Ulla}, A. and {Utrilla}, E. and {van Leeuwen}, M. and {van Reeven}, W. and {Abbas}, U. and {Abreu Aramburu}, A. and {Accart}, S. and {Aerts}, C. and {Aguado}, J.~J. and {Ajaj}, M. and {Altavilla}, G. and {{\'A}lvarez}, M.~A. and {{\'A}lvarez Cid-Fuentes}, J. and {Alves}, J. and {Anderson}, R.~I. and {Anglada Varela}, E. and {Antoja}, T. and {Audard}, M. and {Baines}, D. and {Baker}, S.~G. and {Balaguer-N{\'u}{\~n}ez}, L. and {Balbinot}, E. and {Balog}, Z. and {Barache}, C. and {Barbato}, D. and {Barros}, M. and {Barstow}, M.~A. and {Bartolom{\'e}}, S. and {Bassilana}, J. -L. and {Bauchet}, N. and {Baudesson-Stella}, A. and {Becciani}, U. and {Bellazzini}, M. and {Bernet}, M. and {Bertone}, S. and {Bianchi}, L. and {Blanco-Cuaresma}, S. and {Boch}, T. and {Bombrun}, A. and {Bossini}, D. and {Bouquillon}, S. and {Bragaglia}, A. and {Bramante}, L. and {Breedt}, E. and {Bressan}, A. and {Brouillet}, N. and {Bucciarelli}, B. and {Burlacu}, A. and {Busonero}, D. and {Butkevich}, A.~G. and {Buzzi}, R. and {Caffau}, E. and {Cancelliere}, R. and {C{\'a}novas}, H. and {Cantat-Gaudin}, T. and {Carballo}, R. and {Carlucci}, T. and {Carnerero}, M.~I. and {Carrasco}, J.~M. and {Casamiquela}, L. and {Castellani}, M. and {Castro-Ginard}, A. and {Castro Sampol}, P. and {Chaoul}, L. and {Charlot}, P. and {Chemin}, L. and {Chiavassa}, A. and {Cioni}, M. -R.~L. and {Comoretto}, G. and {Cooper}, W.~J. and {Cornez}, T. and {Cowell}, S. and {Crifo}, F. and {Crosta}, M. and {Crowley}, C. and {Dafonte}, C. and {Dapergolas}, A. and {David}, M. and {David}, P. and {de Laverny}, P. and {De Luise}, F. and {De March}, R. and {De Ridder}, J. and {de Souza}, R. and {de Teodoro}, P. and {de Torres}, A. and {del Peloso}, E.~F. and {del Pozo}, E. and {Delbo}, M. and {Delgado}, A. and {Delgado}, H.~E. and {Delisle}, J. -B. and {Di Matteo}, P. and {Diakite}, S. and {Diener}, C. and {Distefano}, E. and {Dolding}, C. and {Eappachen}, D. and {Edvardsson}, B. and {Enke}, H. and {Esquej}, P. and {Fabre}, C. and {Fabrizio}, M. and {Faigler}, S. and {Fedorets}, G. and {Fernique}, P. and {Fienga}, A. and {Figueras}, F. and {Fouron}, C. and {Fragkoudi}, F. and {Fraile}, E. and {Franke}, F. and {Gai}, M. and {Garabato}, D. and {Garcia-Gutierrez}, A. and {Garc{\'\i}a-Torres}, M. and {Garofalo}, A. and {Gavras}, P. and {Gerlach}, E. and {Geyer}, R. and {Giacobbe}, P. and {Gilmore}, G. and {Girona}, S. and {Giuffrida}, G. and {Gomel}, R. and {Gomez}, A. and {Gonzalez-Santamaria}, I. and {Gonz{\'a}lez-Vidal}, J.~J. and {Granvik}, M. and {Guti{\'e}rrez-S{\'a}nchez}, R. and {Guy}, L.~P. and {Hauser}, M. and {Haywood}, M. and {Helmi}, A. and {Hidalgo}, S.~L. and {Hilger}, T. and {H{\l}adczuk}, N. and {Hobbs}, D. and {Holland}, G. and {Huckle}, H.~E. and {Jasniewicz}, G. and {Jonker}, P.~G. and {Juaristi Campillo}, J. and {Julbe}, F. and {Karbevska}, L. and {Kervella}, P. and {Khanna}, S. and {Kochoska}, A. and {Kontizas}, M. and {Kordopatis}, G. and {Korn}, A.~J. and {Kostrzewa-Rutkowska}, Z. and {Kruszy{\'n}ska}, K. and {Lambert}, S. and {Lanza}, A.~F. and {Lasne}, Y. and {Le Campion}, J. -F. and {Le Fustec}, Y. and {Lebreton}, Y. and {Lebzelter}, T. and {Leccia}, S. and {Leclerc}, N. and {Lecoeur-Taibi}, I. and {Liao}, S. and {Licata}, E. and {Lindstr{\o}m}, E.~P. and {Lister}, T.~A. and {Livanou}, E. and {Lobel}, A. and {Madrero Pardo}, P. and {Managau}, S. and {Mann}, R.~G. and {Marchant}, J.~M. and {Marconi}, M. and {Marcos Santos}, M.~M.~S. and {Marinoni}, S. and {Marocco}, F. and {Marshall}, D.~J. and {Martin Polo}, L. and {Mart{\'\i}n-Fleitas}, J.~M. and {Masip}, A. and {Massari}, D. and {Mastrobuono-Battisti}, A. and {Mazeh}, T. and {McMillan}, P.~J. and {Messina}, S. and {Michalik}, D. and {Millar}, N.~R. and {Mints}, A. and {Molina}, D. and {Molinaro}, R. and {Moln{\'a}r}, L. and {Montegriffo}, P. and {Mor}, R. and {Morbidelli}, R. and {Morel}, T. and {Morris}, D. and {Mulone}, A.~F. and {Munoz}, D. and {Muraveva}, T. and {Murphy}, C.~P. and {Musella}, I. and {Noval}, L. and {Ord{\'e}novic}, C. and {Orr{\`u}}, G. and {Osinde}, J. and {Pagani}, C. and {Pagano}, I. and {Palaversa}, L. and {Palicio}, P.~A. and {Panahi}, A. and {Pawlak}, M. and {Pe{\~n}alosa Esteller}, X. and {Penttil{\"a}}, A. and {Piersimoni}, A.~M. and {Pineau}, F. -X. and {Plachy}, E. and {Plum}, G. and {Poggio}, E. and {Poretti}, E. and {Poujoulet}, E. and {Pr{\v{s}}a}, A. and {Pulone}, L. and {Racero}, E. and {Ragaini}, S. and {Rainer}, M. and {Raiteri}, C.~M. and {Rambaux}, N. and {Ramos}, P. and {Ramos-Lerate}, M. and {Re Fiorentin}, P. and {Regibo}, S. and {Reyl{\'e}}, C. and {Ripepi}, V. and {Riva}, A. and {Rixon}, G. and {Robichon}, N. and {Robin}, C. and {Roelens}, M. and {Rohrbasser}, L. and {Romero-G{\'o}mez}, M. and {Rowell}, N. and {Royer}, F. and {Rybicki}, K.~A. and {Sadowski}, G. and {Sagrist{\`a} Sell{\'e}s}, A. and {Sahlmann}, J. and {Salgado}, J. and {Salguero}, E. and {Samaras}, N. and {Sanchez Gimenez}, V. and {Sanna}, N. and {Santove{\~n}a}, R. and {Sarasso}, M. and {Schultheis}, M. and {Sciacca}, E. and {Segol}, M. and {Segovia}, J.~C. and {S{\'e}gransan}, D. and {Semeux}, D. and {Shahaf}, S. and {Siddiqui}, H.~I. and {Siebert}, A. and {Siltala}, L. and {Slezak}, E. and {Smart}, R.~L. and {Solano}, E. and {Solitro}, F. and {Souami}, D. and {Souchay}, J. and {Spagna}, A. and {Spoto}, F. and {Steele}, I.~A. and {Steidelm{\"u}ller}, H. and {Stephenson}, C.~A. and {S{\"u}veges}, M. and {Szabados}, L. and {Szegedi-Elek}, E. and {Taris}, F. and {Tauran}, G. and {Taylor}, M.~B. and {Teixeira}, R. and {Thuillot}, W. and {Tonello}, N. and {Torra}, F. and {Torra}, J. and {Turon}, C. and {Unger}, N. and {Vaillant}, M. and {van Dillen}, E. and {Vanel}, O. and {Vecchiato}, A. and {Viala}, Y. and {Vicente}, D. and {Voutsinas}, S. and {Weiler}, M. and {Wevers}, T. and {Wyrzykowski}, {\L}. and {Yoldas}, A. and {Yvard}, P. and {Zhao}, H. and {Zorec}, J. and {Zucker}, S. and {Zurbach}, C. and {Zwitter}, T.},
	doi = {10.1051/0004-6361/202039657},
	eid = {A1},
	eprint = {2012.01533},
	journal = {\aap},
	month = may,
	pages = {A1},
	primaryclass = {astro-ph.GA},
	title = {{Gaia Early Data Release 3. Summary of the contents and survey properties}},
	volume = {649},
	year = 2021}

@ARTICLE{y3_1x2pt_2,
       author = "Secco, L.~F. and Samuroff, S. and others",
        title = "{Dark Energy Survey Year 3 results: Cosmology from cosmic shear and robustness to modeling uncertainty}",
      journal = {\prd},
         year = 2022,
        month = jan,
       volume = {105},
       number = {2},
          eid = {023515},
        pages = {023515},
          doi = {10.1103/PhysRevD.105.023515},
archivePrefix = {arXiv},
       eprint = {2105.13544},
 primaryClass = {astro-ph.CO},
       adsurl = {https://ui.adsabs.harvard.edu/abs/2022PhRvD.105b3515S}
}

@ARTICLE{hscy3_1x2pt,
       author = {{Li}, Xiangchong and {Zhang}, Tianqing and {Sugiyama}, Sunao and {Dalal}, Roohi and {Terasawa}, Ryo and {Rau}, Markus M. and {Mandelbaum}, Rachel and {Takada}, Masahiro and {More}, Surhud and {Strauss}, Michael A. and {Miyatake}, Hironao and {Shirasaki}, Masato and {Hamana}, Takashi and {Oguri}, Masamune and {Luo}, Wentao and {Nishizawa}, Atsushi J. and {Takahashi}, Ryuichi and {Nicola}, Andrina and {Osato}, Ken and {Kannawadi}, Arun and {Sunayama}, Tomomi and {Armstrong}, Robert and {Bosch}, James and {Komiyama}, Yutaka and {Lupton}, Robert H. and {Lust}, Nate B. and {MacArthur}, Lauren A. and {Miyazaki}, Satoshi and {Murayama}, Hitoshi and {Nishimichi}, Takahiro and {Okura}, Yuki and {Price}, Paul A. and {Tait}, Philip J. and {Tanaka}, Masayuki and {Wang}, Shiang-Yu},
        title = "{Hyper Suprime-Cam Year 3 results: Cosmology from cosmic shear two-point correlation functions}",
      journal = {\prd},
         year = 2023,
        month = dec,
       volume = {108},
       number = {12},
          eid = {123518},
        pages = {123518},
          doi = {10.1103/PhysRevD.108.123518},
archivePrefix = {arXiv},
       eprint = {2304.00702},
 primaryClass = {astro-ph.CO},
       adsurl = {https://ui.adsabs.harvard.edu/abs/2023PhRvD.108l3518L}
}

@ARTICLE{kids1000_1x2pt,
       author = {{Asgari}, Marika and {Lin}, Chieh-An and {Joachimi}, Benjamin and {Giblin}, Benjamin and {Heymans}, Catherine and {Hildebrandt}, Hendrik and {Kannawadi}, Arun and {St{\"o}lzner}, Benjamin and {Tr{\"o}ster}, Tilman and {van den Busch}, Jan Luca and {Wright}, Angus H. and {Bilicki}, Maciej and {Blake}, Chris and {de Jong}, Jelte and {Dvornik}, Andrej and {Erben}, Thomas and {Getman}, Fedor and {Hoekstra}, Henk and {K{\"o}hlinger}, Fabian and {Kuijken}, Konrad and {Miller}, Lance and {Radovich}, Mario and {Schneider}, Peter and {Shan}, HuanYuan and {Valentijn}, Edwin},
        title = "{KiDS-1000 cosmology: Cosmic shear constraints and comparison between two point statistics}",
      journal = {\aap},
         year = 2021,
        month = jan,
       volume = {645},
          eid = {A104},
        pages = {A104},
          doi = {10.1051/0004-6361/202039070},
archivePrefix = {arXiv},
       eprint = {2007.15633},
 primaryClass = {astro-ph.CO},
       adsurl = {https://ui.adsabs.harvard.edu/abs/2021A&A...645A.104A}
}

@ARTICLE{y3_source_redshift,
       author = {Myles, J. and Alarcon, A. and others},
        title = "{Dark Energy Survey Year 3 results: redshift calibration of the weak lensing source galaxies}",
      journal = {\mnras},
         year = 2021,
        month = aug,
       volume = {505},
       number = {3},
        pages = {4249-4277},
          doi = {10.1093/mnras/stab1515},
archivePrefix = {arXiv},
       eprint = {2012.08566},
 primaryClass = {astro-ph.CO},
       adsurl = {https://ui.adsabs.harvard.edu/abs/2021MNRAS.505.4249M}
}

@article{alonso_namaster,
  title = {A Unified Pseudo-{{C$\ell$}} Framework},
  author = {Alonso, David and Sanchez, Javier and Slosar, An{\v z}e and {LSST Dark Energy Science Collaboration}},
  year = {2019},
  month = apr,
  journal = {Monthly Notices of the Royal Astronomical Society},
  volume = {484},
  pages = {4127--4151},
  publisher = {OUP},
  issn = {0035-8711},
  doi = {10.1093/mnras/stz093},
  urldate = {2024-05-23}
}

@article{hikage_pseudocell,
  title = {Shear Power Spectrum Reconstruction Using the Pseudo-Spectrum Method},
  author = {Hikage, Chiaki and Takada, Masahiro and Hamana, Takashi and Spergel, David},
  year = {2011},
  month = mar,
  journal = {Monthly Notices of the Royal Astronomical Society},
  volume = {412},
  pages = {65--74},
  publisher = {OUP},
  issn = {0035-8711},
  doi = {10.1111/j.1365-2966.2010.17886.x},
  urldate = {2024-05-23}
}

@ARTICLE{lsst_trilegal,
       author = {{Dal Tio}, Piero and {Pastorelli}, Giada and {Mazzi}, Alessandro and {Trabucchi}, Michele and {Costa}, Guglielmo and {Jacques}, Alice and {Pieres}, Adriano and {Girardi}, L{\'e}o and {Chen}, Yang and {Olsen}, Knut A.~G. and {Juric}, Mario and {Ivezi{\'c}}, {\v{Z}}eljko and {Yoachim}, Peter and {Clarkson}, William I. and {Marigo}, Paola and {Rodrigues}, Thaise S. and {Zaggia}, Simone and {Barbieri}, Mauro and {Momany}, Yazan and {Bressan}, Alessandro and {Nikutta}, Robert and {da Costa}, Luiz Nicolaci},
        title = "{Simulating the Legacy Survey of Space and Time Stellar Content with TRILEGAL}",
      journal = {\apjs},
         year = 2022,
        month = sep,
       volume = {262},
       number = {1},
          eid = {22},
        pages = {22},
          doi = {10.3847/1538-4365/ac7be6},
archivePrefix = {arXiv},
       eprint = {2208.00829},
 primaryClass = {astro-ph.GA},
       adsurl = {https://ui.adsabs.harvard.edu/abs/2022ApJS..262...22D}
}

@ARTICLE{y1massmaps,
       author = {{Chang}, C. and {Pujol}, A. and {Mawdsley}, B. and {Bacon}, D. and {Elvin-Poole}, J. and {Melchior}, P. and {Kov{\'a}cs}, A. and {Jain}, B. and {Leistedt}, B. and {Giannantonio}, T. and {Alarcon}, A. and {Baxter}, E. and {Bechtol}, K. and {Becker}, M.~R. and {Benoit-L{\'e}vy}, A. and {Bernstein}, G.~M. and {Bonnett}, C. and {Busha}, M.~T. and {Carnero Rosell}, A. and {Castander}, F.~J. and {Cawthon}, R. and {da Costa}, L.~N. and {Davis}, C. and {De Vicente}, J. and {DeRose}, J. and {Drlica-Wagner}, A. and {Fosalba}, P. and {Gatti}, M. and {Gaztanaga}, E. and {Gruen}, D. and {Gschwend}, J. and {Hartley}, W.~G. and {Hoyle}, B. and {Huff}, E.~M. and {Jarvis}, M. and {Jeffrey}, N. and {Kacprzak}, T. and {Lin}, H. and {MacCrann}, N. and {Maia}, M.~A.~G. and {Ogando}, R.~L.~C. and {Prat}, J. and {Rau}, M.~M. and {Rollins}, R.~P. and {Roodman}, A. and {Rozo}, E. and {Rykoff}, E.~S. and {Samuroff}, S. and {S{\'a}nchez}, C. and {Sevilla-Noarbe}, I. and {Sheldon}, E. and {Troxel}, M.~A. and {Varga}, T.~N. and {Vielzeuf}, P. and {Vikram}, V. and {Wechsler}, R.~H. and {Zuntz}, J. and {Abbott}, T.~M.~C. and {Abdalla}, F.~B. and {Allam}, S. and {Annis}, J. and {Bertin}, E. and {Brooks}, D. and {Buckley-Geer}, E. and {Burke}, D.~L. and {Carrasco Kind}, M. and {Carretero}, J. and {Crocce}, M. and {Cunha}, C.~E. and {D'Andrea}, C.~B. and {Desai}, S. and {Diehl}, H.~T. and {Dietrich}, J.~P. and {Doel}, P. and {Estrada}, J. and {Fausti Neto}, A. and {Fernandez}, E. and {Flaugher}, B. and {Frieman}, J. and {Garc{\'\i}a-Bellido}, J. and {Gruendl}, R.~A. and {Gutierrez}, G. and {Honscheid}, K. and {James}, D.~J. and {Jeltema}, T. and {Johnson}, M.~W.~G. and {Johnson}, M.~D. and {Kent}, S. and {Kirk}, D. and {Krause}, E. and {Kuehn}, K. and {Kuhlmann}, S. and {Lahav}, O. and {Li}, T.~S. and {Lima}, M. and {March}, M. and {Martini}, P. and {Menanteau}, F. and {Miquel}, R. and {Mohr}, J.~J. and {Neilsen}, E. and {Nichol}, R.~C. and {Petravick}, D. and {Plazas}, A.~A. and {Romer}, A.~K. and {Sako}, M. and {Sanchez}, E. and {Scarpine}, V. and {Schubnell}, M. and {Smith}, M. and {Smith}, R.~C. and {Soares-Santos}, M. and {Sobreira}, F. and {Suchyta}, E. and {Tarle}, G. and {Thomas}, D. and {Tucker}, D.~L. and {Walker}, A.~R. and {Wester}, W. and {Zhang}, Y. and {DES Collaboration}},
        title = "{Dark Energy Survey Year 1 results: curved-sky weak lensing mass map}",
      journal = {\mnras},
         year = 2018,
        month = apr,
       volume = {475},
       number = {3},
        pages = {3165-3190},
          doi = {10.1093/mnras/stx3363},
archivePrefix = {arXiv},
       eprint = {1708.01535},
 primaryClass = {astro-ph.CO},
       adsurl = {https://ui.adsabs.harvard.edu/abs/2018MNRAS.475.3165C}
}

@ARTICLE{des_astrometry,
       author = {{Bernstein}, G.~M. and {Armstrong}, R. and {Plazas}, A.~A. and {Walker}, A.~R. and {Abbott}, T.~M.~C. and {Allam}, S. and {Bechtol}, K. and {Benoit-L{\'e}vy}, A. and {Brooks}, D. and {Burke}, D.~L. and {Carnero Rosell}, A. and {Carrasco Kind}, M. and {Carretero}, J. and {Cunha}, C.~E. and {da Costa}, L.~N. and {DePoy}, D.~L. and {Desai}, S. and {Diehl}, H.~T. and {Eifler}, T.~F. and {Fernandez}, E. and {Fosalba}, P. and {Frieman}, J. and {Garc{\'\i}a-Bellido}, J. and {Gerdes}, D.~W. and {Gruen}, D. and {Gruendl}, R.~A. and {Gschwend}, J. and {Gutierrez}, G. and {Honscheid}, K. and {James}, D.~J. and {Kent}, S. and {Krause}, E. and {Kuehn}, K. and {Kuropatkin}, N. and {Li}, T.~S. and {Maia}, M.~A.~G. and {March}, M. and {Marshall}, J.~L. and {Menanteau}, F. and {Miquel}, R. and {Ogando}, R.~L.~C. and {Reil}, K. and {Roodman}, A. and {Rykoff}, E.~S. and {Sanchez}, E. and {Scarpine}, V. and {Schindler}, R. and {Schubnell}, M. and {Sevilla-Noarbe}, I. and {Smith}, M. and {Smith}, R.~C. and {Soares-Santos}, M. and {Sobreira}, F. and {Suchyta}, E. and {Swanson}, M.~E.~C. and {Tarle}, G. and {DES Collaboration}},
        title = "{Astrometric Calibration and Performance of the Dark Energy Camera}",
      journal = {\pasp},
         year = 2017,
        month = jul,
       volume = {129},
       number = {977},
        pages = {074503},
          doi = {10.1088/1538-3873/aa6c55},
archivePrefix = {arXiv},
       eprint = {1703.01679},
 primaryClass = {astro-ph.IM},
       adsurl = {https://ui.adsabs.harvard.edu/abs/2017PASP..129g4503B}
}

@ARTICLE{SourceExtractor1996,
       author = {{Bertin}, E. and {Arnouts}, S.},
        title = "{SExtractor: Software for source extraction.}",
      journal = {\aaps},
         year = "1996",
        month = "Jun",
       volume = {117},
        pages = {393-404},
          doi = {10.1051/aas:1996164},
       adsurl = {https://ui.adsabs.harvard.edu/abs/1996A&AS..117..393B}
}

@inproceedings{scamp2002,
  title={The TERAPIX Pipeline},
  author={Emmanuel Bertin and Yannick Mellier and Mario Radovich and Gilles Missonnier and Pierre Didelon and Bertrand Morin},
  year={2002},
  url={https://api.semanticscholar.org/CorpusID:59911357}
}

%%%%%%%%%%%%%%%%%%%%%%%%%%%%%%%%%%%%%%%%%%%%%%%%%%

%%%%%%%%%%%%%%%%% APPENDICES %%%%%%%%%%%%%%%%%%%%%

\onecolumn

\appendix

\section{Software Versions and Configuration Settings for the Cell-based Coadding and \textsc{Metadetection} Pipelines}\label{app:software}

\begin{table*}
    \centering
    \caption{\label{tab:cellcodever} Software versions and source for key packages used in creating the cell-based coadds.}
    \begin{tabular}[width=\columnwidth]{ccc}
        \hline\hline
        package & version & source \\
        \hline
        \texttt{pizza-cutter} & $0.5.3$ & \url{https://github.com/beckermr/pizza-cutter} \\
        \texttt{numpy} {\citep{numpy}} & $1.19.1$ & \texttt{conda-forge} \\
        \texttt{scipy} {\citep{scipy}} & $1.5.2$ & \texttt{conda-forge} \\
        \texttt{esutil} & $0.6.4$ & \texttt{conda-forge} \\
        \texttt{ngmix} & $2.0.5$ & \texttt{conda-forge} \\
        \texttt{fitsio} & $1.1.5$ & \texttt{conda-forge} \\
        \texttt{piff} & $1.2.0$ & \texttt{conda-forge} \\
        \texttt{pixmappy} & $1.0.0$ & \texttt{conda-forge} \\
        \texttt{desmeds} & $0.9.12$ & \texttt{conda-forge} \\
        \texttt{meds} & $0.9.12$ & \texttt{conda-forge} \\
        \hline\hline
    \end{tabular}
\end{table*}

\begin{table*}
    \centering
    \caption{\label{tab:mdetcodever} Software versions and source for key packages used in running \mdet.}
    \begin{tabular}[width=\columnwidth]{ccc}
        \hline\hline
        package & version & source \\
        \hline
        \texttt{desmeds} & $0.9.16$ & \texttt{conda-forge} \\
        \texttt{easyaccess} & $1.4.11$ & \texttt{conda-forge} \\
        \texttt{pizza-cutter} & $0.8.0$ & \url{https://github.com/beckermr/pizza-cutter} \\
        \texttt{pizza-cutter-metadetect} & $0.9.1$ & \url{https://github.com/beckermr/pizza-cutter-metadetect} \\
        \texttt{sxdes} & $0.3.0$ & \url{https://github.com/esheldon/sxdes} \\
        \texttt{esutil} & $0.6.10$ & \texttt{conda-forge} \\
        \texttt{fitsio} & $1.1.8$ & \texttt{conda-forge} \\
        \texttt{galsim} {\citep{2015A&C....10..121R}} & $2.4.7$ & \texttt{conda-forge} \\
        \texttt{healpy} & $1.16.2$ & \texttt{conda-forge} \\
        \texttt{healsparse} & $1.8.1$ & \texttt{conda-forge} \\
        \texttt{hilbertcurve} & $2.0.5$ & \texttt{conda-forge} \\
        \texttt{hpgeom} & $0.8.2$ & \texttt{conda-forge} \\
        \texttt{lsstdesc.coord} & $1.2.3$ & \texttt{conda-forge} \\
        \texttt{mattspy} & $0.7.2$ & \url{https://github.com/beckermr/mattspy} \\
        \texttt{meds} & $0.9.16$ & \texttt{conda-forge} \\
        \texttt{metadetect} & $0.11.0$ & \url{https://github.com/esheldon/metadetect} \\
        \texttt{ngmix} & $2.3.0$ & \texttt{conda-forge} \\
        \texttt{numba} {\citep{numba}} & $0.56.4$ & \texttt{conda-forge} \\
        \texttt{numpy} {\citep{numpy}} & $1.23.5$ & \texttt{conda-forge} \\
        \texttt{piff} & $1.2.5$ & \texttt{conda-forge} \\
        \texttt{pixmappy} & $1.0.0$ & \texttt{conda-forge} \\
        \texttt{pizza-patches} & $0.10.0$ & \url{https://github.com/esheldon/pizza-patches} \\
        \texttt{scipy} {\citep{scipy}} & $1.10.1$ & \texttt{conda-forge} \\
        \texttt{sep} {\citep{sep}} & $1.2.1$ & \texttt{conda-forge} \\
        \hline\hline
    \end{tabular}
\end{table*}

Tables~\ref{tab:cellcodever} and \ref{tab:mdetcodever} list the software versions for key packages used in creating the cell-based coadds and \mdet\ catalogues. All software is available on \texttt{conda-forge} \citep{condaforge} and/or online as listed in each table. The cell-based coadding and \mdet\ codes require configuration files. For the cell-based coadding code, this file's contents are:
\begin{Verbatim}[fontsize=\tiny, frame=lines, label=\texttt{des-pizza-slices-y6-v15.yaml}, labelposition=topline]
    des_data:
    campaign: Y6A2_COADD
    source_type: finalcut
    piff_campaign: Y6A2_PIFF_V3

  # optional but these are good defaults
  fpack_pars:
    # if you do not set FZTILE, the code sets it to the size of a slice for you
    FZQVALUE: 16
    FZALGOR: "RICE_1"
    # preserve zeros, don't dither them
    FZQMETHD: "SUBTRACTIVE_DITHER_2"
    # do dithering via a checksum
    FZDTHRSD: "CHECKSUM"

  coadd:
    # these are in pixels
    # the total "pizza slice" will be central_size + 2 * buffer_size
    central_size: 100  # size of the central region
    buffer_size: 50  # size of the buffer on each size

    # this should be odd and bigger than any stamp returned by the
    # PSF reconstruction
    psf_box_size: 51

    wcs_type: image
    coadding_weight: 'noise'

  single_epoch:
    # pixel spacing for building various WCS interpolants
    se_wcs_interp_delta: 8
    coadd_wcs_interp_delta: 100

    # fractional amount to increase coadd box size when getting SE region for
    # coadding - set to sqrt(2) for full position angle rotations
    frac_buffer: 1

    # set this to either piff or psfex
    # if using piff in DES and a release earlier than Y6,
    # you need to set the piff_run above too
    psf_type: piff
    psf_kwargs:
      g:
        GI_COLOR: 1.1
      r:
        GI_COLOR: 1.1
      i:
        GI_COLOR: 1.1
      z:
        IZ_COLOR: 0.34
    piff_cuts:
      max_fwhm_cen: 3.6
      min_nstar: 30
      max_exp_T_mean_fac: null
      max_ccd_T_std_fac: null
    mask_piff_failure:
      grid_size: 128
      max_abs_T_diff: 0.15

    # which SE WCS to use - one of piff, pixmappy or image
    wcs_type: pixmappy
    wcs_color: 1.1

    ignored_ccds:
      - 31

    reject_outliers: False
    symmetrize_masking: True
    copy_masked_edges: True
    max_masked_fraction: 0.1
    edge_buffer: 48

    # Y6 already deals with tapebump in a sensible way
    mask_tape_bumps: False

    # DES Y6 bit mask flags
    # "BPM":          1,  #/* set in bpm (hot/dead pixel/column)        */
    # "SATURATE":     2,  #/* saturated pixel                           */
    # "INTERP":       4,  #/* interpolated pixel                        */
    # "BADAMP":       8,  #/* Data from non-functional amplifier        */
    # "CRAY":        16,  #/* cosmic ray pixel                          */
    # "STAR":        32,  #/* bright star pixel                         */
    # "TRAIL":       64,  #/* bleed trail pixel                         */
    # "EDGEBLEED":  128,  #/* edge bleed pixel                          */
    # "SSXTALK":    256,  #/* pixel potentially effected by xtalk from  */
    #                     #/*       a super-saturated source            */
    # "EDGE":       512,  #/* pixel flag to exclude CCD glowing edges   */
    # "STREAK":    1024,  #/* pixel associated with streak from a       */
    #                     #/*       satellite, meteor, ufo...           */
    # "SUSPECT":   2048,  #/* nominally useful pixel but not perfect    */
    # "FIXED":     4096,  # bad coilumn that DESDM reliably fixes       */
    # "NEAREDGE":  8192,  #/* marks 25 bad columns neat the edge        */
    # "TAPEBUMP": 16384,  #/* tape bumps                                */

    spline_interp_flags:
      - 1     # BPM
      - 2     # SATURATE
      - 4     # INTERP. Already interpolated; is this ever set?
      - 16    # CRAY
      - 64    # TRAIL
      - 128   # EDGEBLEED
      - 256   # SSXTALK
      - 512   # EDGE
      - 1024  # STREAK

    noise_interp_flags:
      - 0

    # make the judgment call that it is better to use the somewhat
    # suspect TAPEBUMP/SUSPECT areas than interp, because they are
    # fairly large
    # star areas are ignored for now - GAIA masks will handle them or star-gal sep
    #  - 32    # STAR
    #  - 2048  # SUSPECT
    #  - 4096  # FIXED by DESDM reliably
    #  - 8192  # NEAREDGE 25 bad columns on each edge, removed anyways due to 48 pixel boundry
    #  - 16384 # TAPEBUMP

    bad_image_flags:
      # data from non-functional amplifiers is ignored
      - 8     # BADAMP

    gaia_star_masks:
      poly_coeffs: [1.36055007e-03, -1.55098040e-01,  3.46641671e+00]
      max_g_mag: 18.0
      symmetrize: False
      # interp:
      #   fill_isolated_with_noise: False
      #   iso_buff: 1
      apodize:
        ap_rad: 1
      mask_expand_rad: 16
\end{Verbatim}

For \mdet, this config file is
\begin{Verbatim}[fontsize=\tiny, frame=lines, label=\texttt{metadetect-v10.yaml}, labelposition=topline]
    metacal:
    psf: fitgauss
    types: [noshear, 1p, 1m, 2p, 2m]
    use_noise_image: True

  # use defaults in sxdes package
  sx: null

  fitters:
    - model: pgauss
      weight:
        fwhm: 2.0
      symmetrize: False
    - model: gauss
      weight:
        fwhm: 2.0
      symmetrize: False
      coadd: False

  shear_band_combs: [[1, 2, 3]]
  det_band_combs: [[1, 2, 3]]

  # any regions where the bmask is set with one of these flags will be masked
  # out of detection
  nodet_flags: 33554432  # 2**25 is GAIA stars

  # check for hitting the edge when fitting
  bmask_flags: 1610612736  # 2**29 | 2**30 edge in either MEDS of pizza cutter

  mfrac_fwhm: 2  # arcsec

  meds:
    box_padding: 2
    box_type: iso_radius
    max_box_size: 48
    min_box_size: 48
    rad_fac: 2
    rad_min: 4
    weight_type: uberseg

  pizza-cutter-preprocessing:
    gaia_star_masks:
      poly_coeffs: [ 1.36055007e-03, -1.55098040e-01,  3.46641671e+00]
      max_g_mag: 18.0
      symmetrize: False
      # interp:
      #   fill_isolated_with_noise: False
      #   iso_buff: 1
      apodize:
        ap_rad: 1
      mask_expand_rad: 16

    slice_apodization:
      ap_rad: 1
\end{Verbatim}

\section{Pre-PSF Gaussian (\texttt{pgauss}) Flux and Size Measurements}\label{app:pgauss}

Our primary flux measure for each \mdet\ detection is the \texttt{pgauss} flux measure from the \texttt{ngmix} package. This flux measure has the advantage of matching the effective flux apertures between different bands (up to PSF modeling errors), which is important for the measurement of object colors and photometric redshifts. This flux measure is not adaptive and so can be lower signal-to-noise than a flux measure with an aperture that is better matched to the object. We compute this flux measure, and a measure of object size, by closely following the procedure for the Fourier-space moments from \cite{2016MNRAS.459.4467B} with some key modifications. This technique is also reminiscent of the Gaussian aperture (\texttt{GaaP}) fluxes from \cite{gaap}.

To derive these estimators, we first need a few basic facts about Fourier transforms. These are
\begin{enumerate}
\item The Fourier transform of a quantity like $x^2W(x)$ is $\propto\frac{d^{2}W(k)}{dk^2}$.
\item The integral of the product of two functions in real-space can be computed from their Fourier
        transforms: $\int dx^2 G(x)W(x) = \int dk^2 G(k)W(k)$. This relation is called the Plancherel theorem \citep{plancherel1910}.
\end{enumerate}

In our application, $W(x)$ will be a Gaussian window function that defines the aperture of the flux measure and $G(x)$ will be the pre-PSF galaxy image. If we denote $P(x)$ as the PSF, a convolution as $\circledast$, and $I(x)$ is PSF-convolved profile of the galaxy, then we can compute a pre-PSF, Gaussian weighted flux $F$, via
\begin{eqnarray}
F & = & \int dx^2 W(x) \left[I(x) \circledast P^{-1}(x)\right] \nonumber \\
& = & \int dk^2 \left[I(k) / P(k)\right] W(k)
\end{eqnarray}
where we have used $\dots \circledast P^{-1}(x)$ to indicate a deconvolution, have transformed to Fourier space using Plancherel's Theorem, and converted the deconvolution to a division in Fourier space. Of course, directly deconvolving the PSF from the observed galaxy image, $I(k)/P(k)$, is a numerically unstable operation. To control for this, we implement this set of operations as a multiplication of the Fourier transform of the galaxy image $I(k)$ with a kernel $W(k)/P(k)$. We set the effective real-space size of $W(x)$ to be larger than $P(x)$ so that numerical instabilities from the deconvolution are suppressed.

We can compute higher-order moments as follows. For a moment with a kernel $x_1^nW(x)$, we have
\begin{eqnarray}
    M_i^{(n)} & = & \int dx^2 x_i^2 W(x) \left[I(x) \circledast P^{-1}(x)\right] \nonumber \\
    & \propto & \int dk^2 \frac{d^{n}W(k)}{dk_i^n} \left[I(k) / P(k)\right]
\end{eqnarray}
where we have used the first property enumerated above. Stated more simply, to compute higher-order moments, we use the corresponding derivative of the kernel $W(k)$ in Fourier space. For a Gaussian kernel, these derivates will always carry exponential terms, rendering the process stable as long as the kernel is large enough. With these relationships, we can define the size measure $T^{\rm pgauss}$ as the appropriately normalized sum of $M_1^{(2)} + M_2^{(2)}$. To handle non-trivial WCS transformations, we use a Jacobian WCS approximation about the center of the object and remap the Fourier-space pixel locations to their proper tangent plane locations before evaluating the kernel. Finally, we compute a covariance error matrix for our moments by directly propagating the noise in the pixels through the sums above. Under the assumption that the pixel noise is independent, these covariance error matrices are statistically correct. In practice, for coadds, the pixel noise is not independent, but our simple error measure is a useful approximation nevertheless.

Finally, we apply a few numerical procedures to further stabilize this procedure and make it more efficient. First, we always use a kernel size that is strictly bigger than the PSF size. In our case, this is a Gaussian kernel with FWHM of two arcseconds. Second, we apodize the edges of the object postage stamp images using the same apodization kernel as used in \mdet\ in order to stabilize the results against bright sources near the edge of the image. Third, we zero-pad the image by a factor of four to control for the FFT periodicity. Fourth, we truncate the PSF to zero below an amplitude of $10^{-5}$ relative to its peak value in Fourier space. Finally, we limit the number of FFTs via caching, use \texttt{numba} \citep{numba} to speedup key numerical operations, and use a fast exponential function that works to single precision when computing the kernels. Our final implementation runs at a speed of ${\cal O}(10s)$ of milliseconds per object per image.

\section{Object Selection Details}\label{app:selection}
In this appendix, we provide more details on the object selection cuts in Sec.~\ref{subsec:mdetcuts}.

\subsection{Junk Detections}\label{app:junks}
\begin{figure*}
    \includegraphics[width=0.85\textwidth]{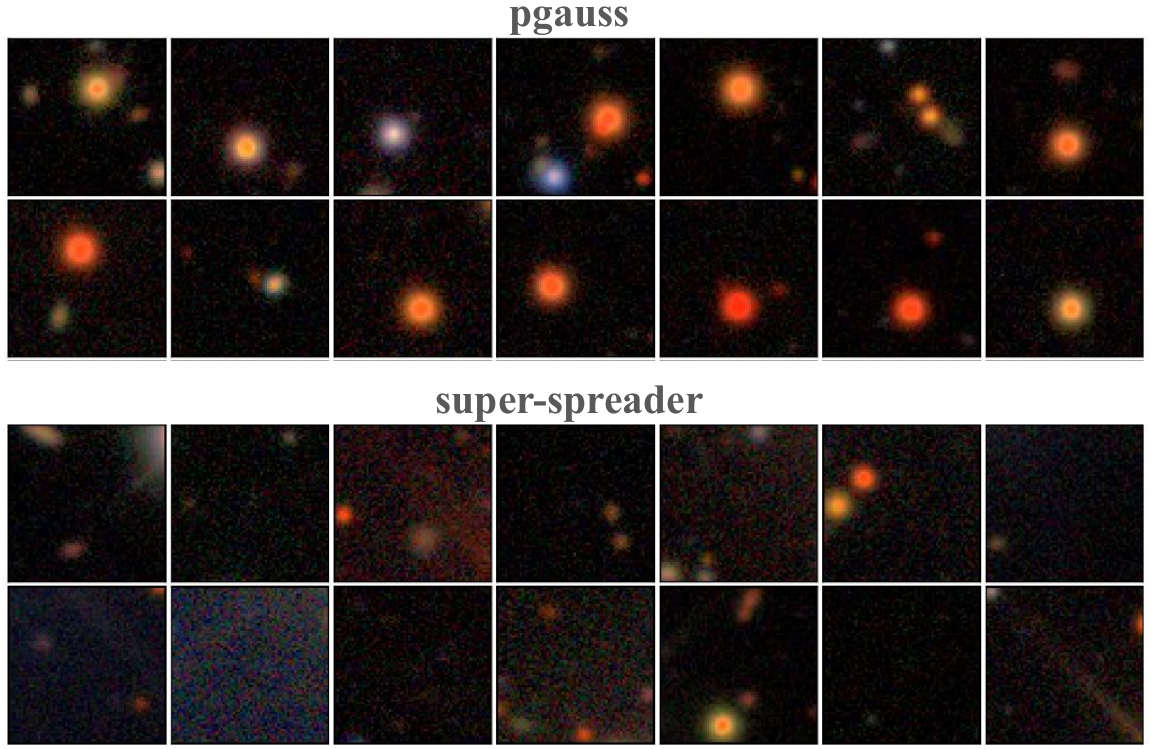}
    \caption{Randomly selected objects that are removed from the catalogue by the ``pgauss'' cut (\textit{top}) and the ``super-spreader'' cut (\textit{bottom}). Each image is a $32\times32$ pixel $gri$ false-color postage stamp around the centroid of the object we cut. In most cases, there is no obvious visible object at the center of the stamp.}
    \label{fig:junks}
\end{figure*}

We visually inspected objects that are removed by the ``junk'' selection cuts.
\begin{enumerate}
    \item \textit{pgauss} cut -- As stated in Sec.~\ref{subsec:mdetcuts}, we removed any objects with $T^{\rm pgauss} > 1.6 - 3.1 \times T^{\rm pgauss}_{\rm err}$. The top panel of Fig.~\ref{fig:junks} shows $32\times32$ pixel $gri$ false-color postage stamp images of randomly selected objects that are removed by this cut. The stamp is centered on the object we cut, which is not visible in many cases. Many of these objects are falsely detected due to background subtraction errors near bright foreground objects.

    \item Super-spreader cut -- Super-spreader objects are objects that have large sizes when fit with bulge+disk models via {\sc ngmix}/{\sc fitvd}. These were initially found in Y3 Balrog studies \citep{y3balrog} and found again as outliers with much too large photometric uncertainty for their brightness in the \gold\ sample \citep{y6gold}. Subsequent studies found that these are the same population of objects. We adjusted the cuts for our slightly different fitting routines and selected objects with $T^{\rm gauss}\times T^{\rm gauss}_{\rm err} < 1$ or $T^{\rm gauss}/T^{\rm gauss}_{\rm err} > 10$. The bottom panel of Fig.~\ref{fig:junks} shows randomly selected objects removed by this cut. Again, the postage stamp images are centered on the object we cut, many of which are not visible in the images.
\end{enumerate}

\subsection{Stellar Contamination}\label{app:stellar}
In this appendix, we present the details of our residual stellar contamination estimates. Due to the fact that the DES Y6 \gold\ catalogue is deeper than the \mdet\ catalogue, we use the fraction of \textit{noshear} \mdet\ objects that match ''likely`` and ''high-confidence`` stars in the \gold\ catalogue as an estimate of the stellar contamination. We first match the entire \mdet\ catalogue to the entire \gold\ catalog, finding the nearest object in the \gold\ catalogue for each \mdet\ catalogue detection. We find that 87\% of \mdet\ objects were matched with a \gold\ object within 0.263 arcseconds, even though the images used to produce the cell-based coadds are a subset of images to produce the \gold\ sample. Investigating the differences by hand, we find that the missing matches come from either few-pixel object centroid offsets between the object centroids or detection differences in the presence of blending. The centroid offsets come from a combination of the larger reconvolution PSF used by \mdet\ coupled with blending and noise. This effect is rather large. If we increase the matching radius to a few pixels, the match rate increases only by $\sim$ 6-7\%. Excluding areas around large foreground galaxies only changed the match rate by $\sim0.1\%$, indicating that these objects are not a major source of the detection differences. Finally, to produce the stellar contamination estimates in Table.~\ref{tab:cuts_fraction}, we compute the fraction of objects that match the ''likely`` and ''high-confidence`` stars from \gold\ within 0.263 arcseconds using their classifications described in Sec.4.2 of \cite{y6gold}.

Unfortunately, the estimates above are too low by a factor of several due to the fact that star-galaxy separation at faint magnitudes is poor in the \gold\ catalogue \citep{y6gold}. Estimates from synthetic-source injection \citep*{y6balrog} indicate a higher fraction of stars around $\sim0.74\%$. This fraction was estimated by computing the fraction of selected \mdet\ objects from the injected sample that matched back to a star in the injection catalogue. However, this estimate is too high since the purity of the star-galaxy classifier for the injected catalogue at faint magnitudes is poor \citep*{y6balrog}. Further, the image simulations also provide an estimate of the effects of stars on the catalogue. The estimates in Sec.~\ref{sec:imagesims} indicate no detectable residual additive bias and an upper bound on the change in the multiplicative bias $m$ of $<3.6\times10^{-3}$ at three-sigma. Thus we conclude that the residual stellar contamination in the catalogue is small enough to be ignored.
 
\section{PSF Leakage and Modeling Errors}\label{app:psf}
In this appendix, we further describe the technique we use to investigate the effects of PSF leakage and modeling errors on the shear catalogue.

As described in Sec.~\ref{subsec:rho}, the systematic contamination due to PSF leakage and modeling errors on the galaxy shape two-point correlation function can be expressed as the two-point correlation functions of observed galaxy shapes and various PSF quantities -- $p_2=e_{\rm PSF}$, $q_2=\Delta e_{\rm PSF}$, $w_2=e_{\rm PSF}\Delta T_{\rm PSF}/T_{\rm PSF}$, $p_4=e^{(4)}_{\rm PSF}$, $q_4=\Delta e^{(4)}_{\rm PSF}$, $w_4=e^{(4)}_{*}\Delta T^{(4)}_{\rm PSF}/T^{(4)}_{*}$, $w_{24}=e_{*}\Delta T^{(4)}_{\rm PSF}/T^{(4)}_{*}$, $w_{42}=e^{(4)}_{*}\Delta T_{\rm PSF}/T_{*}$. Using the contamination model (Eqn.~\ref{eqn:psfcont_xi}, \ref{eqn:model}), the full expressions for these correlations are

\begin{figure*}
\begin{eqnarray}
    \langle g^{\rm obs} p_2 \rangle &=& \alpha \langle p_2 p_2 \rangle + \beta \langle q_2 p_2 \rangle + \eta \langle w_2 p_2 \rangle + \alpha^{(4)} \langle p_4 p_2 \rangle + \beta^{(4)} \langle q_4 p_2 \rangle + \eta^{(4)} \langle w_4 p_2 \rangle + \eta^{(24)} \langle w_{24} p_2 \rangle + \eta^{(42)} \langle w_{42} p_2 \rangle\label{eqn:tau0}
    \\
    \langle g^{\rm obs} q_2 \rangle &=& \alpha \langle p_2 q_2 \rangle + \beta \langle q_2 q_2 \rangle + \eta \langle w_2 q_2 \rangle + \alpha^{(4)} \langle p_4 q_2 \rangle + \beta^{(4)} \langle q_4 q_2 \rangle + \eta^{(4)} \langle w_4 q_2 \rangle + \eta^{(24)} \langle w_{24} q_2 \rangle + \eta^{(42)} \langle w_{42} q_2 \rangle \label{eqn:tau2}
    \\
    \langle g^{\rm obs} w_2 \rangle &=& \alpha \langle p_2 w_2 \rangle + \beta \langle q_2 w_2 \rangle + \eta \langle w_2 w_2 \rangle + \alpha^{(4)} \langle p_4 w_2 \rangle + \beta^{(4)} \langle q_4 w_2 \rangle + \eta^{(4)} \langle w_4 w_2 \rangle + \eta^{(24)} \langle w_{24} w_2 \rangle + \eta^{(42)} \langle w_{42} w_2 \rangle \label{eqn:tau5} \\
    \langle g^{\rm obs} p_4 \rangle &=& \alpha \langle p_2 p_4 \rangle + \beta \langle q_2 p_4 \rangle + \eta \langle w_2 p_4 \rangle + \alpha^{(4)} \langle p_4 p_4 \rangle + \beta^{(4)} \langle q_4 p_4 \rangle + \eta^{(4)} \langle w_4 p_4 \rangle + \eta^{(24)} \langle w_{24} p_4 \rangle + \eta^{(42)} \langle w_{42} p_4 \rangle \label{eqn:tau40} \\
    \langle g^{\rm obs} q_4 \rangle &=& \alpha \langle p_2 q_4 \rangle + \beta \langle q_2 q_4 \rangle + \eta \langle w_2 q_4 \rangle + \alpha^{(4)} \langle p_4 q_4 \rangle + \beta^{(4)} \langle q_4 q_4 \rangle + \eta^{(4)} \langle w_4 q_4 \rangle + \eta^{(24)} \langle w_{24} q_4 \rangle + \eta^{(42)} \langle w_{42} q_4 \rangle \label{eqn:tau42} \\
    \langle g^{\rm obs} w_4 \rangle &=& \alpha \langle p_2 w_4 \rangle + \beta \langle q_2 w_4 \rangle + \eta \langle w_2 w_4 \rangle + \alpha^{(4)} \langle p_4 w_4 \rangle + \beta^{(4)} \langle q_4 w_4 \rangle + \eta^{(4)} \langle w_4 w_4 \rangle + \eta^{(24)} \langle w_{24} w_4 \rangle + \eta^{(42)} \langle w_{42} w_4 \rangle \label{eqn:tau45} \\
    \langle g^{\rm obs} w_{24} \rangle &=& \alpha \langle p_2 w_{24} \rangle + \beta \langle q_2 w_{24} \rangle + \eta \langle w_2 w_{24} \rangle + \alpha^{(4)} \langle p_4 w_{24} \rangle + \beta^{(4)} \langle q_4 w_{24} \rangle + \eta^{(4)} \langle w_4 w_{24} \rangle + \eta^{(24)} \langle w_{24} w_{24} \rangle + \eta^{(42)} \langle w_{42} w_{24} \rangle \label{eqn:tau451} \\
    \langle g^{\rm obs} w_{42} \rangle &=& \alpha \langle p_2 w_{42} \rangle + \beta \langle q_2 w_{42} \rangle + \eta \langle w_2 w_{42} \rangle + \alpha^{(4)} \langle p_4 w_{42} \rangle + \beta^{(4)} \langle q_4 w_{42} \rangle + \eta^{(4)} \langle w_4 w_{42} \rangle + \eta^{(24)} \langle w_{24} w_{42} \rangle + \eta^{(42)} \langle w_{42} w_{42} \rangle \ . \label{eqn:tau452}
\end{eqnarray}
\end{figure*}

\begin{figure*}
	\includegraphics[width=\textwidth]{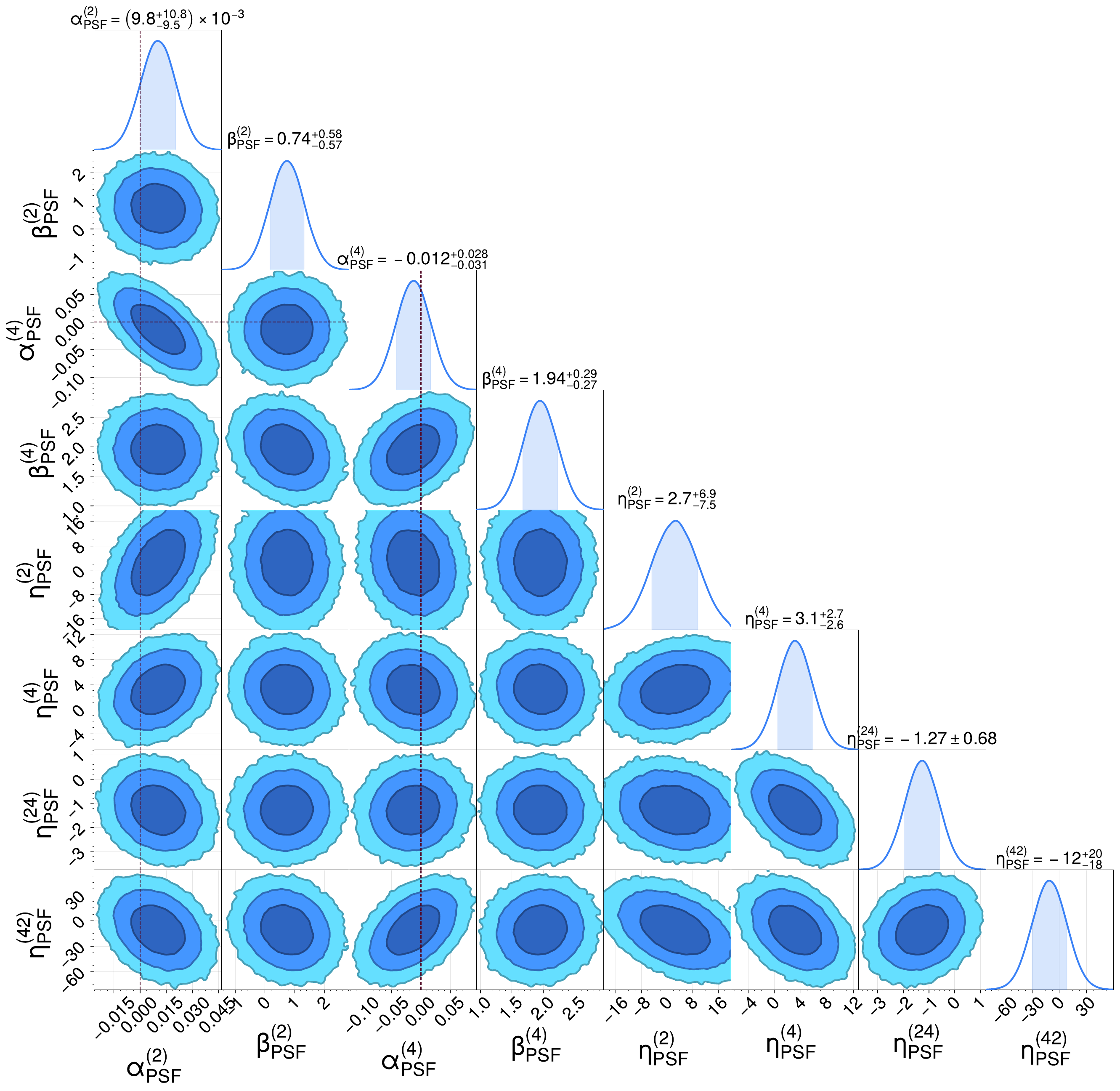}
    \caption{Posterior contours of our PSF contamination model parameter space for the fiducial angular scales and binning. The dotted lines show the parameter values that are expected with no PSF leakage, which are $\alpha^{(2)}, \alpha^{(4)}=0$.}
    \label{fig:mcmc_posterior}
\end{figure*}

We fit the contamination model to our data by first maximizing the likelihood of the two-point cross-correlations above with the Nelder-Mead algorithm \citep{nelder-mead}, and then running Markov Chain Monte Carlo (MCMC) with the {\sc emcee} package \citep{emcee} to obtain the posteriors. The 1D and 2D marginal posterior contours of the parameters are shown in Figure~\ref{fig:mcmc_posterior}. The 2D contour includes $1,2,3 \sigma$ contour lines, and the parameter values indicating no systematic contamination are shown as dotted lines. Most inferred parameters are consistent with zero PSF leakage and modeling errors, except $\beta^{(4)}$ and $\eta^{(24)}$. Each specific analysis using this shear catalogue will need to determine if the residual PSF modeling errors are significant enough to warrant further treatment via marginalizing over the contamination model parameters.

%%%%%%%%%%%%%%%%%%%%%%%%%%%%%%%%%%%%%%%%%%%%%%%%%%

\section{Author Affiliations}
\label{sec:affiliations}
$^{1}$ Department of Astrophysical Sciences, Princeton University, Peyton Hall, Princeton, NJ 08544, USA\\
$^{2}$ Department of Physics, Duke University Durham, NC 27708, USA\\
$^{3}$ Argonne National Laboratory, 9700 South Cass Avenue, Lemont, IL 60439, USA\\
$^{4}$ Brookhaven National Laboratory, Bldg 510, Upton, NY 11973, USA\\
$^{5}$ Department of Physics and Astronomy, University of Pennsylvania, Philadelphia, PA 19104, USA\\
$^{6}$ Center for Astrophysical Surveys, National Center for Supercomputing Applications, 1205 West Clark St., Urbana, IL 61801, USA\\
$^{7}$ Department of Astronomy, University of Illinois at Urbana-Champaign, 1002 W. Green Street, Urbana, IL 61801, USA\\
$^{8}$ Kavli Institute for Particle Astrophysics \& Cosmology, P. O. Box 2450, Stanford University, Stanford, CA 94305, USA\\
$^{9}$ SLAC National Accelerator Laboratory, Menlo Park, CA 94025, USA\\
$^{10}$ Department of Physics, Stanford University, 382 Via Pueblo Mall, Stanford, CA 94305, USA\\
$^{11}$ Kavli Institute for Cosmological Physics, University of Chicago, Chicago, IL 60637, USA\\
$^{12}$ University Observatory, Faculty of Physics, Ludwig-Maximilians-Universit\"at, Scheinerstr. 1, 81679 Munich, Germany\\
$^{13}$ Jet Propulsion Laboratory, California Institute of Technology, 4800 Oak Grove Dr., Pasadena, CA 91109, USA\\
$^{14}$ Physics Department, 2320 Chamberlin Hall, University of Wisconsin-Madison, 1150 University Avenue Madison, WI  53706-1390\\
$^{15}$ Fermi National Accelerator Laboratory, P. O. Box 500, Batavia, IL 60510, USA\\
$^{16}$ Cerro Tololo Inter-American Observatory, NSF's National Optical-Infrared Astronomy Research Laboratory, Casilla 603, La Serena, Chile\\
$^{17}$ Laborat\'orio Interinstitucional de e-Astronomia - LIneA, Av. Pastor Martin Luther King Jr, 126 Del Castilho, Nova Am\'erica Offices, Torre 3000/sala 817 CEP: 20765-000, Brazil\\
$^{18}$ Institute of Space Sciences (ICE, CSIC),  Campus UAB, Carrer de Can Magrans, s/n,  08193 Barcelona, Spain\\
$^{19}$ Physik-Institut, University of Zürich, Winterthurerstrasse 190, CH-8057 Zürich, Switzerland\\
$^{20}$ Department of Physics, Northeastern University, Boston, MA 02115, USA\\
$^{21}$ Department of Physics \& Astronomy, University College London, Gower Street, London, WC1E 6BT, UK\\
$^{22}$ Instituto de Astrofisica de Canarias, E-38205 La Laguna, Tenerife, Spain\\
$^{23}$ Universidad de La Laguna, Dpto. Astrofísica, E-38206 La Laguna, Tenerife, Spain\\
$^{24}$ Institut de F\'{\i}sica d'Altes Energies (IFAE), The Barcelona Institute of Science and Technology, Campus UAB, 08193 Bellaterra (Barcelona) Spain\\
$^{25}$ Department of Astronomy and Astrophysics, University of Chicago, Chicago, IL 60637, USA\\
$^{26}$ NASA Goddard Space Flight Center, 8800 Greenbelt Rd, Greenbelt, MD 20771, USA\\
$^{27}$ Astronomy Unit, Department of Physics, University of Trieste, via Tiepolo 11, I-34131 Trieste, Italy\\
$^{28}$ INAF-Osservatorio Astronomico di Trieste, via G. B. Tiepolo 11, I-34143 Trieste, Italy\\
$^{29}$ Institute for Fundamental Physics of the Universe, Via Beirut 2, 34014 Trieste, Italy\\
$^{30}$ Institut d'Estudis Espacials de Catalunya (IEEC), 08034 Barcelona, Spain\\
$^{31}$ School of Mathematics and Physics, University of Queensland,  Brisbane, QLD 4072, Australia\\
$^{32}$ Centro de Investigaciones Energ\'eticas, Medioambientales y Tecnol\'ogicas (CIEMAT), Madrid, Spain\\
$^{33}$ Department of Physics, IIT Hyderabad, Kandi, Telangana 502285, India\\
$^{34}$ Universit\'e Grenoble Alpes, CNRS, LPSC-IN2P3, 38000 Grenoble, France\\
$^{35}$ Instituto de Fisica Teorica UAM/CSIC, Universidad Autonoma de Madrid, 28049 Madrid, Spain\\
$^{36}$ Institute of Cosmology and Gravitation, University of Portsmouth, Portsmouth, PO1 3FX, UK\\
$^{37}$ Department of Astronomy, University of Geneva, ch. d'\'Ecogia 16, CH-1290 Versoix, Switzerland\\
$^{38}$ Santa Cruz Institute for Particle Physics, Santa Cruz, CA 95064, USA\\
$^{39}$ Center for Cosmology and Astro-Particle Physics, The Ohio State University, Columbus, OH 43210, USA\\
$^{40}$ Department of Physics, The Ohio State University, Columbus, OH 43210, USA\\
$^{41}$ Department of Physics, University of Michigan, Ann Arbor, MI 48109, USA\\
$^{42}$ Department of Astronomy/Steward Observatory, University of Arizona, 933 North Cherry Avenue, Tucson, AZ 85721-0065, USA\\
$^{43}$ Australian Astronomical Optics, Macquarie University, North Ryde, NSW 2113, Australia\\
$^{44}$ Lowell Observatory, 1400 Mars Hill Rd, Flagstaff, AZ 86001, USA\\
$^{45}$ Departamento de F\'isica Matem\'atica, Instituto de F\'isica, Universidade de S\~ao Paulo, CP 66318, S\~ao Paulo, SP, 05314-970, Brazil\\
$^{46}$ George P. and Cynthia Woods Mitchell Institute for Fundamental Physics and Astronomy, and Department of Physics and Astronomy, Texas A\&M University, College Station, TX 77843,  USA\\
$^{47}$ LPSC Grenoble - 53, Avenue des Martyrs 38026 Grenoble, France\\
$^{48}$ Instituci\'o Catalana de Recerca i Estudis Avan\c{c}ats, E-08010 Barcelona, Spain\\
$^{49}$ Max Planck Institute for Extraterrestrial Physics, Giessenbachstrasse, 85748 Garching, Germany\\
$^{50}$ Department of Physics, University of Cincinnati, Cincinnati, Ohio 45221, USA\\
$^{51}$ Perimeter Institute for Theoretical Physics, 31 Caroline St. North, Waterloo, ON N2L 2Y5, Canada\\
$^{52}$ Observat\'orio Nacional, Rua Gal. Jos\'e Cristino 77, Rio de Janeiro, RJ - 20921-400, Brazil\\
$^{53}$ Ruhr University Bochum, Faculty of Physics and Astronomy, Astronomical Institute, German Centre for Cosmological Lensing, 44780 Bochum, Germany\\
$^{54}$ Nordita, KTH Royal Institute of Technology and Stockholm University, Hannes Alfv\'ens v\"ag 12, SE-10691 Stockholm, Sweden\\
$^{55}$ Department of Physics, University of Genova and INFN, Via Dodecaneso 33, 16146, Genova, Italy\\
$^{56}$ Physics Department, Lancaster University, Lancaster, LA1 4YB, UK\\
$^{57}$ Computer Science and Mathematics Division, Oak Ridge National Laboratory, Oak Ridge, TN 37831\\
$^{58}$ Central University of Kerala, Kasaragod, Kerala, India\\
$^{59}$ Department of Astronomy, University of California, Berkeley,  501 Campbell Hall, Berkeley, CA 94720, USA\\
$^{60}$ Lawrence Berkeley National Laboratory, 1 Cyclotron Road, Berkeley, CA 94720, USA\\
$^{61}$ School of Physics and Astronomy, University of Southampton,  Southampton, SO17 1BJ, UK\\

% Don't change these lines
\bsp	% typesetting comment
\label{lastpage}
\end{document}